\documentclass[11pt]{article}
\usepackage{amsfonts}
\usepackage{amssymb}
\usepackage{amsmath}
\usepackage{amsxtra}
\usepackage{graphicx}
\usepackage{psfrag}
\usepackage{mathrsfs}
\usepackage{natbib}
\usepackage{stmaryrd}
\usepackage{subfigure}
\usepackage{tikz}
\usepackage{empheq}
\usepackage[normalem]{ulem}
\usepackage{color}

\begin{document}
\def\BGamma{\mbox{\boldmath$\Gamma$}}
\def\BDelta{\mbox{\boldmath$\Delta$}}
\def\BTheta{\mbox{\boldmath$\Theta$}}
\def\BLambda{\mbox{\boldmath$\Lambda$}}
\def\BXi{\mbox{\boldmath$\Xi$}}
\def\BPi{\mbox{\boldmath$\Pi$}}
\def\BSigma{\mbox{\boldmath$\Sigma$}}
\def\BUpsilon{\mbox{\boldmath$\Upsilon$}}
\def\BPhi{\mbox{\boldmath$\Phi$}}
\def\BPsi{\mbox{\boldmath$\Psi$}}
\def\BOmega{\mbox{\boldmath$\Omega$}}
\def\Balpha{\mbox{\boldmath$\alpha$}}
\def\Bbeta{\mbox{\boldmath$\beta$}}
\def\Bgamma{\mbox{\boldmath$\gamma$}}
\def\Bdelta{\mbox{\boldmath$\delta$}}
\def\Bepsilon{\mbox{\boldmath$\epsilon$}}
\def\Bzeta{\mbox{\boldmath$\zeta$}}
\def\Beta{\mbox{\boldmath$\eta$}}
\def\Btheta{\mbox{\boldmath$\theta$}}
\def\Biota{\mbox{\boldmath$\iota$}}
\def\Bkappa{\mbox{\boldmath$\kappa$}}
\def\Blambda{\mbox{\boldmath$\lambda$}}
\def\Bmu{\mbox{\boldmath$\mu$}}
\def\Bnu{\mbox{\boldmath$\nu$}}
\def\Bxi{\mbox{\boldmath$\xi$}}
\def\Bpi{\mbox{\boldmath$\pi$}}
\def\Brho{\mbox{\boldmath$\rho$}}
\def\Bsigma{\mbox{\boldmath$\sigma$}}
\def\Bepsilon{\mbox{\boldmath$\epsilon$}}
\def\Btau{\mbox{\boldmath$\tau$}}
\def\Bupsilon{\mbox{\boldmath$\upsilon$}}
\def\Bphi{\mbox{\boldmath$\phi$}}
\def\Bchi{\mbox{\boldmath$\chi$}}
\def\Bpsi{\mbox{\boldmath$\psi$}}
\def\Bomega{\mbox{\boldmath$\omega$}}
\def\Bvarepsilon{\mbox{\boldmath$\varepsilon$}}
\def\Bvartheta{\mbox{\boldmath$\vartheta$}}
\def\Bvarpi{\mbox{\boldmath$\varpi$}}
\def\Bvarrho{\mbox{\boldmath$\varrho$}}
\def\Bvarsigma{\mbox{\boldmath$\varsigma$}}
\def\Bvarphi{\mbox{\boldmath$\varphi$}}
\def\bone{\mbox{\boldmath$1$}}
\def\bzero{\mbox{\boldmath$0$}}
\def\bnabla{\mbox{\boldmath$\nabla$}}
\def\bvarepsilon{\mbox{\boldmath$\varepsilon$}}
\def\bA{\mbox{\boldmath$ A$}}
\def\bB{\mbox{\boldmath$ B$}}
\def\bC{\mbox{\boldmath$ C$}}
\def\bD{\mbox{\boldmath$ D$}}
\def\bE{\mbox{\boldmath$ E$}}
\def\bF{\mbox{\boldmath$ F$}}
\def\bG{\mbox{\boldmath$ G$}}
\def\bH{\mbox{\boldmath$ H$}}
\def\bI{\mbox{\boldmath$ I$}}
\def\bJ{\mbox{\boldmath$ J$}}
\def\bK{\mbox{\boldmath$ K$}}
\def\bL{\mbox{\boldmath$ L$}}
\def\bM{\mbox{\boldmath$ M$}}
\def\bN{\mbox{\boldmath$ N$}}
\def\bO{\mbox{\boldmath$ O$}}
\def\bP{\mbox{\boldmath$ P$}}
\def\bQ{\mbox{\boldmath$ Q$}}
\def\bR{\mbox{\boldmath$ R$}}
\def\bS{\mbox{\boldmath$ S$}}
\def\bT{\mbox{\boldmath$ T$}}
\def\bU{\mbox{\boldmath$ U$}}
\def\bV{\mbox{\boldmath$ V$}}
\def\bW{\mbox{\boldmath$ W$}}
\def\bX{\mbox{\boldmath$ X$}}
\def\bY{\mbox{\boldmath$ Y$}}
\def\bZ{\mbox{\boldmath$ Z$}}
\def\ba{\mbox{\boldmath$ a$}}
\def\bb{\mbox{\boldmath$ b$}}
\def\bc{\mbox{\boldmath$ c$}}
\def\bd{\mbox{\boldmath$ d$}}
\def\be{\mbox{\boldmath$ e$}}
\def\bff{\mbox{\boldmath$ f$}}
\def\bg{\mbox{\boldmath$ g$}}
\def\bh{\mbox{\boldmath$ h$}}
\def\bi{\mbox{\boldmath$ i$}}
\def\bj{\mbox{\boldmath$ j$}}
\def\bk{\mbox{\boldmath$ k$}}
\def\bl{\mbox{\boldmath$ l$}}
\def\bm{\mbox{\boldmath$ m$}}
\def\bn{\mbox{\boldmath$ n$}}
\def\bo{\mbox{\boldmath$ o$}}
\def\bp{\mbox{\boldmath$ p$}}
\def\bq{\mbox{\boldmath$ q$}}
\def\br{\mbox{\boldmath$ r$}}
\def\bs{\mbox{\boldmath$ s$}}
\def\bt{\mbox{\boldmath$ t$}}
\def\bu{\mbox{\boldmath$ u$}}
\def\bv{\mbox{\boldmath$ v$}}
\def\bw{\mbox{\boldmath$ w$}}
\def\bx{\mbox{\boldmath$ x$}}
\def\by{\mbox{\boldmath$ y$}}
\def\bz{\mbox{\boldmath$ z$}}
\newcommand*\mycirc[1]{%
  \begin{tikzpicture}
    \node[draw,circle,inner sep=1pt] {#1};
  \end{tikzpicture}
}

\title{A three dimensional field formulation, and isogeometric solutions to point and line defects using Toupin's theory of gradient elasticity at finite strains}
\author{Z.~Wang\thanks{Department of Mechanical Engineering}, S.~Rudraraju\thanks{Department of Mechanical Engineering} \& K.~Garikipati\thanks{Departments of Mechanical Engineering, and Mathematics. Corresponding Author, \tt{krishna@umich.edu}} \\ University of Michigan, Ann Arbor}
\maketitle
\abstract{We present a field formulation for defects that draws from the classical representation of the cores as force dipoles. We write these dipoles as singular distributions. Exploiting the key insight that the variational setting is the only appropriate one for the theory of distributions, we arrive at universally applicable weak forms for  defects in nonlinear elasticity. Remarkably, the standard, Galerkin finite element method yields numerical solutions for the elastic fields of defects that, when parameterized suitably, match very well with classical, linearized elasticity solutions. The true potential of our approach, however, lies in its easy extension to generate solutions to elastic fields of defects in the regime of nonlinear elasticity, and even more notably for Toupin's theory of gradient elasticity at finite strains (\emph{Arch. Rat. Mech. Anal.}, \textbf{11}, 385, 1962). In computing these solutions we adopt recent numerical work on an isogeometric analytic framework that enabled the first three-dimensional solutions to general boundary value problems of Toupin's theory (Rudraraju et al. \emph{Comp. Meth. App. Mech. Engr.}, \textbf{278}, 705, 2014). We first present exhaustive solutions to point defects, edge and screw dislocations, and a study on the energetics of interacting dislocations. Then, to demonstrate the generality and potential of our treatment, we apply it to other complex dislocation configurations, including loops and low-angle grain boundaries.}

%
%
\section{Introduction}
\label{sec:introduction} 
The elastic fields of crystal defects have proven difficult to represent. As is well appreciated, the large distortion in the core places the elasticity problem in the nonlinear (i.e., finite strain) regime. Furthermore, due to the very non-uniform distortion in the core, measures that go beyond the deformation gradient exert a magnified influence. Generalized \citep{Cosserats1909}, higher-order \citep{Toupin1962} and nonlocal \citep{EringenEdelen1972} theories of elasticity that also preserve geometric nonlinearities therefore provide the appropriate treatment. Here, we work with Toupin's theory of gradient elasticity at finite strains \citep{Toupin1962} motivated primarily by the emergence of strain gradients as second-order terms in the Taylor series expansion of the elastic free energy density \citep{Rudrarajuetal2014}.

A number of different approaches have been adopted to obtain analytic solutions to classical (i.e., without higher-order measures of deformation), nonlinearly elastic, boundary value problems of defect configurations. \cite{Rosakis1988} developed anti-plane shear field solutions to screw dislocations in nonlinear, homogeneous, isotropic, incompressible solids. Their work defines a stress-based measure of nonlinearity, motivated by the differences in total stored energy that arise, depending on the chosen strain energy function. \cite{Zubov2004} developed a treatment of distributed dislocation and disclination densities at finite strains by invoking internal degrees of freedom and couple stresses. Analytic solutions were obtained in the setting of plane elasticity. Working with distributed defects within the setting of differential geometry, \cite{YavariGoriely2012} described analytic solutions to a number of screw and edge dislocation configurations, finding, in some cases, that the singularity depends on the material model. Later, these authors extended their analysis to point defects \citep{YavariGoriely2012a}. By representing the cores with a continuous distribution of defects, the authors avoided the singularities that emerge in the classical linear solutions. Using a continuum theory of geometrically necessary dislocations, \cite{Acharya2001} obtained analytic solutions to screw dislocations in a neo-Hookean solid. Working with the continuum theory of dislocations based on a weak definition of the dislocation density tensor, \cite{Acharya2004} subsequently developed singularity-free field solutions in a partial differential equation framework. Subsequently, the author presented a statistical mechanics-based thermodynamic theory of dislocations, also using the dislocation density tensor  \citep{Acharya2011}. A number of other significant works have been cited in the above references; we are constrained only by space from listing more of them here.

Significant as the above works are, there remains room for a numerical treatment of the problem of defect fields in classical, nonlinear elasticity, especially in finite bodies, and for more complex defect configurations. To develop such a framework is a goal of our work, but more crucially, we seek a framework that can be extended to gradient elasticity at finite strains as a means to regularize the core fields. This is not a straightforward task as we aim to argue.

Such solutions that do exist of higher-order elasticity theories applied to defect structures are restricted to a linearized treatment. Typical, are those based on Mindlin's formulation of linearized gradient elasticity \citep{Mindlin1964}, further reduced to lower dimensions and simplified boundary value problems. There are a number of such treatments, of which we point the reader to \cite{GutkinAifantis1999,LazarMaugin2005}. Also see \cite{Kessel1970,Lazaretal2005,Lazaretal2006} for applications of generalized theories of elasticity to defect fields in the linear regime. Higher-order elasticity theories, because of their complexity, have continued to resist general solution in three dimensions. This challenge has kept even numerical, field solutions out of reach to higher-order, finite strain elasticity treatments of defects in general, three-dimensional boundary value problems.

The present work relies on a numerical framework, developed in \cite{Rudrarajuetal2014}, which enables solutions of Toupin's gradient elasticity theory at finite strain. It is based on isogeometric analysis \citep{HughesCottrellBazilevs2005,CottrellHughesBazilevs2009} and, in particular, exploits the ease of developing spline basis functions with arbitrary degree of continuity in this framework. This is critical because, in the weak form of strain gradient elasticity, second order spatial derivatives appear on the trial solutions as well as the weighting functions, requiring functions that lie in $\mathscr{H}^2$. This requirement is satisfied by $C^1$-continuous spline functions. The framework in \cite{Rudrarajuetal2014} demonstrated the first three-dimensional solutions to general boundary value problems of Toupin's theory of gradient elasticity at finite strains. We seek here to exploit this numerical treatment and extend the catalogue of three-dimensional solutions of Toupin's theory to include defect fields.

Classical descriptions of defect fields, such as the Volterra dislocation \citep{Volterra1907}, are based on a representation of the defect field away from the core (the far-field) that incorporates a displacement discontinuity at the core. Absent a direct model of the core, the elastic fields are not accurate near the core. A different approach is to represent the core by dipole arrangements of forces, which model the tractions experienced by the atoms immediately surrounding the core. Such a model has been employed for point defects to derive the Green's function solution \citep{HirthLothe1982}. For dislocations, the potential-based methods of \cite{Eshelby1953} have been applied to determine the linear elastic fields in \cite{Gehlenetal1972} and \cite{HirthLothe1973} by modelling the force dipole as an ellipsoidal center of expansion. In \cite{SinclairHirth1978} lattice Green's functions were used in conjunction with the analytic expressions of Eshelby and co-workers to couple the far-field elasticity solution with the core field obtained from molecular statics. More recently, this class of approaches was extended to compute the elastic energies of dislocations by parameterizing the linear elastic fields against \emph{ab initio} calculations \citep{Clouet2011,Clouetetal2011}. These works rely on analytic expressions or series expansions of the elastic fields. Here, we aim to incorporate dipole force representations within large scale, mesh-based numerical methods such as finite element or isogeometric methods. The key insight that allows us to realize this goal is that dipole force arrangements have a singular distributional character. In fact, dipole force arrangements give rise to singular dipole distributions. Furthermore, the variational setting is the only correct one for singular distributions. Therefore, variationally based numerical techniques are particularly well-suited to compute the fields resulting from the singular dipole distributions. This holds for finite element and isogeometric methods, with the only distinguishing feature being that isogeometric methods ease the representation of $C^1$-functions for strain gradient elasticity.

We first review Toupin's theory of gradient elasticity at finite strain in Section \ref{sec:sgElast}. This is followed by a derivation of the singular distributional representation for dipole arrangements of forces in Section \ref{sec:math}. The numerical framework, consisting of isogeometric analysis and including quadrature rules, appears in Section \ref{sec:numerical}. An extensive set of numerical results for the elastic fields of point and line defects is presented in Section \ref{sec:Numerical Simulations}. We conclude by placing our work in perspective in Section \ref{concl}.

%
%
\section{Toupin's theory of strain gradient elasticity at finite strains}
\label{sec:sgElast}

\subsection{Weak and strong forms}
Our treatment is posed in the Cartesian coordinate system, with basis vectors $\be_i$, $i = 1,\dots 3$, $\be_i\cdot\be_j = \delta_{ij}$. The reference configuration, its boundary and the surface normal at any boundary point are denoted by $\Omega_0$, $\partial \Omega_0$ and $\bN$, respectively, with $\vert\bN\vert = 1$. The corresponding entities in the current configuration are denoted by $\Omega$, $\partial \Omega$ and $\bn$, respectively. We work mostly with coordinate notation. Upper case subscript indices are used to denote the components of vectors and tensors in the reference configuration and lower case subscript indices are reserved for those in the current configuration. Working in the reference configuration, we consider the boundary to be the union of a finite number of smooth surfaces $\Gamma_0$, smooth edges $\Upsilon_0$ and corners $\Xi_0$: $\partial \Omega_0 = \Gamma_0 \cup \Upsilon_0 \cup \Xi_0$, for full generality.  For functions defined on $\partial \Omega_0$, when necessary, the gradient operator is decomposed into the normal gradient operator $D$ and the surface gradient operator $D_{K}$, 
\begin{align}
\psi_{,K} &= D \psi N_{K} + D_{K} \psi\nonumber\\
\textrm{where}\quad D \psi N_{K} &= \psi_{,I} N_{I}N_{K}\;\textrm{and}\; D_{K} \psi = \psi_{,K} - \psi_{,I} N_{I}N_{K}
\label{surfacenormalgradient}
\end{align}

\noindent A material point is denoted by $\bX\, \in\, \Omega_0$. The deformation map between $\Omega_0$ and $\Omega$ is given by $\Bvarphi(\bX,t) =  \bX + \bu = \bx$, where $\bu$ is the displacement field. The deformation gradient is $\bF = \partial\Bvarphi/\partial\bX = \bone + \partial\bu/\partial\bX$, which in coordinate notation is expressed as $F_{iJ} = \partial \varphi_{i} / \partial X_{J} = \delta_{iJ} + \partial u_{i}/\partial X_{J}$. The Green-Lagrange strain tensor is given in coordinate notation by $E_{IJ} = \frac{1}{2}(F_{kI}F_{kJ} - \delta_{IJ})$. 

The strain energy density function is $W(\bE,\textrm{Grad}\bE)$. We recall that the dependence on $\bE$ and $\textrm{Grad}\bE$ renders $W$ a frame invariant elastic free energy density function for materials of grade two \citep{Toupin1962}. Constitutive relations follow for the first Piola-Kirchhoff stress and the higher-order stress tensors:

\begin{align}
P_{iJ} &= \frac{\partial W}{\partial F_{iJ}}\label{eqn:stressP}\\
B_{iJK} &= \frac{\partial W}{\partial F_{iJ,K}}\label{eqn:stressB}
\end{align}

We note that instead of Equations (\ref{eqn:stressP}) and (\ref{eqn:stressB}) the single stress tensor that combines them could be used as in \cite{Toupin1964}. Our treatment \citep{Rudrarajuetal2014, Rudrarajuetal2015} has been based on \cite{Toupin1962}, and in a future communication we will consider a comparison of the two approaches for the representation of defects. One  obtains the same governing equations, constitutive prescriptions and jump conditions across internal interfaces as in \cite{Toupin1964} with a strain energy density function of the form $W(\bE,\textrm{Grad}\bE)$, but no higher-order stress in the theory, and a global statement of the second law. See \cite{Fressengeas2015} in this regard, where the authors recover the formulation of \cite{Toupin1964} with couple stress, but not higher-order stress. 

In the most general case, we have a body force distribution $\bff$, a surface traction $\bT$, a surface moment $\bM$ and a line force $\bL$. For $i = 1,2,3$ denoting the Cartesian coordinates, the smooth surfaces of the boundary are decomposed as $\Gamma_0= \Gamma_{0^i}^u \cup  \Gamma_{0^i}^T = \Gamma_{0^i}^m \cup  \Gamma_{0^i}^M$, and the smooth edges of the boundary are decomposed as $\Upsilon_0= \Upsilon_{0^i}^l \cup ~\Upsilon_{0^i}^L$. Here, Dirichlet boundary subsets are identified by superscripts $u, m ~\textrm{and} ~l$ and Neumann boundary subsets are identified by superscripts $T, M ~\textrm{and} ~L$. 

We begin with the weak form of the problem. We seek a displacement field of the form
\begin{equation}
u_i \in \mathscr{S}, \;\textrm{with}\; u_i = \bar{u}_i\;\mathrm{on}\; \Gamma_{0^i}^u;\quad u_i = \bar{l}_i;\mathrm{on}\; \Upsilon_{0^i}^l;\quad Du_i = \bar{m}_i;\mathrm{on}\; \Gamma_{0^i}^m
\label{dirbcsu}
\end{equation}

\noindent such that for all variations of the form 
\begin{equation}
 w_i\in\mathscr{V}, \;\textrm{with}\;w_i = 0 ;\mathrm{on}\;  \Gamma_{0^i}^u \cup  \Upsilon_{0^i}^l, ~Dw_i = 0 ;\mathrm{on}\;  \Gamma_{0^i}^m 
\label{dirbcsw}
\end{equation}

\noindent the following equation holds:
\begin{equation}
  \int_{\Omega_0} \left( w_{i,J} P_{iJ} +  w_{i,JK} B_{iJK}  \right) ~\mathrm{d}V -\int_{\Omega_0}w_i f_i ~\mathrm{d}V - \int_{\Gamma_{0^i}^T} w_i T_i \, ~\mathrm{d}S  - \int_{\Gamma_{0^i}^M} Dw_i M_i \, ~\mathrm{d}S  - \int_{\Upsilon_{0^i}^L} w_i L_i \, \mathrm{d}C = 0.
\label{eqn:weakform}
\end{equation}

\noindent The problem has a fourth-order character, which resides in products of $B_{iJK}$ and $w_{i,JK}$, each of which involves second-order spatial gradients. Standard variational arguments lead to the strong form of the problem:
\begin{equation}
\begin{array}{lll}
P_{iJ,J} - B_{iJK,JK} + f_i &= 0 &\mathrm{in} ~\Omega_0\\
u_{i}  &= \bar{u}_i   &\mathrm{on} ~\Gamma_{0^i}^u\\
P_{iJ}N_J - DB_{iJK}N_KN_J - 2D_J(B_{iJK})N_K - B_{iJK}D_JN_K + (b^L_LN_JN_K-b_{JK})B_{iJK} & = T_{i} &\mathrm{on} ~\Gamma_{0^i}^T\\
Du_i  &= \bar{m}_i &\mathrm{on} ~\Gamma_{0^i}^m\\
B_{iJK}N_JN_K &= M_{i} &\mathrm{on}  ~\Gamma_{0^i}^M\\
u_{i}  &= \bar{l}_i  &\mathrm{on} ~\Upsilon_{0^i}^l\\
\llbracket N^{\Gamma}_{J} N_{K} B_{iJK} \rrbracket &= L_{i} &\text{\small on} ~\Upsilon_{0^i}^L\\ \\
\end{array}
\label{eqn:strongformgradelasticity}
\end{equation}

Here, $b_{IJ}=-D_{I}N_J=-D_{J}N_I$ are components of the second fundamental form of the smooth parts of the boundary and $\bN^\Gamma = \BXi\times\bN$, where $\BXi$ is the unit tangent to the curve $\Upsilon_0$ \citep{Toupin1962}. If $\Upsilon_0$ is a curve separating smooth surfaces $\Gamma_0^+ \subset \Gamma_0$ and $\Gamma_0^-\subset \Gamma_0$,  with $\bN^{\Gamma^+}$ being the unit outward normal to $\Upsilon_0$ from $\Gamma_0^+$ and $\bN^{\Gamma^-}$ being the unit outward normal to $\Upsilon_0$ from $\Gamma_0^-$ we define $\llbracket N^{\Gamma}_{J} N_{K} B_{iJK} \rrbracket := N^{\Gamma^+}_{J} N_{K} B_{iJK} + N^{\Gamma^-}_{J} N_{K} B_{iJK}$. The (nonlinear) fourth-order nature of the governing partial differential equation above is now visible in the term $B_{iJK,JK}$, which introduces $F_{aB,CJK}$ via Equation \eqref{eqn:stressB}. The Dirichlet boundary condition in \eqref{eqn:strongformgradelasticity}$_2$ has the same form as for conventional elasticity. However, its dual Neumann boundary condition \eqref{eqn:strongformgradelasticity}$_3$ is notably more complex than its conventional counterpart, which would have only the first term on the left hand-side. Equation \eqref{eqn:strongformgradelasticity}$_4$ is the higher-order Dirichlet boundary condition applied to the normal gradient of the displacement field, and Equation \eqref{eqn:strongformgradelasticity}$_5$ is the higher-order Neumann boundary condition on the higher-order stress, $\bB$. Adopting the physical interpretation of $\bB$ as a couple stress \citep{Toupin1962}, the homogeneous form of this boundary condition, if extended to the atomic scale, states that there is no boundary mechanism to impose a generalized moment across atomic bonds. Finally, Equation \eqref{eqn:strongformgradelasticity}$_6$ is the Dirichlet boundary condition on the smooth edges of the boundary and Equation \eqref{eqn:strongformgradelasticity}$_7$ is its conjugate Neumann boundary condition. Following \cite{Toupin1962}, the homogeneous form of this condition requires that there be no discontinuity in the higher order (couple) stress traction across a smooth edge $\Upsilon_0^L$ in the absence of a balancing line traction along $\Upsilon_0^L$. In \cite{Rudrarajuetal2014} we have detailed the variational treatment leading to Equations (\ref{eqn:weakform}) and (\ref{eqn:strongformgradelasticity}), as well as the corresponding statements in the current configuration.

\section{Representation of defects as force dipole distributions}
\label{sec:math}
\subsection{Point defects}
\label{sec:Pointdipole}
In $\Omega_0$, consider a discrete dipole formed of force vectors $-\bR$ and $\bR$ located at distinct points $\bX^\prime_{1}$ and $\bX^\prime_{2}$, respectively. The force field is a singular distribution. While the theory of distributions, rigorously applied, holds in a variational setting, it permits a formal representation as classical functions \citep{Stakgold1979}. This allows a manipulation of the force fields as follows:
\begin{align}
\bff(\bX;\bX^\prime_1,\bX^\prime_2)=\bR\delta^3(\bX;\bX^\prime_2)-\bR\delta^3(\bX;\bX^\prime_1),
\label{eq:discreteForces}
\end{align}
\noindent where $\delta^3(\bX;\bX^\prime)$ is the three-dimensional Dirac delta distribution. Expanding it in a Taylor's series around $\bY^\prime=(\bX^\prime_1+\bX^\prime_2)/2$.
\begin{align}
\bR\delta^3(\bX;\bX^\prime_1)=\bR[\delta^3(\bX;\bY^\prime)]+\bR\left(\frac{\partial}{\partial\bX^\prime}\delta^3(\bX;\bX^\prime)\right)\bigg\vert_{\bY^\prime}\cdot(\bX^\prime_1-\bY^\prime)+\mathcal{O}(\vert\bX^\prime_1-\bY^\prime\vert^2)\\
\bR\delta^3(\bX;\bX^\prime_2)=\bR[\delta^3(\bX;\bY^\prime)]+\bR\left(\frac{\partial}{\partial\bX^\prime}\delta^3(\bX;\bX^\prime)\right)\bigg\vert_{\bY^\prime}\cdot(\bX^\prime_2-\bY^\prime)+\mathcal{O}(\vert\bX^\prime_2-\bY^\prime\vert^2),
\end{align}
\noindent To first order, therefore, the force field is a dipole distribution
\begin{align}
\bff(\bX;\bY^\prime)=\bR\left(\frac{\partial}{\partial\bX^\prime}\delta^3(\bX;\bX^{'})\right)\bigg\vert_{\bY^\prime}\cdot(\bX^\prime_2-\bX^\prime_1).
\end{align}
\noindent Letting $\Bxi = \bX^\prime_2 - \bX^\prime_1$ for brevity, we have, in coordinate notation, 
\begin{align}
f_{i}=R_{i}\frac{\partial\delta^3(\bX;\bX^\prime)}{\partial X_{J}^{'}}\bigg\vert_{\bY^\prime}\xi_J
\end{align}
\noindent We define the dipole tensor,
 \begin{align}
 \bD=\bR\otimes\Bxi,
 \label{dipoleTensor}
 \end{align}
 \noindent and use $\partial\delta^3({\bX;\bX^\prime})/\partial\bX=-\partial\delta^3({\bX;\bX^\prime})/\partial\bX^\prime$ to write the force distribution,
\begin{align}
\bff(\bX;\bY^\prime)=-\bD\frac{\partial\delta^3(\bX;\bY^\prime)}{\partial\bX}
\label{eq:dipoleForceDistPointDefect}
\end{align}
\begin{figure}[hbtp]
\centering
\includegraphics[scale=0.5]{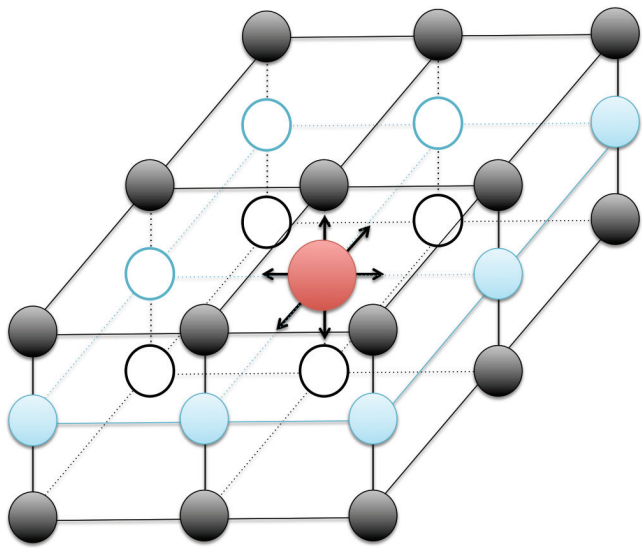}
\caption{A dipole of forces representing a substitional point defect.}
\label{fig:dipolePointDefect}
\end{figure}
A direct application of the above development is the representation of a point defect as a center of expansion or contraction. In this case, the dipole is a diagonal tensor (Figure \ref{fig:dipolePointDefect}). Using the orthonormal Euclidean basis $\{\be_1,\be_2,\be_3\}$, we have
\begin{equation}
\bD = \sum\limits_{i,I = 1}^3R_i \xi_I \be_i\otimes\be_I,
\label{eq:dipoleTensorPointDefect}
\end{equation}
\noindent where $R_1 \neq R_2 \neq R_3$ and $\xi_1 \neq \xi_2 \neq \xi_3$ to model anisotropic point defects.

\subsection{Line defects}
\label{sec:Surface dipole}
\begin{figure}[hbtp]
\centering
\includegraphics[scale=0.5]{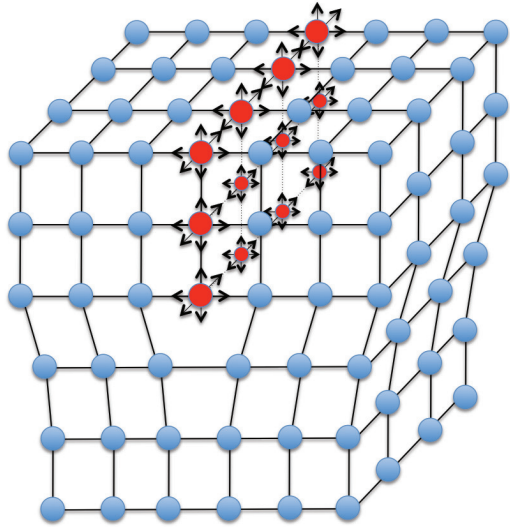}
\caption{An edge dislocation represented as an interfacial distribution of force dipoles.}
\label{fig:dislocationDipole}
\end{figure}

We retrace the above development from Equation (\ref{eq:discreteForces}) through (\ref{eq:dipoleForceDistPointDefect}), but with interfacial force densities $-\bR^\mathrm{i}$ and $\bR^\mathrm{i}$ that are uniform over parallel planes $\Gamma_1^\prime$ and $\Gamma_2^\prime$, respectively. In this case, we arrive at relations with the same form as Equations (\ref{eq:dipoleForceDistPointDefect}) and (\ref{eq:dipoleTensorPointDefect}). However,  the one-dimensional Dirac-delta distribution, $\delta^1(\bX;\bX^\prime)$ replaces the three-dimensional Dirac-delta distribution in the expression for the force distribution,
\begin{align}
\bff(\bX;\bY^\prime)=-\bD^\mathrm{i}\frac{\partial\delta^1(\bX;\bY^\prime)}{\partial \bX}.
\label{eq:dipoleForceDistLineDefect}
\end{align}

\noindent  The resulting interfacial distribution of force dipoles is shown in Figure \ref{fig:dislocationDipole} for an edge dislocation, where the dipole tensor is defined as
\begin{equation}
\bD^\mathrm{i} = \sum\limits_{i,I = 1}^3R^\mathrm{i}_i \xi_I \be_i\otimes\be_I.
\label{eq:dipoleTensorLineDefect}
\end{equation}
\noindent Here, $\Bxi$ denotes the vector pointing from $\Gamma^\prime_1$ toward $\Gamma^\prime_2$, $\bR^\mathrm{i} = \alpha\Bxi$ for edge dislocations, and $\bR^\mathrm{i}\cdot\Bxi = 0$ for screw dislocations.

\subsection{Linearized elasticity solutions for the dipole tensor}
\label{sec:linElastDipoleTensor}

In the limit of infinitesimal elasticity it can be shown that the dipole tensor of a point defect in an infinite domain is \citep{Garikipatietal2006}.
\begin{equation}
D_{IJ} = \mathbb{C}_{IJKL} V^\mathrm{r}_{KL},
\label{eq:dipoleTensorRelaxVol}
\end{equation}

\noindent where $\bV^\mathrm{r}$ is the relaxation volume tensor of the point defect. Note that coordinate notation in this section uses uppercase superscript indices because of the coincidence of $\Omega$ and $\Omega_0$ in the infinitesimal limit.

A related result is possible for line defects by invoking Volterra's linearized elasticity solution for the displacement field of a dislocation in an infinite domain:
\begin{equation}
u^{\infty}_{M}(\bX^\prime)=-\int_{\Gamma^\prime}\frac{\partial G_{MK}(\bX;\bX^\prime)}{\partial X_{L}}C_{IJKL}N_{I}b_{J}\mathrm{d}S
\label{eq:volterrasSoln}
\end{equation}
\noindent where $G_{MK} = G_{KM}$ is the infinite space Green's function for elasticity, and gives the displacement in the $\be_M$ direction for a unit force in the $\be_K$ direction. The surface of integration is the half plane of the line defect, $\bN$ is the unit normal to $\Gamma^\prime$ and $\bb$ is the Burgers vector of the dislocation. The definition of the Green's function implies that, $u^\infty_M$ also can be written using the force distribution introduced in (\ref{eq:dipoleForceDistLineDefect}).
\begin{align}
u^{\infty}_{M}(\bX^\prime)=-\int_{\Omega}G_{MK}(\bX;\bX^\prime)f_{K}(\bX,\bX^{\prime\prime})\mathrm{d}V
\end{align}
Using Equations (\ref{eq:dipoleForceDistLineDefect}) and (\ref{eq:dipoleTensorLineDefect}), and a standard result on the gradient of the Dirac-delta distribution
\begin{align}
u^{\infty}_{M}(\bX^\prime)&=-\int_{\Omega}G_{MK}(\bX;\bX^\prime)(-D^\mathrm{i}_{KL}\frac{\partial\delta^1(\bX;\bX^{\prime\prime})}{\partial X_{L}})\mathrm{d}V\nonumber\\
                     &=-\int_{\Omega}\frac{\partial G_{MK}(\bX;\bX^\prime)}{\partial X_{L}}(D^\mathrm{i}_{KL}\delta^1(\bX;\bX^{\prime\prime}))\mathrm{d}V\nonumber\\
                     &=-\int_{\Gamma^{\prime\prime}}\frac{\partial G_{MK}(\bX;\bX^\prime)}{\partial X_{L}}\bigg\vert_{\bX=\bX^{\prime\prime}}D^\mathrm{i}_{KL}\mathrm{d}S^{\prime\prime}
                     \label{eq:dispFromDipole}
\end{align}
Comparing Equations (\ref{eq:volterrasSoln}) and (\ref{eq:dispFromDipole}) we obtain
\begin{align}
D^\mathrm{i}_{KL}=C_{IJKL}N_{I}b_{J}
\label{eq:dipleStrength}
\end{align}
\\
Using $C_{IJKL}=\lambda\delta_{IJ}\delta_{KL} + \mu(\delta_{IK}\delta_{JL}+\delta_{IL}\delta_{JK})$ in the isotropic case, and $\bb = \beta\bN$ for an edge dislocation, we have
\begin{align}
D^\mathrm{ie}_{KL}=\lambda\delta_{KL}N_{i}b_{i} + \mu(N_K b_L + N_L b_K).
\label{eq:dipoleTensorEdgeDisloc}
\end{align}
\noindent Using $\bb\cdot\bN =0$ for a screw dislocation, leads to
\begin{align}
D^\mathrm{is}_{KL}=\mu(N_K b_L + N_L b_K).
\label{eq:dipoleTensorScrewDisloc}
\end{align}

Equations (\ref{eq:dipoleTensorRelaxVol}), (\ref{eq:dipoleTensorEdgeDisloc}) and (\ref{eq:dipoleTensorScrewDisloc}) allow a direct comparison of our approach with linearized elastic fields for the corresponding defects. In the remainder of this communication, we will continue to use these relations as approximations to the respective point and line defect dipole tensors outside the linearized elastic regime of infinitesimal strain; i.e., for nonlinear elasticity with and without gradient effects at finite strains. In drawing conclusions at the end of the manuscript, we outline our approaches for better estimates of these defect dipole tensors.

\subsection{A field formulation for defects in gradient elasticity at finite strains}
\label{sec:defectsForceDistribSGElast}

On substituting Equation (\ref{eq:dipoleForceDistPointDefect}) and (\ref{eq:dipoleForceDistLineDefect}) in the weak form (\ref{eqn:weakform}), we have, for point defects:
\begin{align}
  \int_{\Omega_0} \left( P_{iJ} w_{i,J} +  B_{iJK} w_{i,JK} \right) ~\mathrm{d}V +\int_{\Omega_0}w_i(\bX) D_{iJ} \frac{\partial\delta^3(\bX;\bY^\prime)}{\partial X_J}~\mathrm{d}V -&\nonumber\\
   \int_{\Gamma_{0^i}^T} w_i T_i \, ~\mathrm{d}S  - \int_{\Gamma_{0^i}^M} Dw_i M_i \, ~\mathrm{d}S  - \int_{\Upsilon_{0^i}^L} w_i L_i \, \mathrm{d}C &= 0,
   \end{align}
\noindent and for line defects,
   \begin{align}
   \int_{\Omega_0} \left( P_{iJ} w_{i,J} +  B_{iJK} w_{i,JK} \right) ~\mathrm{d}V +\int_{\Omega_0}w_i(\bX) D^\mathrm{i}_{iJ} \frac{\partial\delta^1(\bX;\bY^\prime)}{\partial X_J}~\mathrm{d}V -&\nonumber\\ \int_{\Gamma_{0^i}^T} w_i T_i \, ~\mathrm{d}S  - \int_{\Gamma_{0^i}^M} Dw_i M_i \, ~\mathrm{d}S  - \int_{\Upsilon_{0^i}^L} w_i L_i \, \mathrm{d}C &= 0.
\end{align}
On again using the result for the gradient of distributions, and the fundamental definition of the Dirac-delta distribution these simplify to

\begin{align}
  \int_{\Omega_0} \left( P_{iJ} w_{i,J} +  B_{iJK} w_{i,JK} \right) ~\mathrm{d}V  -w_{i,J}(\bY^\prime) D_{iJ}~\mathrm -&\nonumber\\
   \int_{\Gamma_{0^i}^T} w_i T_i \, ~\mathrm{d}S  - \int_{\Gamma_{0^i}^M} Dw_i M_i \, ~\mathrm{d}S  - \int_{\Upsilon_{0^i}^L} w_i L_i \, \mathrm{d}C &= 0,\label{eq:weakformDipolePointDefect}
   \end{align}
\noindent for point defects, and   
   \begin{align}
   \int_{\Omega_0} \left( P_{iJ} w_{i,J} +  B_{iJK} w_{i,JK} \right) ~\mathrm{d}V -\int_{\Gamma^\prime}w_{i,J} D^\mathrm{i}_{iJ} ~\mathrm{d}S -&\nonumber\\
    \int_{\Gamma_{0^i}^T} w_i T_i \, ~\mathrm{d}S  - \int_{\Gamma_{0^i}^M} Dw_i M_i \, ~\mathrm{d}S  - \int_{\Upsilon_{0^i}^L} w_i L_i \, \mathrm{d}C &= 0
   \label{eq:weakformDipoleLineDefect}
\end{align}
for line defects, with $D_{iJ}$ given by (\ref{eq:dipoleTensorRelaxVol}), and $D^\mathrm{i}_{iJ}$ given by(\ref{eq:dipoleTensorEdgeDisloc}) or (\ref{eq:dipoleTensorScrewDisloc}).

It bears emphasizing that the rather transparent form of (\ref{eq:weakformDipolePointDefect}) and (\ref{eq:weakformDipoleLineDefect}) is a direct consequence of the variational nature of the theory of distributions. The singular, Dirac-delta and dipole distributions are meaningful only in the variational setting. With a weak form at hand, their actions are simply transferred to the variations---or the test functions, $\bw$. We proceed to show that this leads to a surprisingly effective representation of point and line defect fields using variationally based numerical methods.

%
%
\section{Numerical treatment}
\label{sec:numerical}
\subsection{Galerkin formulation}
\label{sec:galerkin}

As always, the Galerkin weak form is obtained by restriction to finite dimensional functions $(\bullet)^h$: Find $u^h_i \in \mathscr{S}^h \subset \mathscr{S}$, where $\mathscr{S}^h= \{ u^h_i \in \mathscr{H}^2(\Omega_0) ~\vert  ~u^h_{i} = ~\bar{u}_i\; \mathrm{on}\;  \Gamma_{0^i}^u, ~Du^h_i = ~\bar{m}_i \;\mathrm{on}\;  \Gamma_{0^i}^m, ~u^h_{i} = ~\bar{l}_i  \mathrm{on}\;  \Upsilon_{0^i}^l \}$,  such that $\forall ~w^h_i \in \mathscr{V}^h \subset \mathscr{V}$, where $\mathscr{V}^h= \{ w^h_i \in\mathscr{H}^2(\Omega_0)~\vert  ~w^h_{i} = ~0 \;\mathrm{on}\;  \Gamma_{0^i}^u \cup \Upsilon_{0^i}^l, ~Dw^h_i = 0\; \mathrm{on}\; \Gamma_{0^i}^m\}$
\begin{align}
  \int_{\Omega_0} \left( P^h_{iJ} w^h_{i,J} +  B^h_{iJK} w^h_{i,JK} \right) ~\mathrm{d}V -\phantom{\int}w^h_{i,J} D_{iJ}\Big\vert_{\bY^\prime} \phantom{\mathrm{d}V}&\nonumber\\
   -\int_{\Gamma_{0^i}^T} w^h_i T_i \, ~\mathrm{d}S  - \int_{\Gamma_{0^i}^M} Dw^h_i M_i \, ~\mathrm{d}S  - \int_{\Upsilon_{0^i}^L} w^h_i L_i \, \mathrm{d}C &= 0,
\label{eqn:galerkinformPointDefect}
\end{align}
for point defect fields, and 
\begin{align}
  \int_{\Omega_0} \left( P^h_{iJ} w^h_{i,J} +  B^h_{iJK} w^h_{i,JK} \right) ~\mathrm{d}V -\int_{\Gamma^\prime}w^h_{i,J} D^\mathrm{i}_{iJ} ~\mathrm{d}S &\nonumber\\
   -\int_{\Gamma_{0^i}^T} w^h_i T_i \, ~\mathrm{d}S  - \int_{\Gamma_{0^i}^M} Dw^h_i M_i \, ~\mathrm{d}S  - \int_{\Upsilon_{0^i}^L} w^h_i L_i \, \mathrm{d}C & = 0,
\label{eqn:galerkinformLineDefect}
\end{align}
for line defect fields.

Our adoption of Toupin's theory of gradient elasticity at finite strains is motivated by an interest in obtaining fields that are accurate at large strains, while remaining singularity-free through the core. As is well-appreciated, the second-order gradients in the weak form require the solutions to lie in $\mathscr{H}^2(\Omega_0)$, a more restrictive condition than the formulation of finite strain elasticity for materials of grade one, where the solutions are drawn from the larger space $\mathscr{H}^1(\Omega_0) \supset \mathscr{H}^2(\Omega_0)$. The variations, $\bw^h$ and trial solutions $\bu^h$ are defined component-wise using a finite number of basis functions,
\begin{equation}
\bw^h = \sum_{a=1}^{n_\mathrm{b}} \bc^a N^a, \quad \qquad \bu^h = \sum_{a=1}^{n_\mathrm{b}} \bd^a N^a 
\label{eq:basisdef}
\end{equation}
\noindent where $n_\mathrm{b}$ is the dimensionality of the function spaces $\mathscr{S}^h$ and $\mathscr{V}^h$, and $N^a$ represents the basis functions. Since $\mathscr{S}^h \subset \mathscr{H}^2$, $C^0$ basis functions do not provide the degree of regularity demanded by the problem; however, it suffices to consider $C^1$ basis functions in $\mathscr{S}^h$. One possibility is the use of $C^1$ Hermite elements as in \cite{Papanicolopulos2009}. Alternately, one could invoke the class of continuous/discontinuous Galerkin methods \cite{Engeletal2002, Wellsetal2004, Molarietal2006, WellsKuhlGarikipati2006}, in which the displacement field is $C^0$-continuous, but the strains are discontinuous  across element interfaces.  A mixed formulation of finite strain gradient elasticity could be constructed by introducing an independent kinematic field for the deformation gradient or another strain measure. These last two approaches, however, incur additional stability requirements. We prefer to avoid the complexities of Hermite elements in three dimensions, and seek to circumvent the challenges posed by discontinuous Galerkin methods and mixed formulations by turning to Isogeometric Analysis introduced by \citet{HughesCottrellBazilevs2005}. Also see \citet{CottrellHughesBazilevs2009} for details.\\

\subsubsection{Isogeometric Analysis}
\label{sec:iga}
As is now well-appreciated in the computational mechanics community, Isogeomeric Analysis (IGA) is a mesh-based numerical method with NURBS (Non-Uniform Rational B-Splines) basis functions. The NURBS basis leads to many desirable properties, chief among them being the exact representation of the problem geometry. Like the Lagrange polynomial basis functions traditionally used in the Finite Element Method (FEM), the NURBS basis functions are partitions of unity with compact support, satisfy affine covariance (i.e an affine transformation of the basis is obtained by the affine transformation of its nodes/control points) and support an isoparametric formulation, thereby making them suitable for a Galerkin framework. They enjoy advantages over Lagrange polynomial basis functions in being able to ensure $C^n$-continuity, in possessing the positive basis and convex hull properties, and being variation diminishing. A detailed discussion of the NURBS basis and IGA is beyond the scope of this article and interested readers are referred to \citet{CottrellHughesBazilevs2009}. However, we briefly present the construction of the basis functions. \\

The building blocks of the NURBS basis functions are univariate B-spline functions that are defined as follows: Consider two positive integers $p$ and $n$, and a non-decreasing sequence of values $\chi=[\xi_1, \xi_2,...., \xi_{n+p+1}]$, where p is the polynomial order, n is the number of basis functions, the $\xi_i$ are coordinates in the parametric space referred to as knots (equivalent to nodes in FEM) and $\chi$ is the knot vector. The B-spline basis functions $B_{i,p}(\xi)$ are defined starting with the zeroth order basis functions
\begin{align}
B^i_0(\xi) &= \left\{\begin{array}{ll}
1 &\mathrm{if}\;\xi_i \le \xi < \xi_{i+1},\\
0 &\mathrm{otherwise}
\end{array}\right.
\end{align}
and using the Cox-de Boor recursive formula for $p \geq 1$ \citep{Piegl1997}
\begin{align}
 B^i_p (\xi) &=
  \frac{\xi-\xi_i}{\xi_{i+p}-\xi_i} B^i_{p-1} (\xi) + \frac{\xi_{i+p+1}-\xi}{\xi_{i+p+1}-\xi_{i+1}} B^{i+1}_{p-1} (\xi)
\end{align}

\begin{figure}[hbt]
\centering
\includegraphics[width=0.6\textwidth]{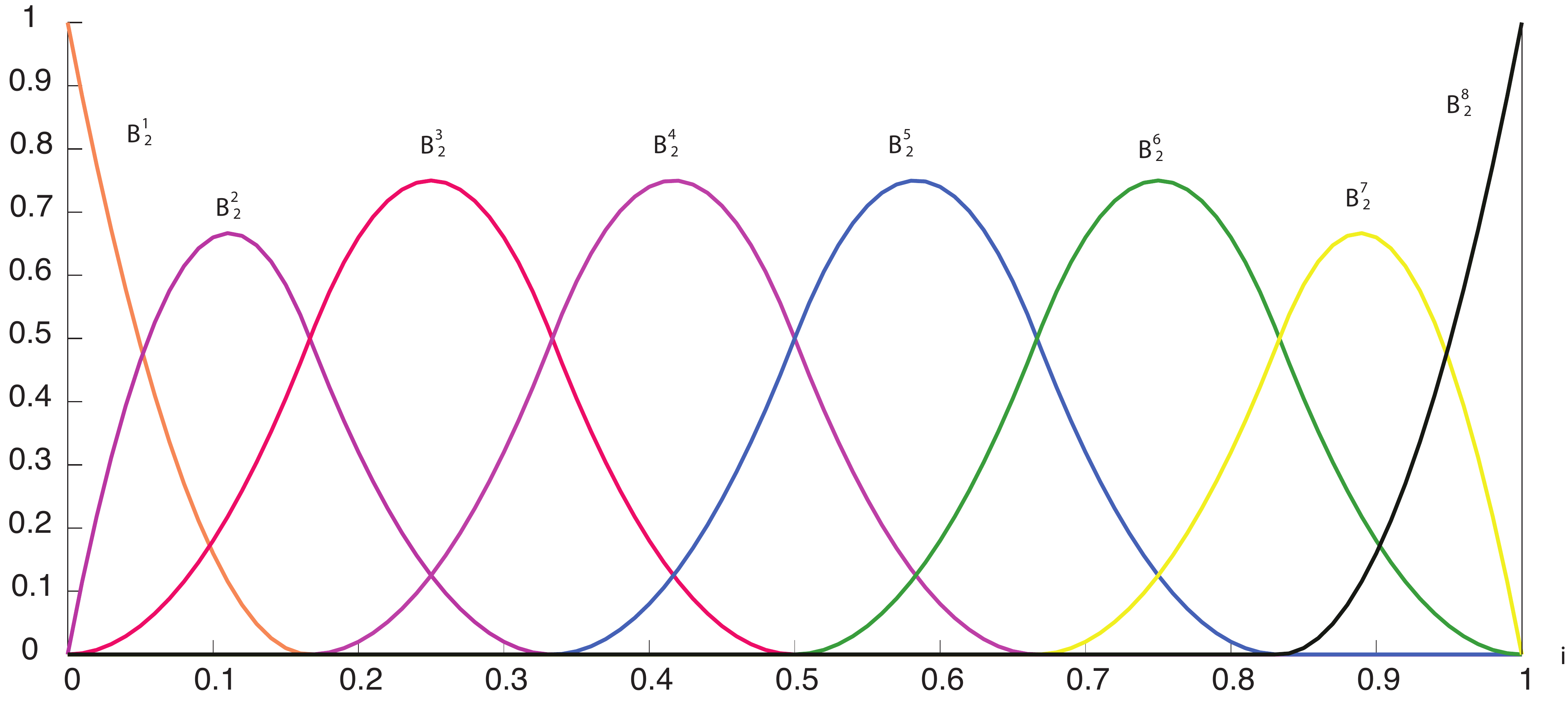}
\caption{A quadratic B-spline basis constructed from the knot vector $\chi = [0, 0, 0, 1/6, 1/3, 1/2, 2/3, 5/6, 1, 1, 1]$.}
\label{fig:bsplines}
\end{figure}

The knot vector divides the parametric space into intervals referred to as knot spans (equivalent to elements in FEM). A B-spline basis function is $C^{\infty}$-continuous inside knot spans and $C^{p-1}$-continuous at the knots. If an interior knot value repeats, it is referred to as a multiple knot. At a knot of multiplicity $k$, the continuity is $C^{p-k}$. Now, using a quadratic B-spline basis (Figure (\ref{fig:bsplines})), a $C^1$-continuous one dimensional NURBS basis can be constructed. \footnote{The boundary value problems that follow in Section \ref{sec:Numerical Simulations} consider only simple geometries. For this reason, we have used the simpler B-spline basis functions instead of the NURBS basis. However, we have included the latter in this discussion for the sake of completeness, noting that the numerical formulation as presented is valid for any single-patch NURBS geometry.}
\begin{align}
N^{i}_p (\xi) =
  \frac{B^i_p (\xi) \textit{w}_{i}}{\sum_{i=1}^{n_b} B^i_p (\xi) \textit{w}_{i}}
  \label{eq:onednurbs}
\end{align}
where $w_i$ are the weights associated with each of the B-spline functions. In higher dimensions, NURBS basis functions are constructed as a tensor product of the one dimensional basis functions:
\begin{align}
 N^{ij}_p (\xi,\eta) &=
  \frac{B^i_p (\xi) B^j_p (\eta) \textit{w}_{ij}}{\sum_{i=1}^{n_{b1}} \sum_{j=1}^{n_{b2}} B^i_p(\xi) B^j_p (\eta) \textit{w}_{ij}} &\mathrm{(2D)} \label{eq:higherordernurbs2D} \\
 N^{ijk}_p (\xi,\eta, \zeta) &=
  \frac{B^i_p (\xi) B^j_p (\eta) B^k_p (\zeta) \textit{w}_{ijk}}{\sum_{i=1}^{n_{b1}} \sum_{j=1}^{n_{b2}} \sum_{k=1}^{n_{b3}} B^i_p(\xi) B^j_p (\eta) B^k_2 (\zeta) \textit{w}_{ijk}} &\mathrm{(3D)}
\label{eq:higherordernurbs3D}
\end{align}

\subsection{Numerical integration of singular force distributions}
\label{sec:planeintegration}
\begin{figure}[hbtp]
\centering
\includegraphics[width=0.6\textwidth]{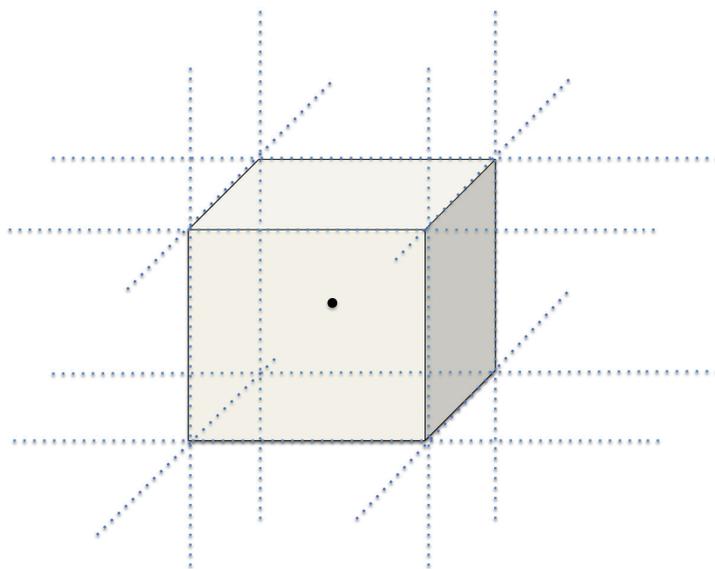}
\caption{Quadrature for a point defect.}
\label{fig:quadPointDefect}
\end{figure}

\begin{figure}[hbtp]
\centering
\includegraphics[width=0.6\textwidth]{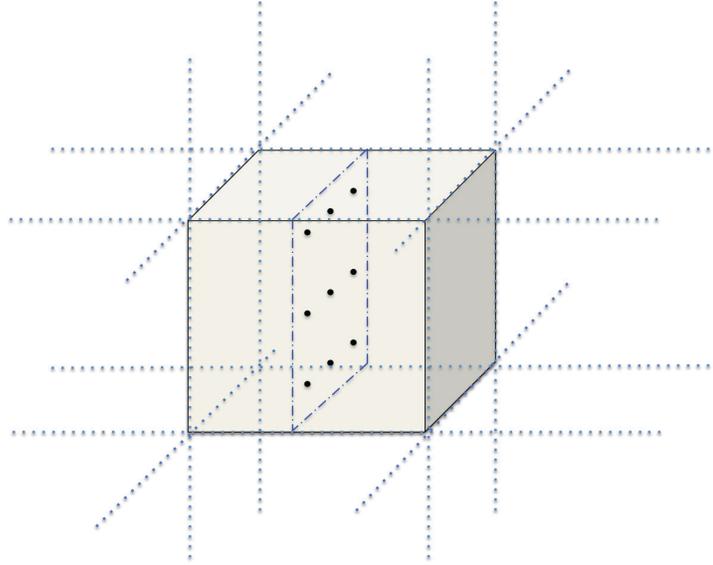}
\caption{Quadrature points on a dislocation plane.}
\label{fig:quadDislocation}
\end{figure}

From the theory of distributions, the forcing term is applied at the point defect in equation (\ref{eqn:galerkinformPointDefect}) and along the plane of the dislocation in equation (\ref{eqn:galerkinformLineDefect}). Special quadrature points must be introduced to numerically integrate these terms. For the point defect in Equation (\ref{eqn:galerkinformPointDefect}) this is accomplished with a single quadrature point as shown in Figure (\ref{fig:quadPointDefect}) and Equation (\ref{eqn:quadPointDefect}).
\begin{align}
w^h_{i,J}D_{iJ}\Big\vert_{\bY^\prime} =\sum_{a=1}^{n_\mathrm{b}}w_i^aN^a_{,J}\Big\vert_{\bY^\prime}D_{iJ}
\label{eqn:quadPointDefect}
\end{align}
\noindent The scheme for integration along the dislocation plane in Equation (\ref{eqn:galerkinformLineDefect}) is similar to that for Neumann boundary conditions. An internal surface of quadrature points is introduced on the plane as shown in Figure \ref{fig:quadDislocation} and Equation (\ref{eqn:quadDislocation}). Two-dimensional Gaussian quadrature is found to be sufficient since the B-spline-derived functions are polynomials. The forcing term is,
\begin{align}
\int_{\Gamma^\prime}w^h_{i,J} D^\mathrm{i}_{iJ}dS=\sum_{q=1}^{n_q}\sum_{a=1}^{n_\mathrm{b}} w_i^a N^a_{,J}(\zeta_q)D^\mathrm{i}_{iJ}(\zeta_q)\varpi_q=\sum_{a=1}^{n_\mathrm{b}}w_i^a\sum_{q=1}^{n_q}  N^a_{,J}(\zeta_q)D^\mathrm{i}_{iJ}(\zeta_q)\varpi_q
\label{eqn:quadDislocation}
\end{align}

\noindent where $n_q$ is the number of quadrature points on the plane $\Gamma^\prime$, and $\zeta_q$ and $\varpi_q$ are coordinates and weights for the corresponding quadrature points.




%
%
\section{Numerical results}
\label{sec:Numerical Simulations}
For the full generality of Toupin's theory, we consider an elastic free energy density function $W$, that incorporates gradient effects at finite strain. As is well-known, material frame invariance is guaranteed by requiring $W$ to be a function of the Green-Lagrange strain tensor, $\bE = \frac{1}{2}(\bF^\mathrm{T}\bF - \bone)$, and its gradient. We choose a simple extension of the St. Venant-Kirchhoff function by a quadratic term in $\mathrm{Grad}\bE$, and write it in coordinate notation:
\begin{align}
W=(\bE,\mathrm{Grad}\bE)=\frac{\lambda}{2}(E_{AA})^{2}+\mu(E_{AB}E_{AB})+\frac{1}{2}\mu l^{2}E_{AB,C}E_{AB,C},
\end{align}
\noindent where $l$ is a gradient length scale parameter. The Lam\'{e} parameters   have been normalized so that $\lambda=1$ and $\mu=1$. All computations were carried out on the unit cube $(0,1)^3$, relative to which the magnitude of the Burgers vector $b$,, and gradient length scale $l$ have been normalized.

To validate the dipole representation of defects, we first simplify our model to linearized elasticity (infinitesimal strain) and ignore the gradient effects. This enables a comparison between the classical, analytical solutions in this regime and numerical solutions for the point defect, edge dislocation and screw dislocation. We also include in this study, comparisons with solutions obtained with gradient elasticity at finite strains, which demonstrate the expected regularization of displacement and stress fields. Following this establishment of the fundamental solution characteristics, we present a study of the strain energy of a single dislocation, and of two interacting dislocations in the setting of classical, linearized elasticity as well as gradient elasticity at finite strains. Finally, to demonstrate the potential for extension to more complex defect configurations, we compute the solutions for an edge dislocation near a traction-free surface, a non-planar dislocation loop, and a grain boundary represented by an arrangement of edge dislocations.

\subsection{Comparison of analytic and numerical solutions for point defects}
\label{sec:point defect} 

In the regime of linearized elasticity, the analytical solution to the displacement field around a point defect in an infinite medium is \citep{HirthLothe1982}
\begin{equation}
\begin{array}{rcl}
u_1^\mathrm{an}=\frac{D}{4\pi(\lambda+2\mu)}\frac{X_1}{(X_1^2+X_2^2+X_3^2)^{3/2}} \\
u_2^\mathrm{an}=\frac{D}{4\pi(\lambda+2\mu)}\frac{X_2}{(X_1^2+X_2^2+X_3^2)^{3/2}}  \\
u_3^\mathrm{an}=\frac{D}{4\pi(\lambda+2\mu)}\frac{X_3}{(X_1^2+X_2^2+X_3^2)^{3/2}}
\end{array}
\end{equation}
 where the dipole tensor is $\bD = D\bone$, with the normalization $D=1\times 10^{-6}$. To relate $\bD$ to the relaxation volume tensor, we have from (\ref{eq:dipoleTensorRelaxVol}) in the isotropic case, 
 \begin{align}
 D_{IJ}=\lambda\delta_{IJ} V^\mathrm{r}_{KK} + \mu( V^\mathrm{r}_{IJ}+ V^\mathrm{r}_{JI})
 \end{align}
In this 
 \begin{align}
 V^\mathrm{r}_{IJ}=\frac{1}{3\lambda+2\mu}D\delta_{IJ}
 \end{align}
All our numerical solutions are over a unit cube $\Omega_0 = (0,1)^3$. For the point defect, we consider an interstitial located at $\bX^\prime = \{0.5, 0.5, 0.5\}^\mathrm{T}$. For consistency, we apply the analytical solution on the boundary surfaces $\Gamma_0$ as Dirichlet boundary conditions. The weak form (\ref{eq:weakformDipolePointDefect}) is then re-written as:
 \begin{align}
 \int_{\Omega_0}w_{I,J}\sigma_{IJ}dV-w_{i,i}\Big\vert_{\bX^\prime}D&=0\\
 \bu&=\bu^\mathrm{an} \quad \mathrm{on} \;\Gamma_0.
 \end{align}

\begin{figure}[hbtp]
\centering
\includegraphics[scale=0.5]{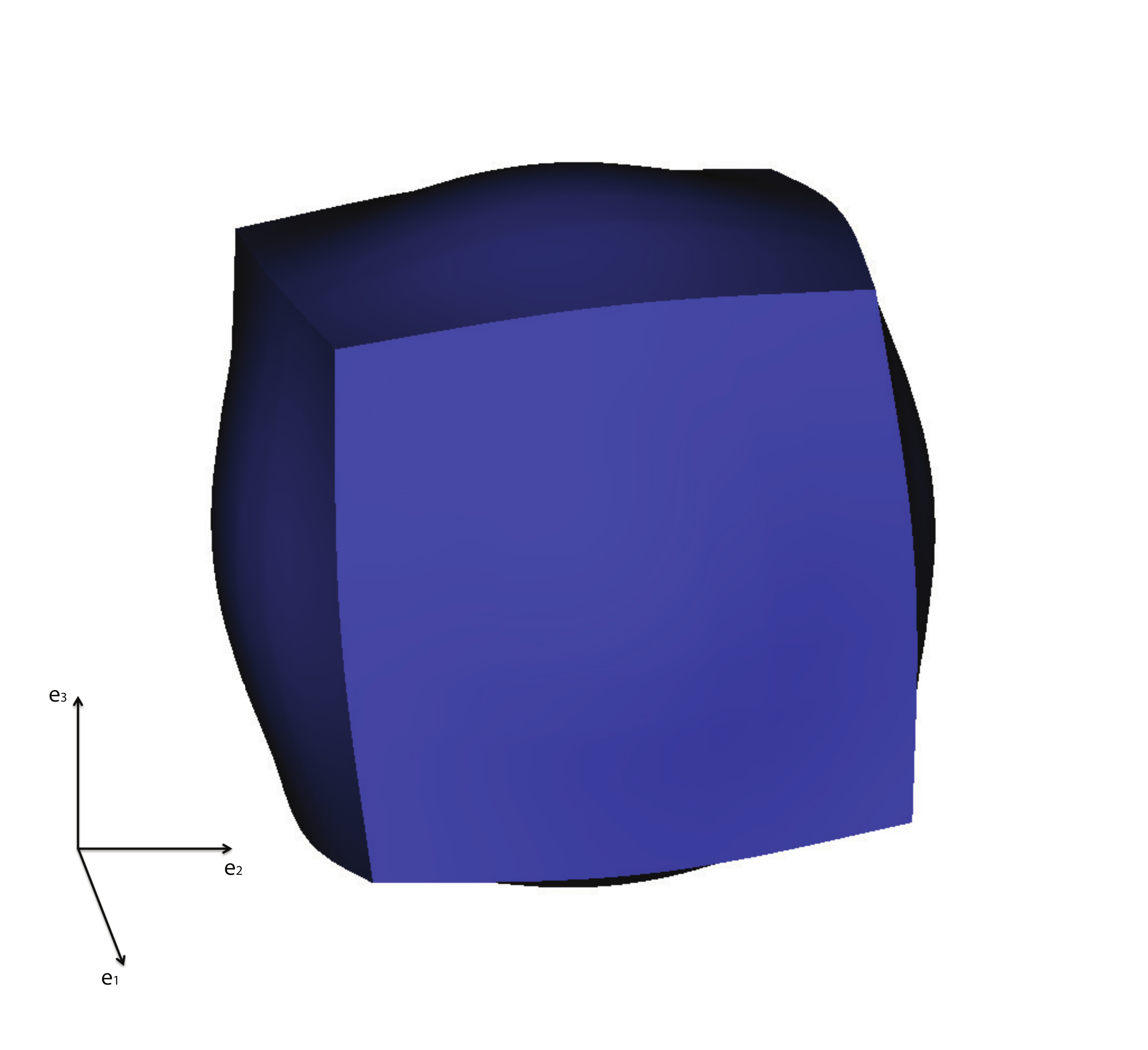}
\caption{The deformed configuration of a solid around an interstitial point defect.}
\label{fig:pointDefect}
\end{figure}

Numerical solutions are obtained using isogeometric analysis as described in Section \ref{sec:numerical}. Figure \ref{fig:pointDefect} is the deformed configuration of the cube around the interstitial. Displacements have been scaled by a factor of $2\times 10^6$ for ease of visualization.

\begin{figure}[hbtp]
\centering
\includegraphics[scale=0.5]{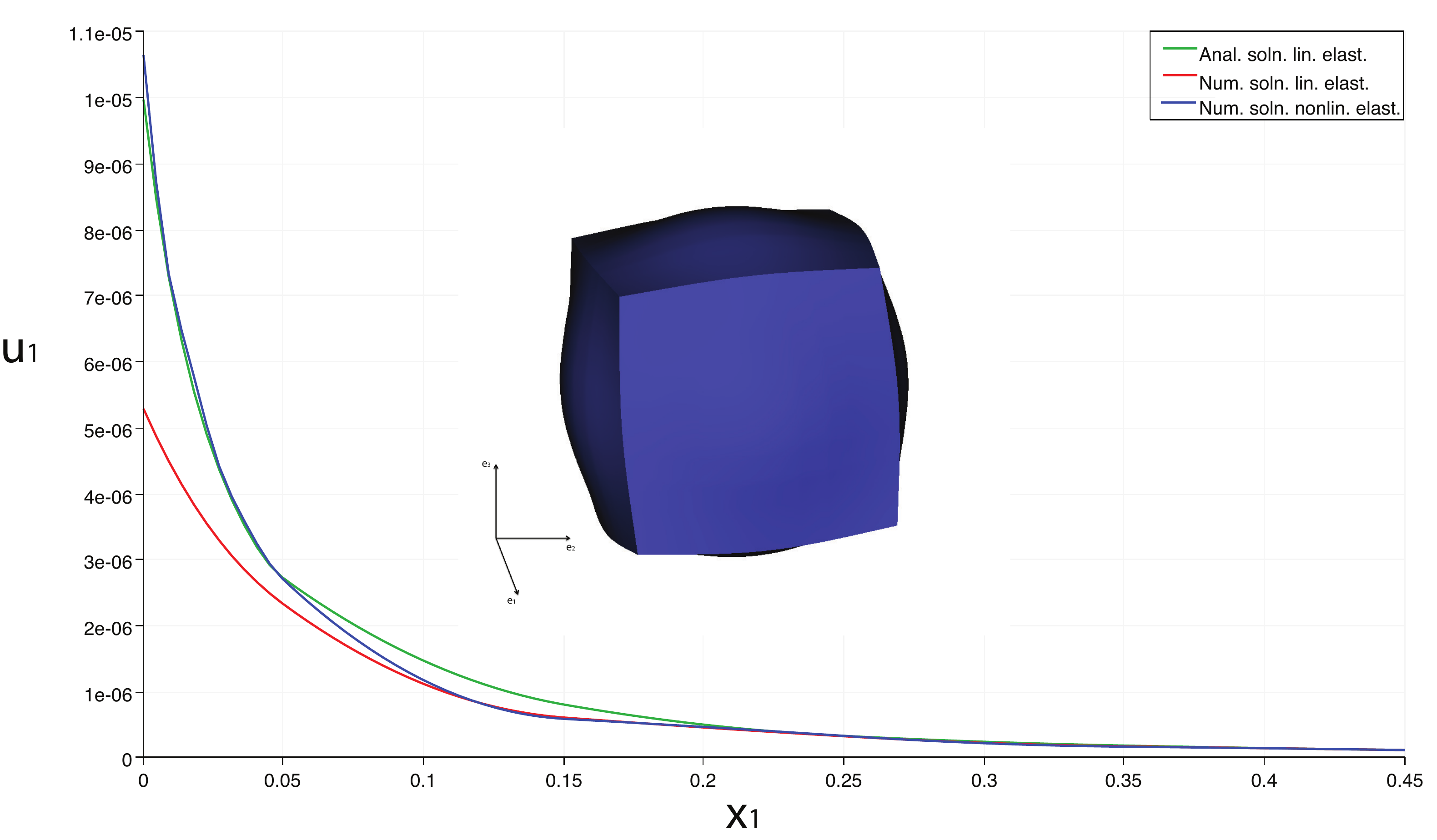}
\caption{Point defect, $u_1$ field along the segment from $\bX = \{0.55,0.5,0.5\}$ to $\bX = \{1,0.5,0.5\}$. The core is located at $\bX = \{0.5,0.5,0.5\}$. The gradient elasticity result was obtained for finite strain with $l = 0.05$.}
\label{fig:point1}
\end{figure}

Figure \ref{fig:point1} shows the $u_1$ displacement component along a segment starting outside the core, to avoid the singularity that exists in the linearized elasticity solution. Note the close agreement between the analytic and numerical solutions for linearized elasticity. This is our first demonstration of the viability of the proposed approach to model defects via dipole tensors, and exploit the native, variational structure of the theory of distributions. More such demonstrations follow in the coming sections. Also shown is the regularization of the displacement field obtained with the full, gradient theory of elasticity at finite strains. Note the significantly more gentle increase in $\vert u_1\vert$ approaching the core.

\begin{figure}[hbtp]
\centering
\includegraphics[scale=0.5]{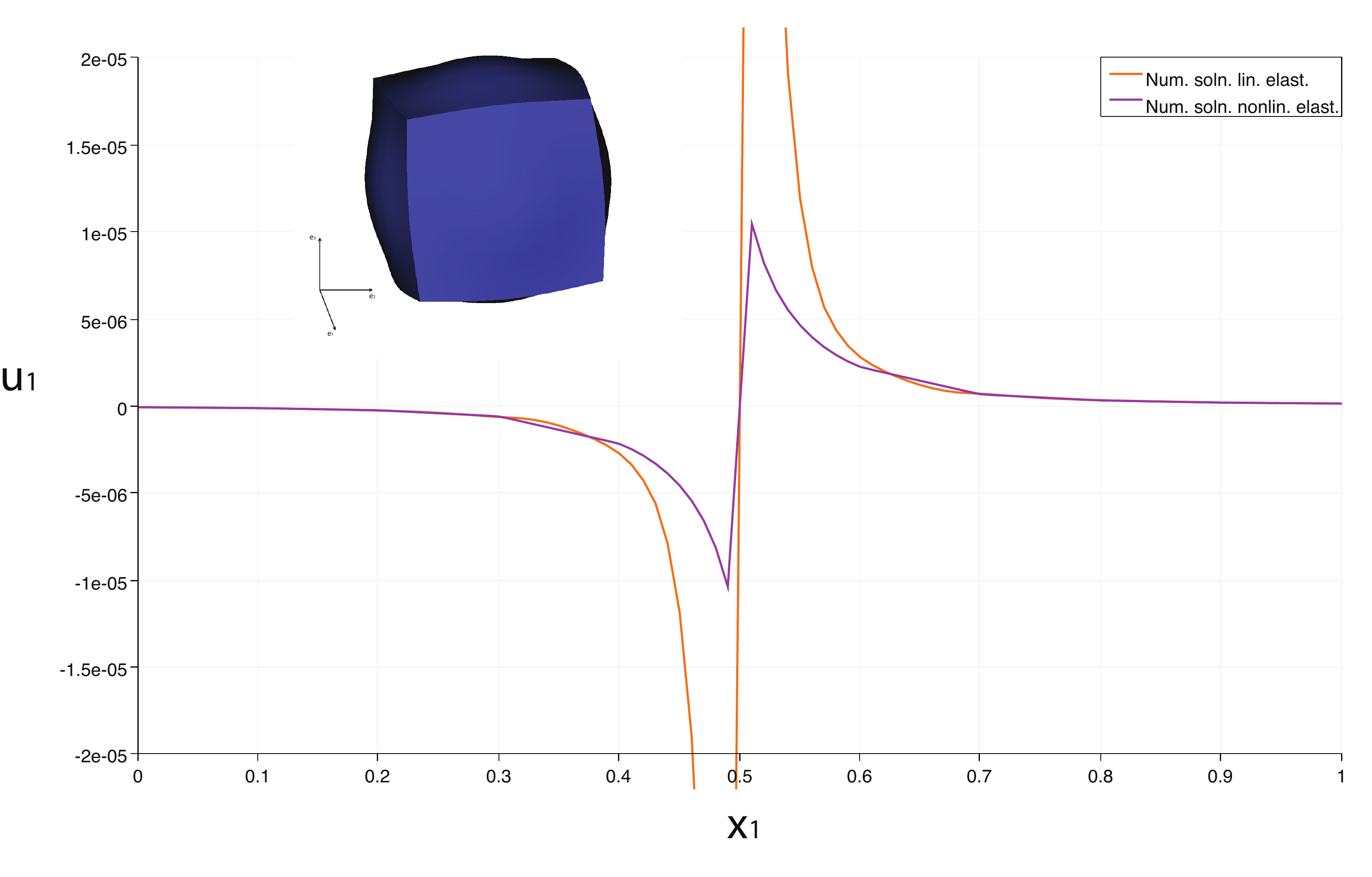}
\caption{The $u_1$ displacement field of a point defect along the segment from $\bX = \{0,0.5,0.5\}$ to $\bX = \{1,0.5,0.5\}$. The analytical field diverges at the core. The gradient elasticity result was obtained for finite strain with $l = 0.1$.}
\label{fig:point2}
\end{figure}

Figure \ref{fig:point2} shows the solution, $u_1$ through the core, highlighting the completely regularized gradient elastic solution at finite strain in comparison with the diverging, singular analytic solution.



\begin{figure}[hbtp]
\centering
\includegraphics[scale=0.75]{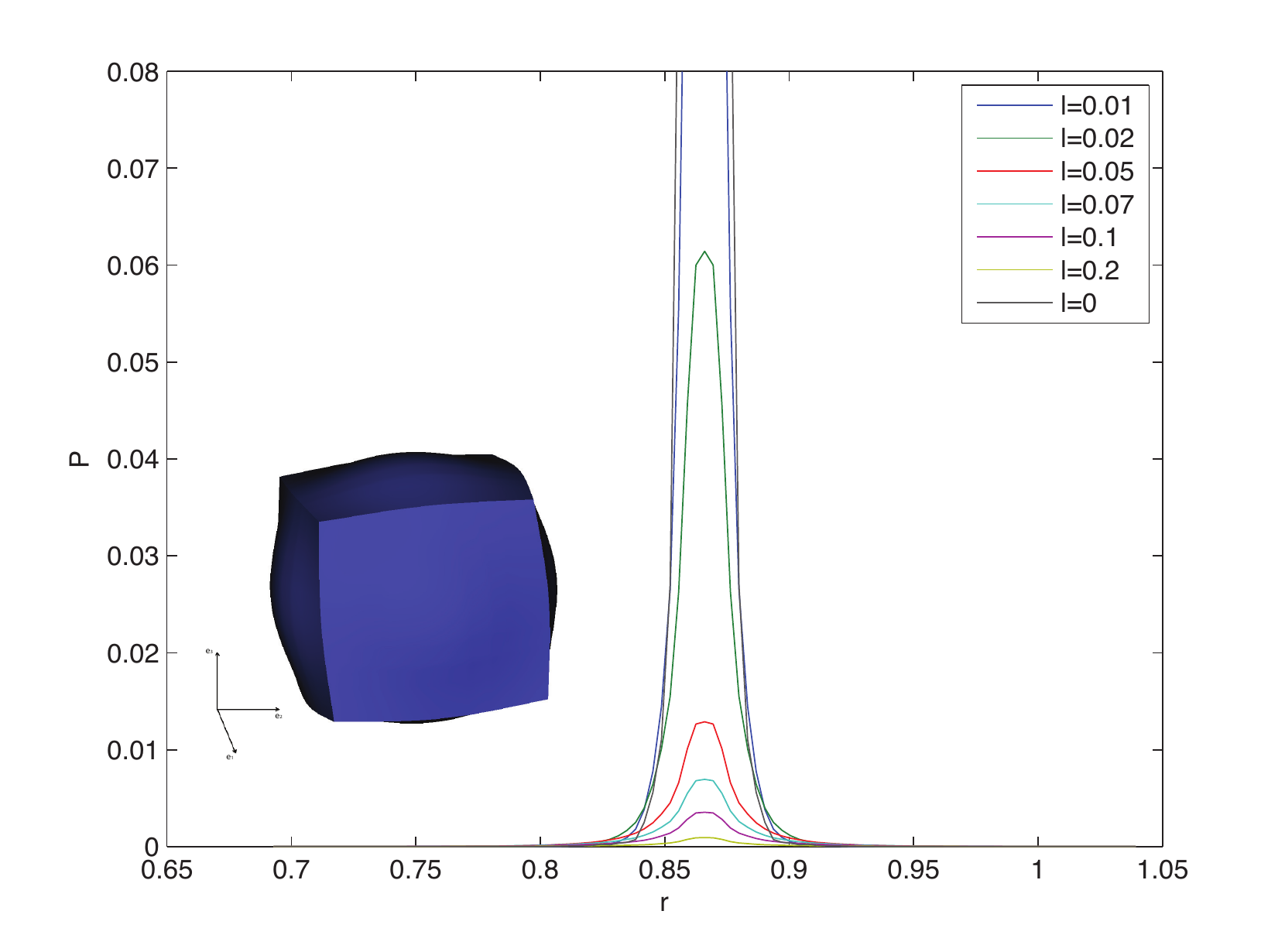}
\caption{The $\mathrm{tr}[\bP]$ stress field of a point defect along the segment from $\bX = \{0.4,0.4,0.4\}$ to $\bX = \{0.6,0.6,0.6\}$. The classical, non-gradient, finite strain elasticity solution diverges, in comparison with the regularization obtained with gradient elasticity at finite strain.}
\label{fig:point3}
\end{figure}

Figure \ref{fig:point3} shows the trace of the first Piola-Kirchhoff stress tensor through the core of the point defect. Note the divergence of the classical, non-gradient, finite strain elasticity solution, in comparison with the regularization attained with even a very small gradient elastic length scale parameter, $l$. We also draw attention to the progressively lower variation in the stress field as the gradient elastic effect is strengthened. 

Figures \ref{fig:point1}-\ref{fig:point3} demonstrate the verstaility of our numerical approach to representing defect fields via dipole tensors in the variational, distributional setting. To the best of our knowledge, these figures also show the first numerical solutions to point defect fields with Toupin's theory of gradient elasticity at finite strains.

\clearpage
\subsection{Comparison of analytic and numerical solutions for the edge dislocation}
\label{sec:edge dislocation} 

Consider an edge dislocation in the unit cube $\Omega_0 = (0,1)^3$, with half plane $\Gamma^\prime$ a subset of the $X_{2}-X_{3}$ plane. The dislocation line and core are aligned with $\be_3$, lie at $X_2 = 0.5$ and have Burgers vector $\bb = b\be_1$. We recall the analytic solution in the regime of linearized elasticity for such an edge dislocation in an isotropic, infinite medium \citep{HirthLothe1982}:
\begin{align}
u_1^\mathrm{an} &=\frac{b}{2\pi}[\tan^{-1}\frac{X_2}{X_1}+\frac{X_1X_2}{2(1-\nu)(X_1^2+X_2^2)}] \nonumber\\
u_2^\mathrm{an} &=-\frac{b}{2\pi}[\frac{1-2\nu}{4(1-\nu)}\ln(X_1^2+X_2^2)+\frac{X_1^2-X_2^2}{4(1-\nu)(X_1^2+X_2^2)}] \nonumber\\
u_3^\mathrm{an} &=0
\label{edgeDislocationAnal}
\end{align}

The dipole tensor obtained by comparing our approach with Volterra's solution (see Equation (\ref{eq:dipoleTensorEdgeDisloc})) is 
\begin{equation}
\bD=
\left[
\begin{array}{ccc}
  2\mu b+\lambda b & 0 &0   \nonumber \\
 0 & \lambda b &0   \nonumber \\
  0   & 0 &\lambda b
\end{array}
\right]
\end{equation}

\noindent We apply the analytic displacement field (\ref{edgeDislocationAnal})  on the boundary surfaces $\Gamma_0$ as Dirichlet conditions. The weak form (\ref{eqn:galerkinformLineDefect}) is re-written as:
 \begin{align}
\int_{\Omega}w_{I,J}\sigma_{IJ}dV-\int_{\Gamma^\prime}w_{I,J}D_{IJ}\mathrm{d}S &=0\\
 \bu&=\bu^\mathrm{an} \quad \mathrm{on} \; \Gamma_0
 \end{align}
 
\begin{figure}[hbtp]
\centering
\includegraphics[scale=0.5]{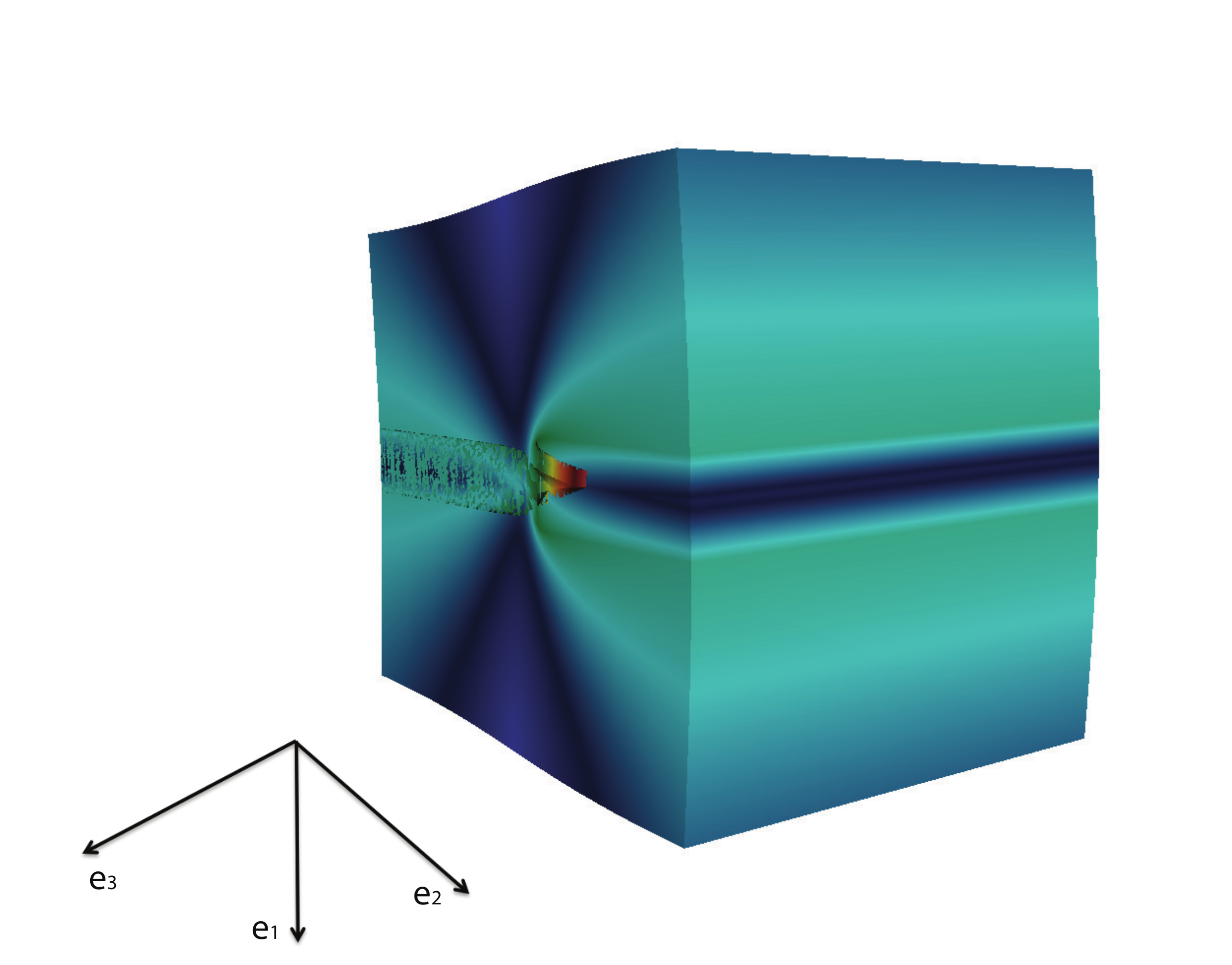}
\caption{The deformed configuration of an isotropic, linear elastic solid around an edge dislocation.}
\label{fig:edgeDislocation}
\end{figure}

\noindent Figure \ref{fig:edgeDislocation} shows the distortion around such an edge dislocation, with displacements scaled by a factor of 20.
 
\begin{figure}[hbtp]
\centering
\includegraphics[scale=0.5]{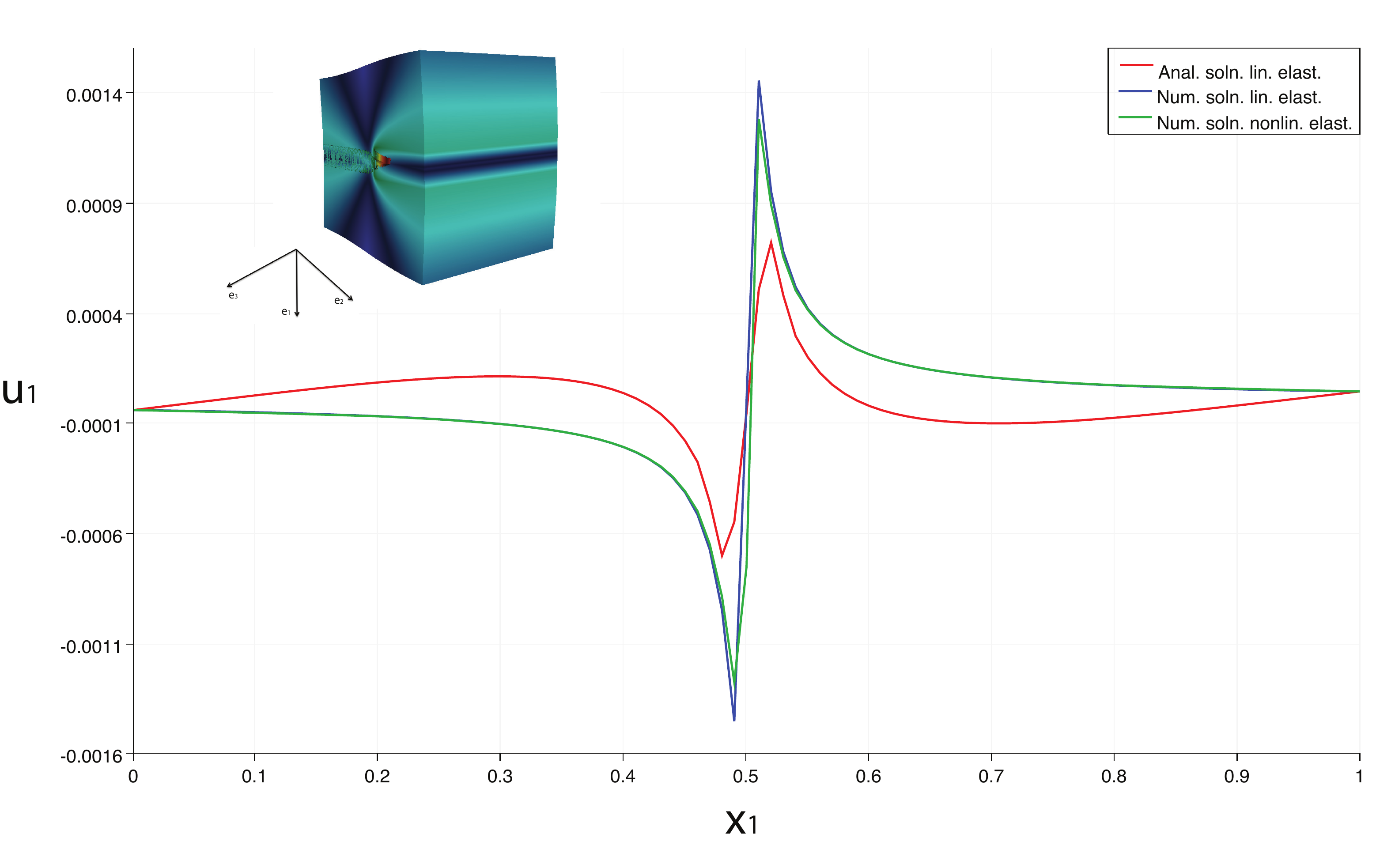}
\caption{The $u_1$ displacement field for the edge dislocation along the line segment from $\bX = \{0,0.51, 0.5\}$ to $\bX = \{1,0.51,0.5\}$. The gradient elasticity result was obtained for finite strain with $l = 0.1$.}
\label{fig:edge1}
\end{figure}

Figure \ref{fig:edge1} shows the trace of the $u_1$ displacement component along a segment oriented with $\be_1$, but positioned slightly away from the core to avoid the singularity that exists in the Volterra solution (\ref{edgeDislocationAnal}). The Burgers vector has magnitude $b = 0.008$. The numerical solution with linearized elasticity was computed with linear, $C^0$, basis functions and a knot span $h = 0.016$. We draw attention to the close match between this numerical solution and the analytic solution. Also shown is a trace of the $u_1$ field computed with gradient elasticity at finite strain, using quadratic, $C^1$, basis functions and the same knot span, $h = 0.016$. The gradient elastic length scale is $l = 0.01$. Note the regularization of the displacement field, which is the anticipated result of using a higher-order theory.

\begin{figure}[hbtp]
\centering
\includegraphics[scale=0.75]{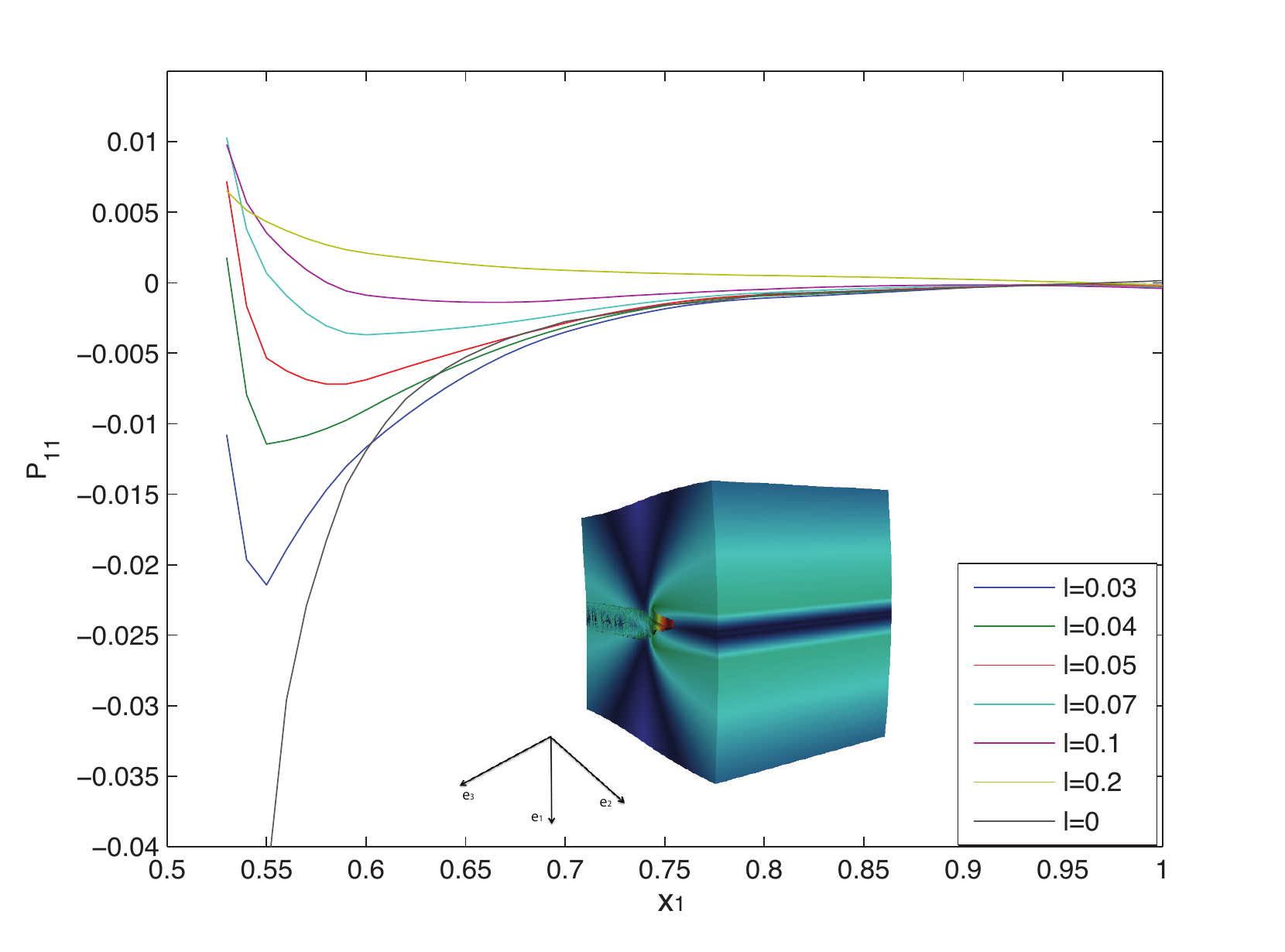}
\caption{Line plots of the $P_{11}$ stress component along a line segment from $\bX = \{0.53, 0.52, 0.5\}$ to $\bX = \{1, 0.52, 0.5\}$, as the gradient length scale parameter, $l$ is varied.}
\label{fig:edge8}
\end{figure}

Figure \ref{fig:edge8} shows the $P_{11}$ component of the Piola-Kirchhoff stress tensor computed with gradient elasticity at finite strain as the length scale parameter is varied. Note the increasing degree of regularization of solutions as $l$ increases from zero. For $l = 0$, the classical, non-gradient, finite strain elasticity solution is obtained, and the stress diverges at the core. This singularity is reflected in the sharp increase in stress magnitude approaching the left end of the interval. These solutions were obtained with the Dirichlet boundary condition $\bu = \bzero$ on the boundary $X_2 = 0$, the Burgers vector magnitude $b = 0.01$, and for quadratic, $C^1$ basis functions with knot span $h$ such that $b/h = 11.5$.
\clearpage
 
 \subsection{Comparison of analytic and numerical solutions for the screw dislocation}
 \label{sec:screw dislocation} 

Consider a screw dislocation in the unit cube $\Omega_0 = (0,1)^3$, with half plane $\Gamma^\prime$ a subset of the $X_{2}-X_{3}$ plane. The dislocation line and its core are aligned with the $\be_3$ direction, lie at $X_2 = 0.5$, and the Burgers vector is $\bb=b\be_3$. We recall the analytic solution in the regime of linearized elasticity for such a screw dislocation in an isotropic, infinite medium \citep{HirthLothe1982}:
\begin{align}
u_1^\mathrm{an}&=0\nonumber\\
u_2^\mathrm{an}&=0\nonumber\\
u_3^\mathrm{an}&=\frac{b}{2\pi}\tan^{-1}\frac{X_2}{X_1}.
\label{screwDislocationAnal}
\end{align}

\noindent The dipole tensor obtained by comparing our approach with Volterra's solution (see Equation (\ref{eq:dipoleTensorEdgeDisloc})) is 
\begin{equation}
\bD=
\left[
\begin{array}{ccc}
  0 & 0 &\mu b    \\
 0 & 0 &0\\
 \mu b   & 0 &0
\end{array}
\right]
\end{equation}
\noindent We apply the analytic displacement field (\ref{screwDislocationAnal})  on the boundary surfaces $\Gamma_0$ as Dirichlet conditions. The weak form (\ref{eqn:galerkinformLineDefect}) is re-written as:
 \begin{align}
\int_{\Omega_0}w_{I,J}\sigma_{IJ}\mathrm{d}V-\int_{\Gamma^\prime}(w_{1,3}D_{13}+w_{3,1}D_{31})\mathrm{d}S &=0\\
 \bu &=\bu^\mathrm{an} \quad \mathrm{on} \; \Gamma_0
 \end{align}
 
\begin{figure}[hbtp]
\centering
\includegraphics[scale=0.5]{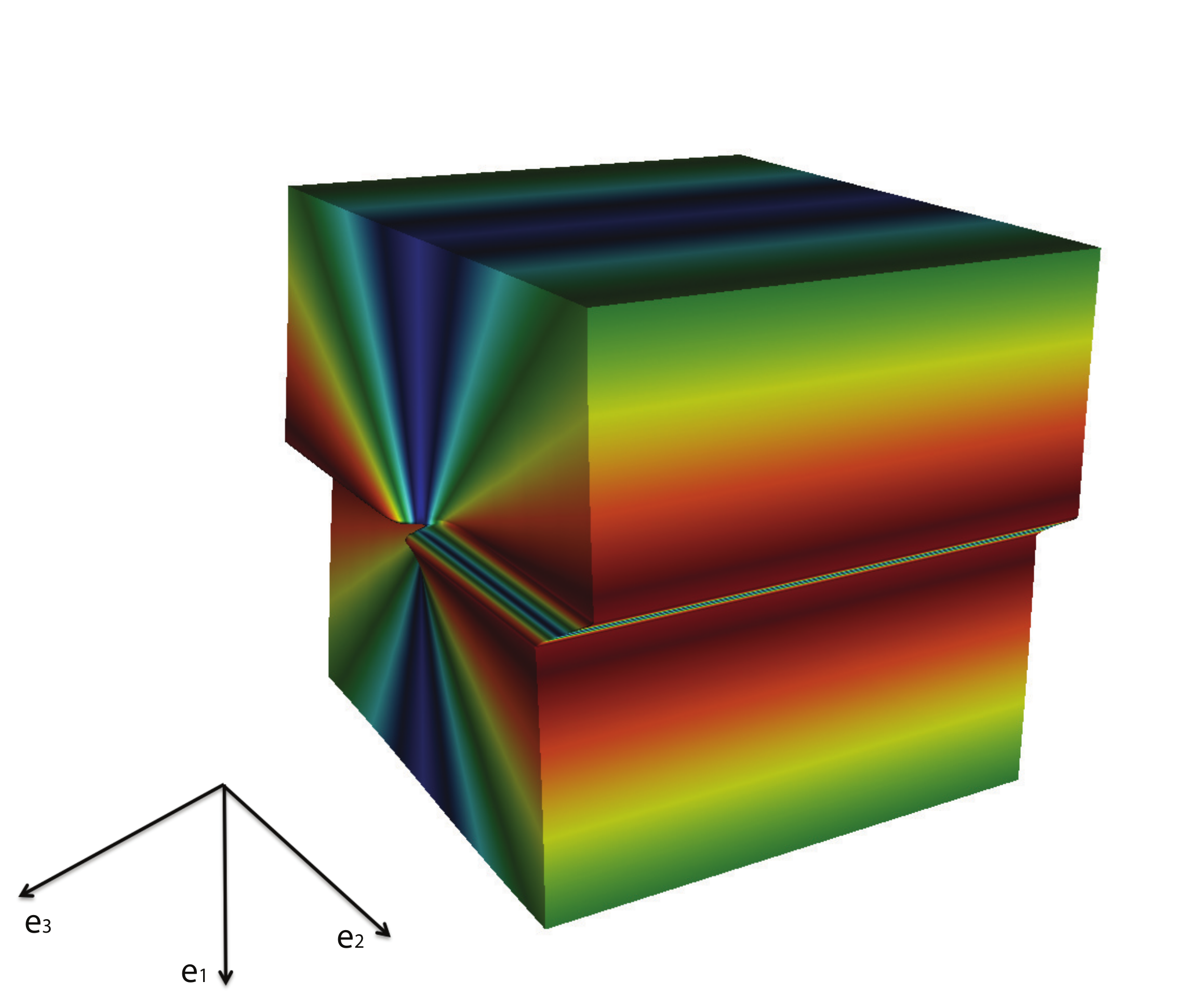}
\caption{The deformed configuration of an isotropic, linear elastic solid around a screw dislocation.}
\label{fig:screwDislocation}
\end{figure}

 Figure \ref{fig:screwDislocation} shows the distortion around such a screw dislocation, with displacements scaled by a factor of 20.

\begin{figure}[hbtp]
\centering
\includegraphics[scale=0.5]{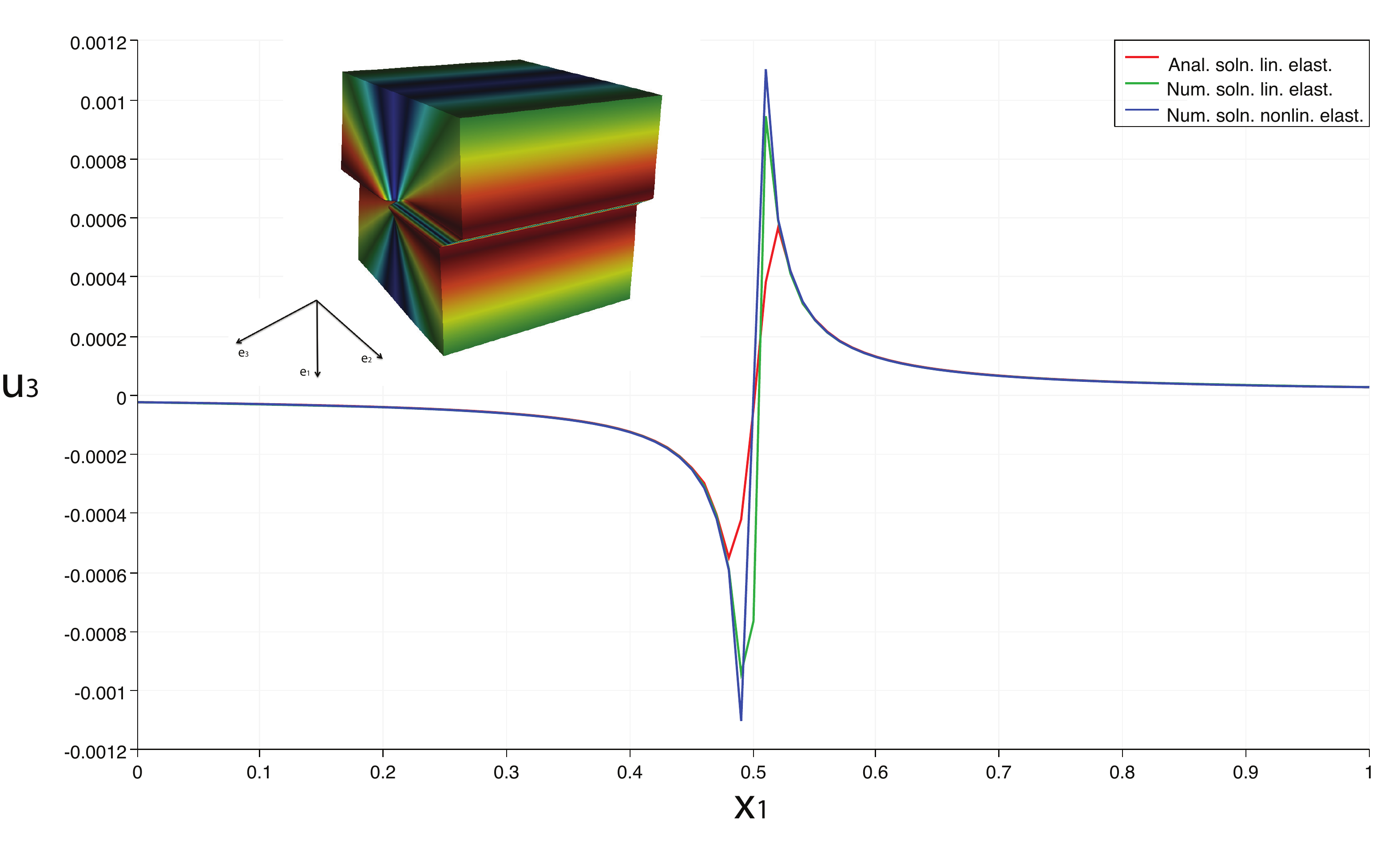}
\caption{The $u_3$ displacement field for the screw dislocation along the line segment from $\bX = \{0,0.51, 0.5\}$ to $\bX = \{1,0.51,0.5\}$. The gradient elasticity result was obtained for finite strain with $l = 0.01$.}
\label{fig:edge6}
\end{figure}

Figure \ref{fig:edge6} shows the trace of the $u_3$ displacement component along a segment oriented with $\be_1$, but positioned slightly away from the core to avoid the singularity that exists in the Volterra solution (\ref{screwDislocationAnal}). The Burgers vector has magnitude $b = 0.008$. The numerical solution with linearized elasticity was computed with linear, $C^0$ basis functions and a knot span $h = 0.016$. We draw attention to the close match between this numerical solution and the analytic solution. Also shown is a trace of the $u_3$ field computed with gradient elasticity at finite strain, using quadratic, $C^1$ basis functions and the same knot span, $h = 0.016$. The gradient elastic length scale $l = 0.01$. Note the regularization of the displacement field, which is the anticipated result of using a higher-order theory.



\begin{figure}[hbtp]
\centering
\includegraphics[scale=0.75]{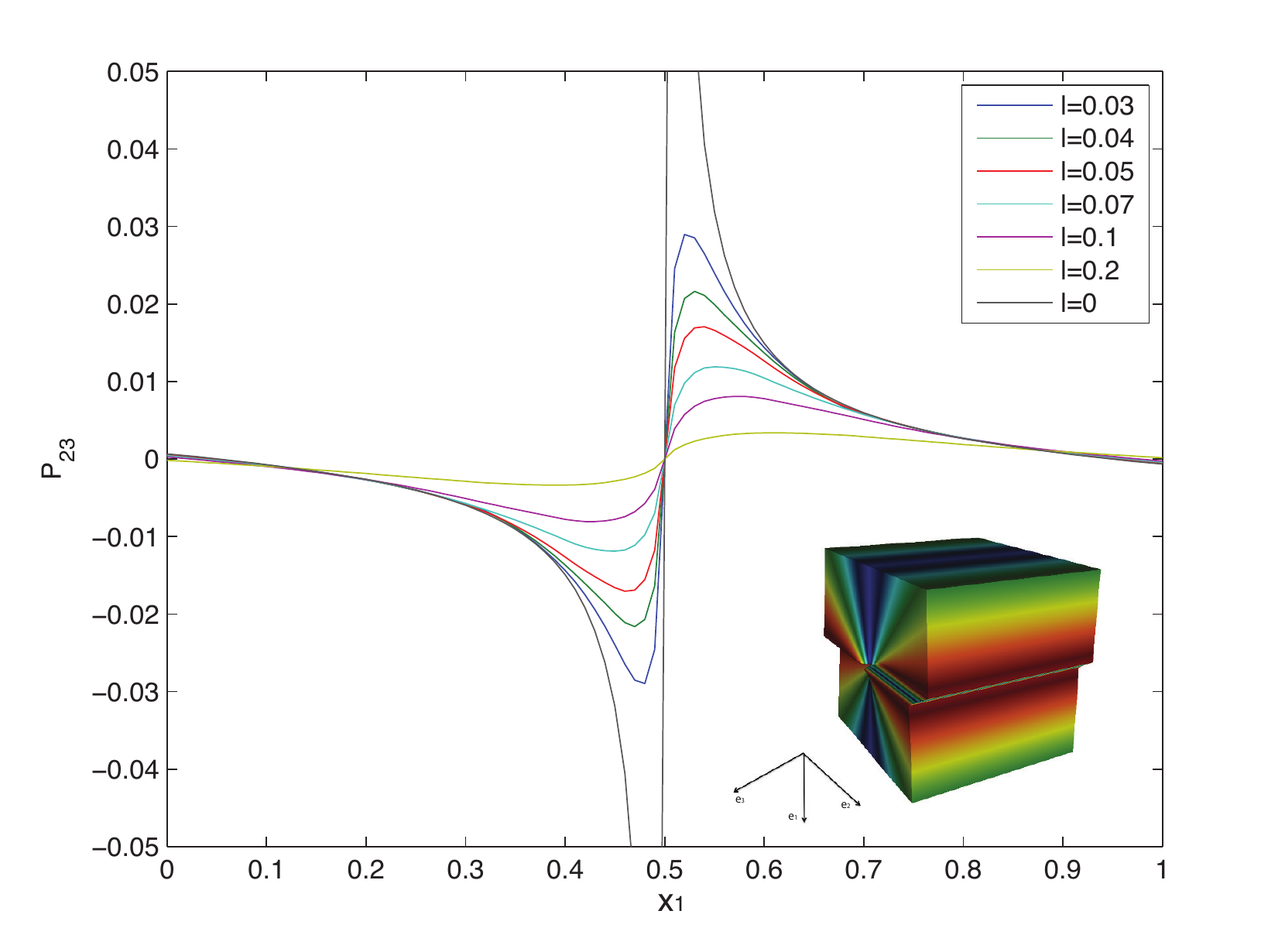}
\caption{Line plots of the $P_{23}$ stress component along a line segment from $\bX = \{0, 0.5, 0.5\}$ to $\bX = \{1,0.5,0.5\}$, as the gradient length scale parameter, $l$ is varied.}
\label{fig:edge7}
\end{figure}

Figure \ref{fig:edge7} shows traces of the $P_{23}$ component of the Piola-Kirchhoff stress tensor computed with gradient elasticity at finite strain as the length scale parameter is varied. Note the increasing degree of regularization of solutions as $l$ increases from zero. For $l = 0$, the classical, non-gradient, finite strain elasticity solution is obtained, and the stress diverges at the core. This singularity is reflected in the stress magnitude approaching the left end of the interval. These solutions were obtained with the Dirichlet boundary condition $\bu = \bzero$ on the boundary $X_2 = 0$, the Burgers vector magnitude $b = 0.01$, and with quadratic, $C^1$ basis functions with knot span $h$ such that $b//h = 11.5$.

%
%

\subsection{Energy of a single edge dislocation}
\label{sec:singleDislocationEnergy}

From classical, linearized elasticity we have the expression for the self energy of a single edge dislocation that is coaxial with a cylinder of radius R and incorporates a core cutoff, $r_0$, to eliminate the singularity \citep{HirthLothe1982}:
\begin{align}
W_s&=\frac{\mu b^2}{4\pi(1-\nu)}\ln\frac{R}{r_0} 
\label{energySingleEdgeDisloc}
\end{align}

Usually, this expression is presented as a logarithmic divergence with $R$. Instead, we have computed the self-energy of a single edge dislocation in the unit cube and carried out mesh refinement studies. As the knot span shrinks, quadrature points move closer to the core, and the computed self energy can be expected to diverge for fixed $b$ as $h\to 0$ in the same manner that $\ln(R/r_0)$ diverges for fixed $R$ as $r_0 \to 0$. The boundary conditions for this computation, and for the remaining energy studies are $\bu_1 = \bzero$ on $X_2 = 0$, with homogeneous traction and higher-order traction on the remaining boundaries.

%
%

\begin{figure}[hbtp]
\centering
\includegraphics[scale=0.65]{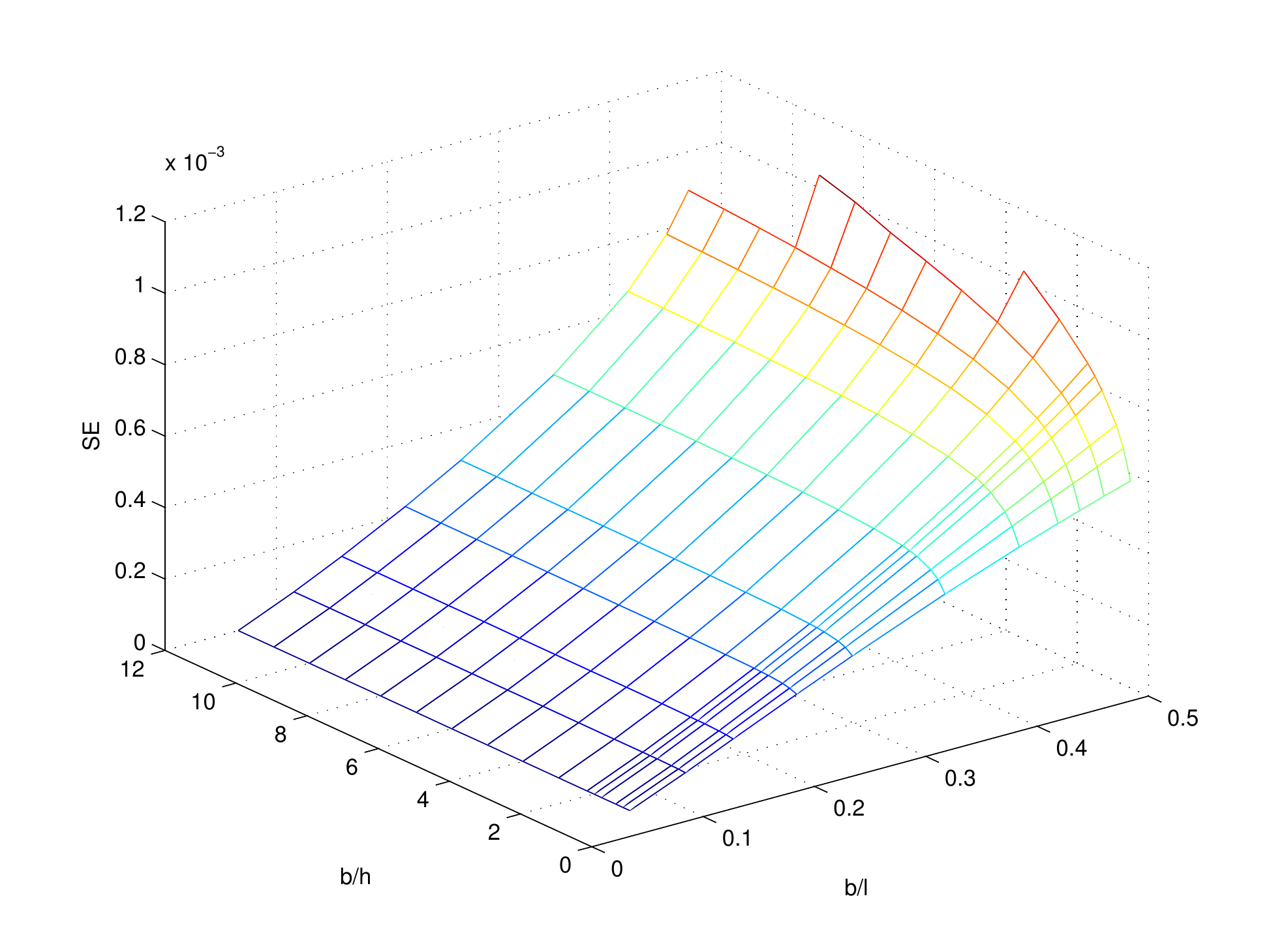}\\
\includegraphics[scale=0.65]{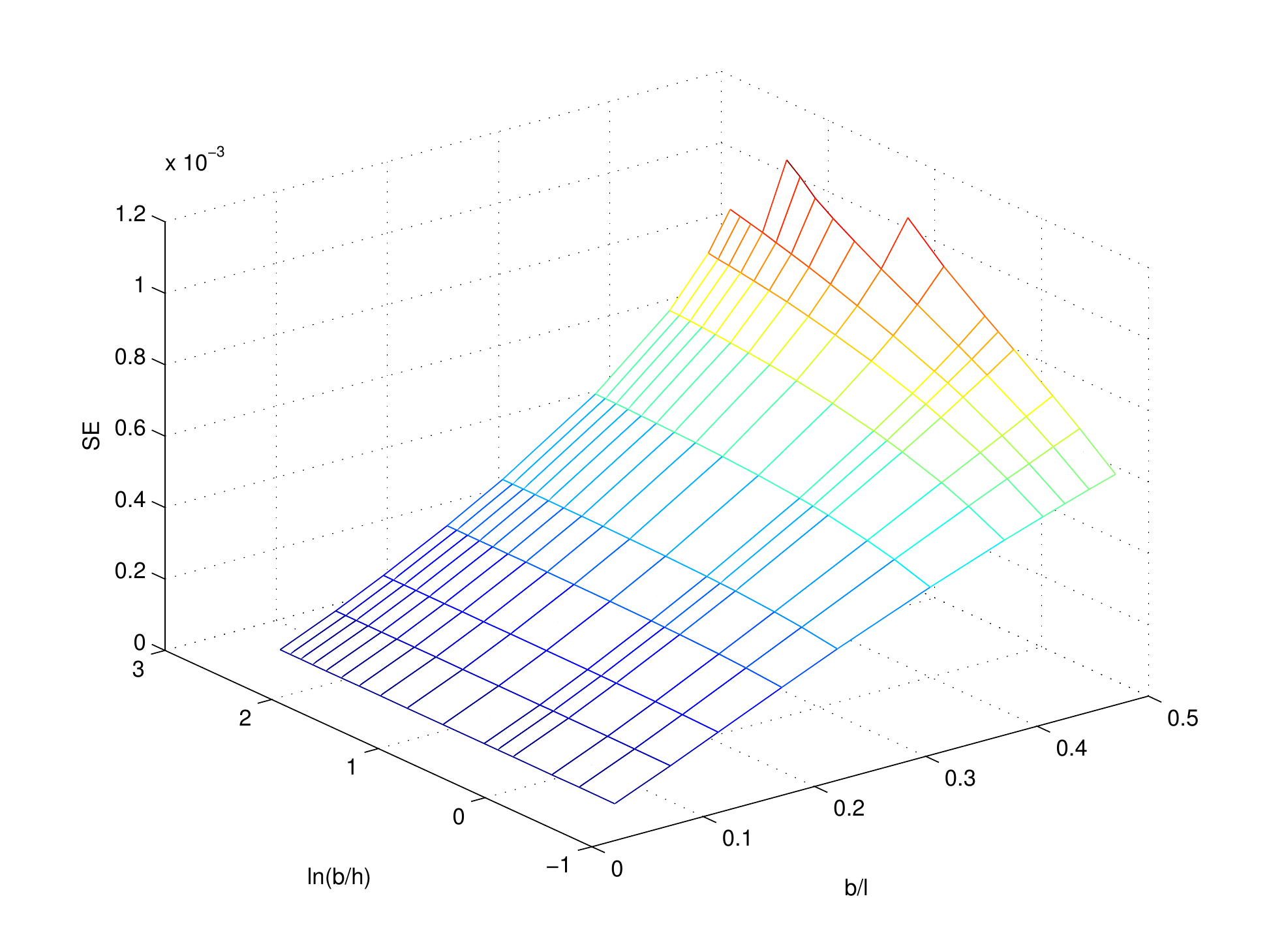}
\caption{The strain energy of a single dislocation, computed with gradient elasticity at finite strain plotted on (a) a linear scale for $b/h$, and (b) a logarithmic scale for $b/h$ to emphasize the logarithmic divergence in the $l\to 0$ regime (large $b/l$). Note the convergence with mesh refinement, and the decrease in strain energy with an increase of gradient length scale, $l$ (small $b/l$).}
\label{fig:energyE2}
\end{figure}

\begin{figure}[hbtp]
\centering
\includegraphics[scale=0.65]{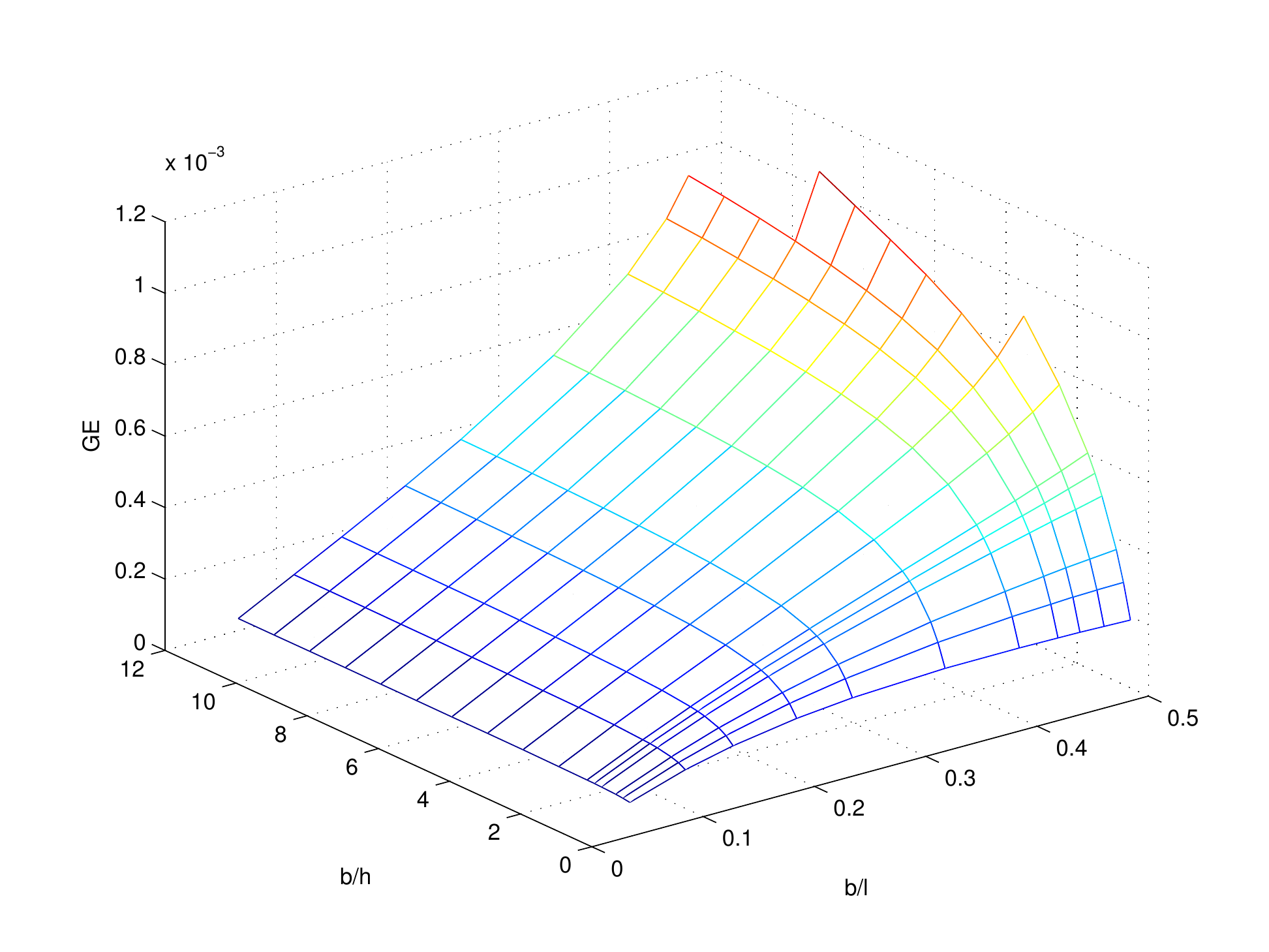}\\
\includegraphics[scale=0.65]{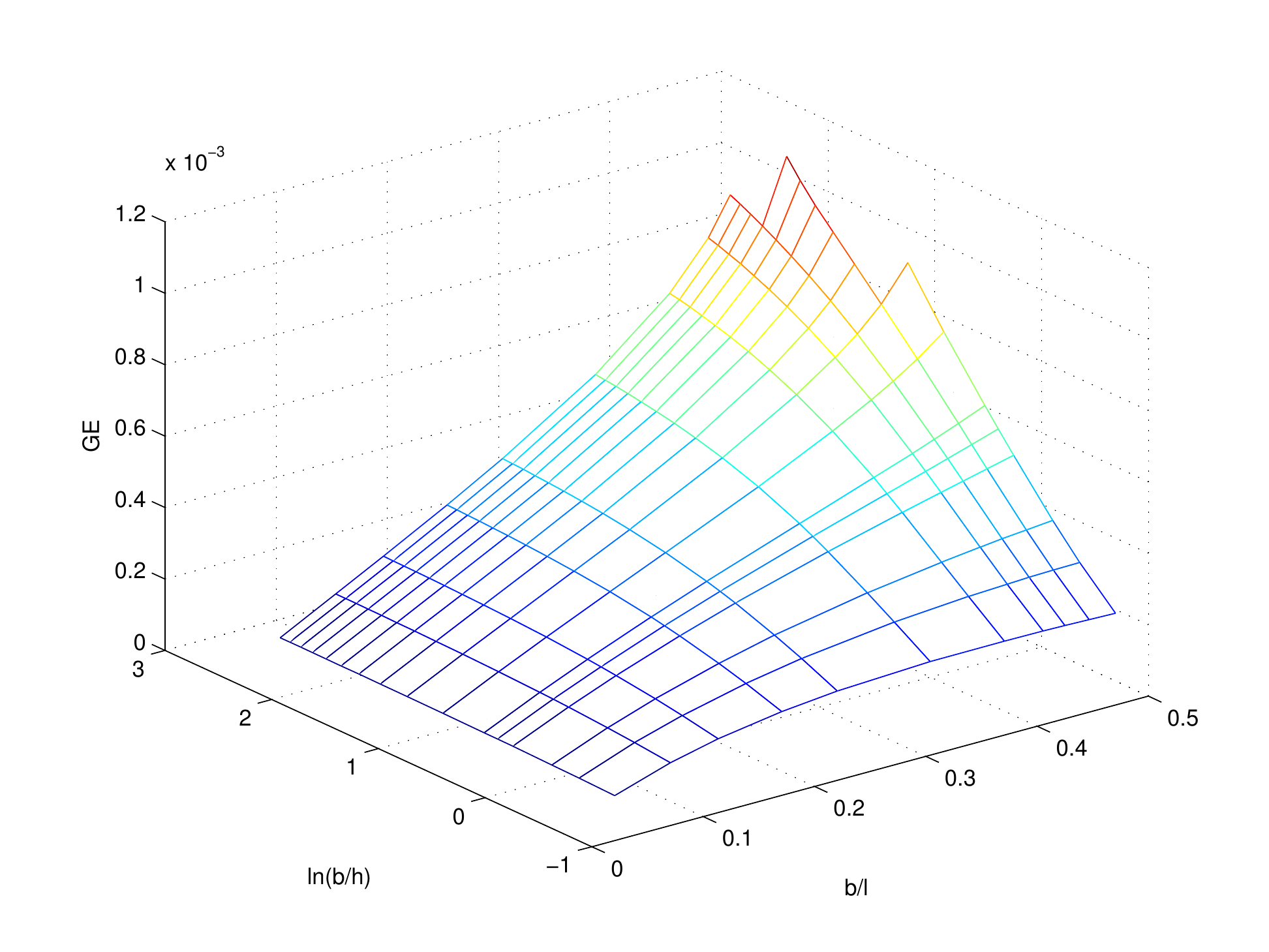}
\caption{The strain gradient energy of a single dislocation, computed with gradient elasticity at finite strain plotted on (a) a linear scale for $b/h$, and (b) a logarithmic scale for $b/h$ to emphasize the logarithmic divergence in the $l\to 0$ regime (large $b/l$). Note the convergence with mesh refinement, and the decrease in strain energy with an increase of gradient length scale, $l$ (small $b/l$).}
\label{fig:energyE3}
\end{figure}

Figures \ref{fig:energyE2} and \ref{fig:energyE3}, respectively, show the strain energy and strain gradient energies of a single edge dislocation computed with gradient elasticity at finite strains. These computations use $b = 0.01$. Note that for the gradient length scale $l \to 0$, i.e. for large values of $b/l$, the strain energy diverges logarithmically with $b/h$. This represents the regime wherein gradient effects are vanishingly present, and the classical, non-gradient response may be expected. Interestingly, the strain gradient energy also displays the same logarithmic behavior with $b/h$. For larger values of $l$, i.e. for $b/l \to 0$ the strain gradients strongly regularize the problem and the logarithmic divergence with $b/h$ is mollified for the strain energy as well as the strain gradient energy. We note that as $l$ increases, both components of the elastic free energy decrease. This is due to the increasing stiffness of the displacement fields as $l$ increases, which also is reflected in the above field solutions around defects.

\subsection{Studies of the elastic free energy of pairs of interacting, parallel dislocations}

\begin{figure}[!hbtp]
\centering
\includegraphics[scale=0.7]{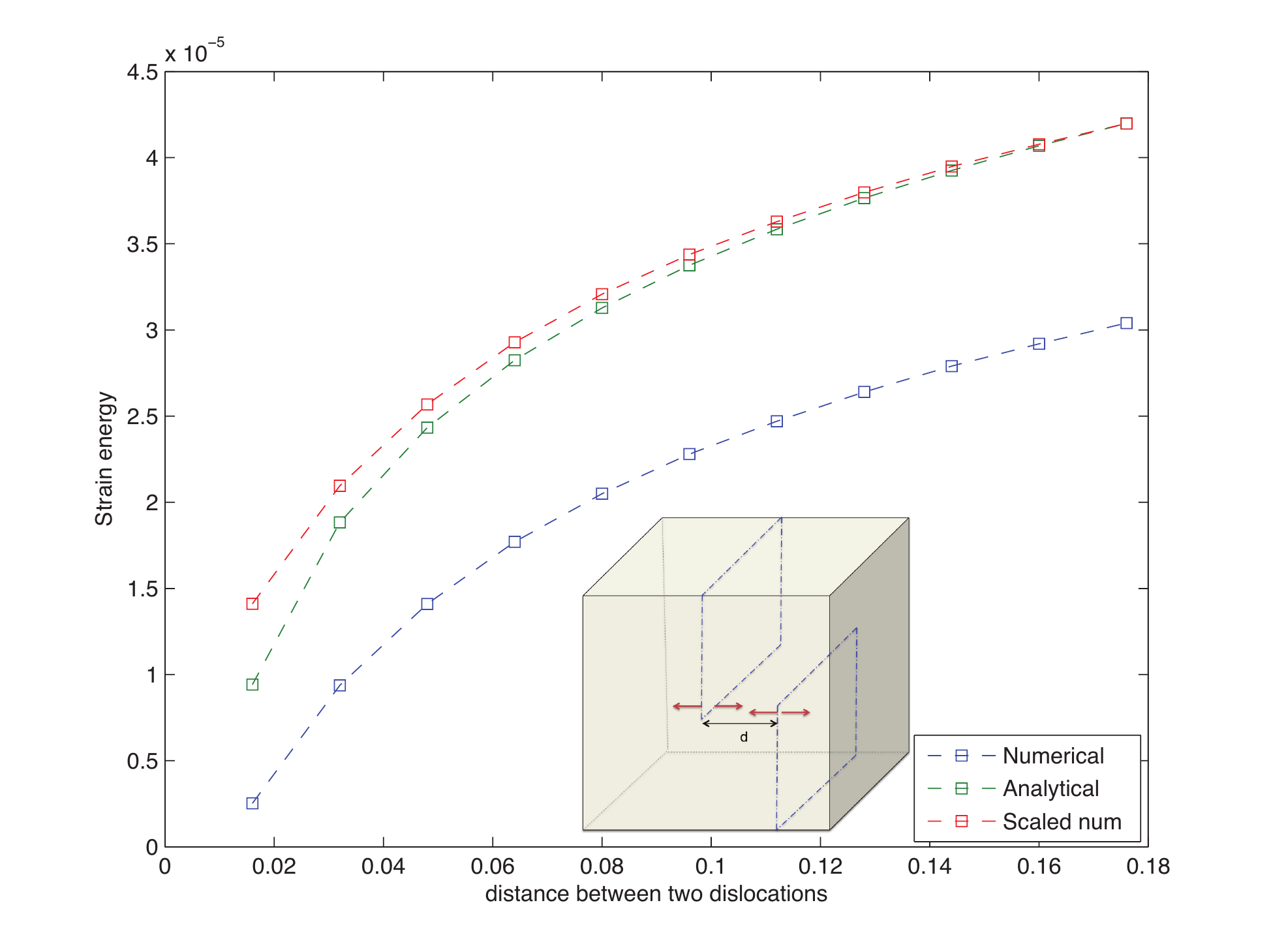}
\caption{Energy of two edge dislocations with opposite signs; classical, linearized elasticity.}
\label{fig:DenergyE1}
\end{figure}

In a cylindrical domain of radius $R$, the interaction energy per unit length between pairs of parallel dislocations, each with Burgers vector $b$ and radial separation $r$, when calculated using classical, linearized elasticity, is
 
\begin{align}
\frac{W}{L}&=\frac{\mu b^2}{2\pi(1-\nu)}\ln\frac{\alpha r}{b} \quad \text{for two edge dislocations of opposite signs}\label{oppEdge}\\
\frac{W}{L}&=\frac{\mu b^2}{2\pi(1-\nu)}\ln\frac{\alpha R^2}{br} \quad \text{for two edge dislocations with like signs}\label{likeEdge}\\
\frac{W}{L}&=\frac{\mu b^2}{2\pi}\ln\frac{\alpha r}{b} \quad \text{for two screw dislocations of opposite signs}\label{oppScrew}\\
\frac{W}{L}&=\frac{\mu b^2}{2\pi}\ln\frac{\alpha R^2}{br} \quad \text{for two screw dislocations with like signs}\label{likeScrew}
\end{align}

\noindent where $\alpha$ is a numerical parameter controlling the size of the core cutoff. 
%
%
Figure \ref{fig:DenergyE1} compares the above interaction energy \emph{versus} $r$ for a pair of oppositely signed edge dislocations with computations of the total energy of a pair of similarly interacting edge dislocations using our formulation for classical, linearized elasticity. Shown are the analytic result (\ref{oppEdge}), the numerically computed result, and a shifted numerical result that accounts for the fact that the numerical domain is a unit cube with a square cylindrical core cutoff in comparison with the analytic solution, which uses infinitely long circular cylinders for the domain and the core. This shifting is achieved by requiring the solutions to match for large separations between the dislocations. Without this shifting, although the numerical solution, based on classical, linearized elasticity, has the correct trend, the values are systematically in error because of the difference in representation of the domain and core, and the combination of interaction and self energies in the numerical solution. We note that the energy of interacting dislocations decreases as the fields of the oppositely signed dislocations compensate with decreasing separation $r$.

\begin{figure}[hbtp]
\centering
\includegraphics[scale=0.7]{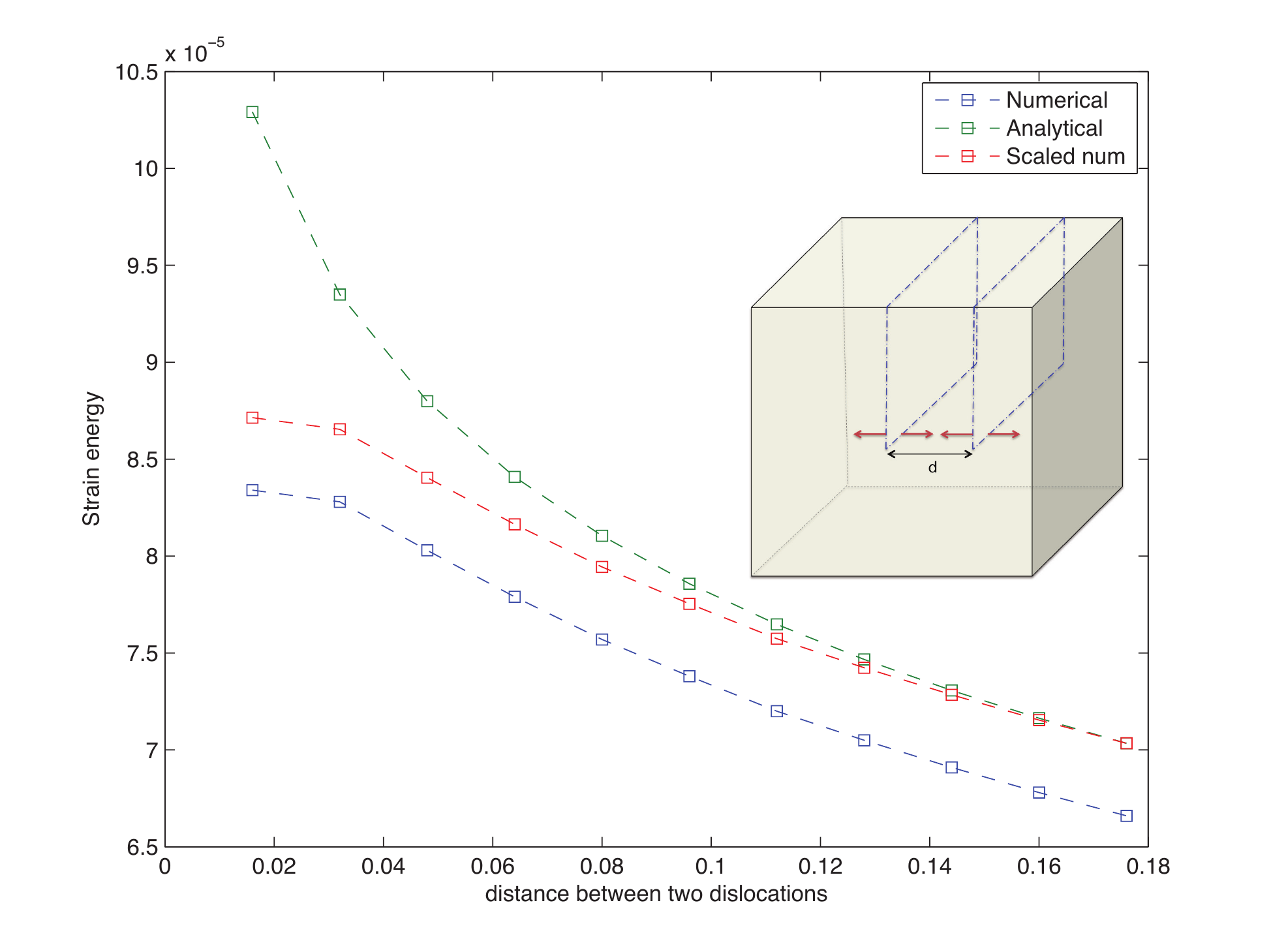}
\caption{Energy of two edge dislocations with like signs; classical, linearized elasticity.}
\label{fig:DenergyE2}
\end{figure}

\begin{figure}[hbtp]
\centering
\includegraphics[scale=0.7]{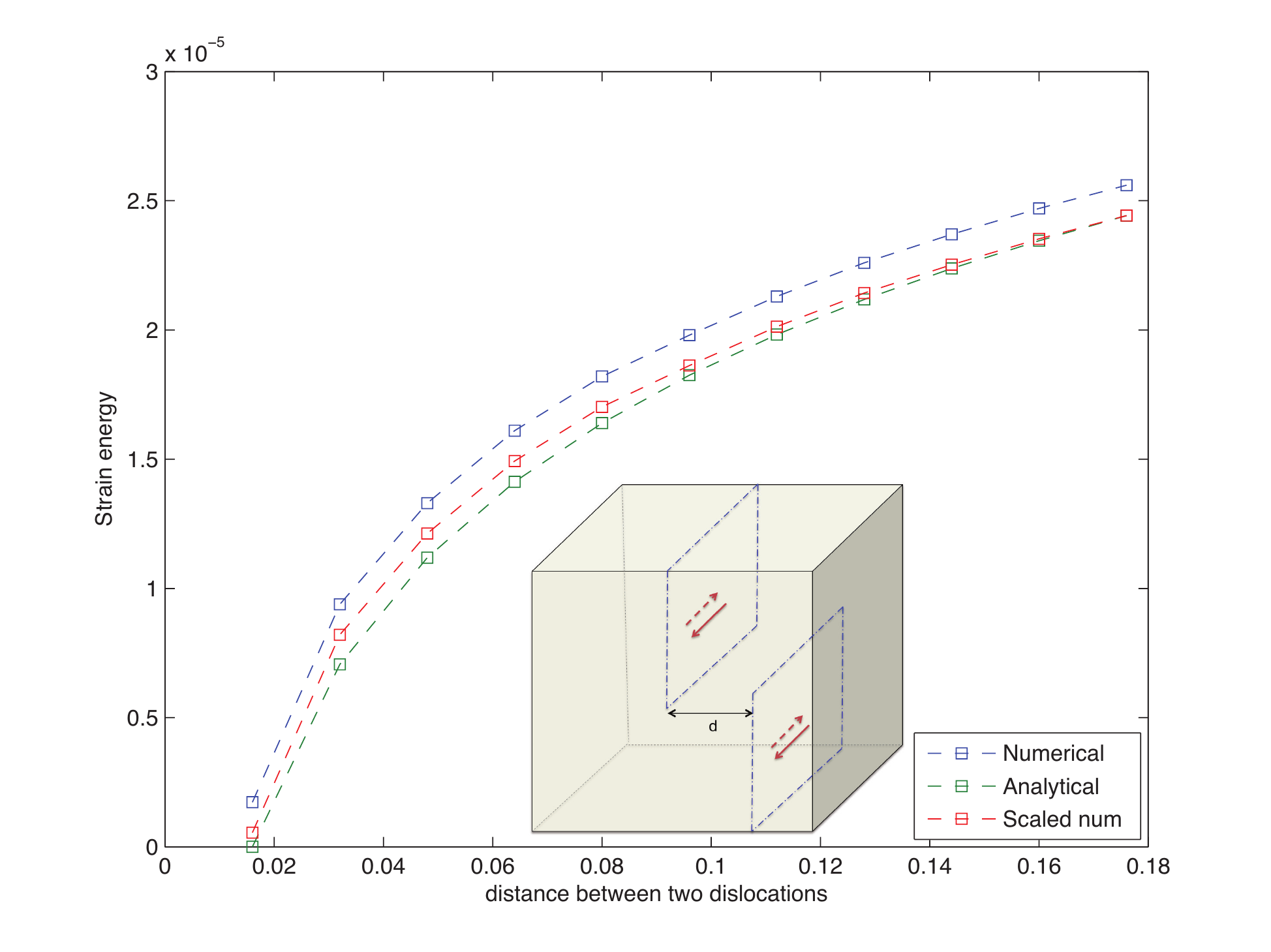}
\caption{Energy of two screw dislocations with opposite signs; classical, linearized elasticity.}
\label{fig:DenergyE3}
\end{figure}

\begin{figure}[hbtp]
\centering
\includegraphics[scale=0.7]{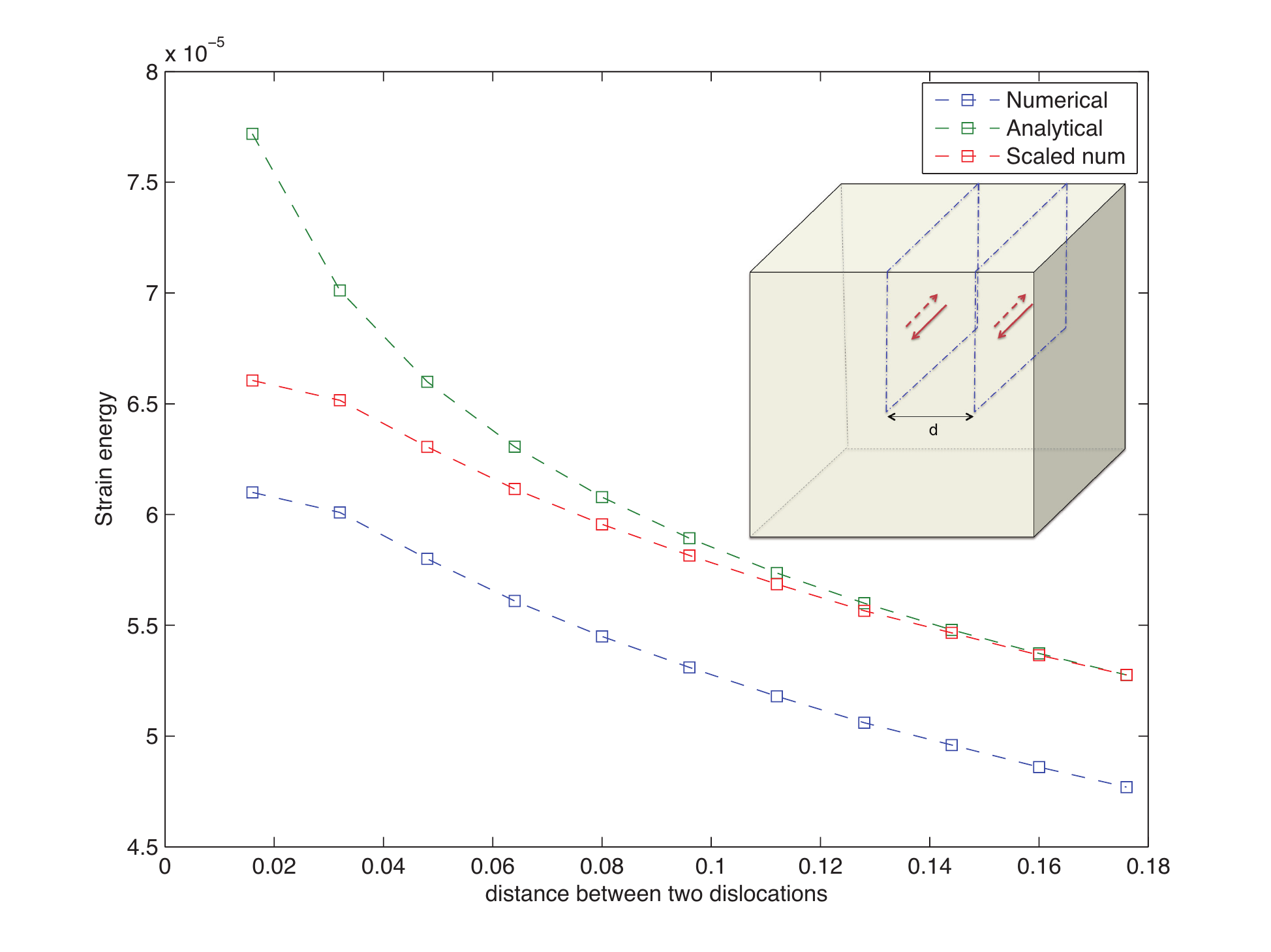}
\caption{Energy of two screw dislocations with like signs; classical, linearized elasticity.}
\label{fig:DenergyE4}
\end{figure}

Figures \ref{fig:DenergyE2}--\ref{fig:DenergyE4} show corresponding results for like signed edge dislocations, oppositely signed screw dislocations and for like signed screw dislocations, respectively. In each case the trend shown by the numerical solution is correct, and the values improve upon the shifting as explained above to account for the domain and core shapes, and combination of interaction and self energies in the numerical solutions. In general, interacting oppositely signed dislocations show a decrease in total energy while like signed dislocations show an increase in energy. The results in Figures \ref{fig:DenergyE1}--\ref{fig:DenergyE4} have been generated with $b = 0.008$ and a knot span $h = 0.016$ for linear, $C^0$  basis functions. The analytic comparisons are with Equations (\ref{oppEdge}--\ref{likeScrew}) using $\alpha = 1$.
\clearpage
%
%

\begin{figure}[hbtp]
\centering
\includegraphics[scale=0.7]{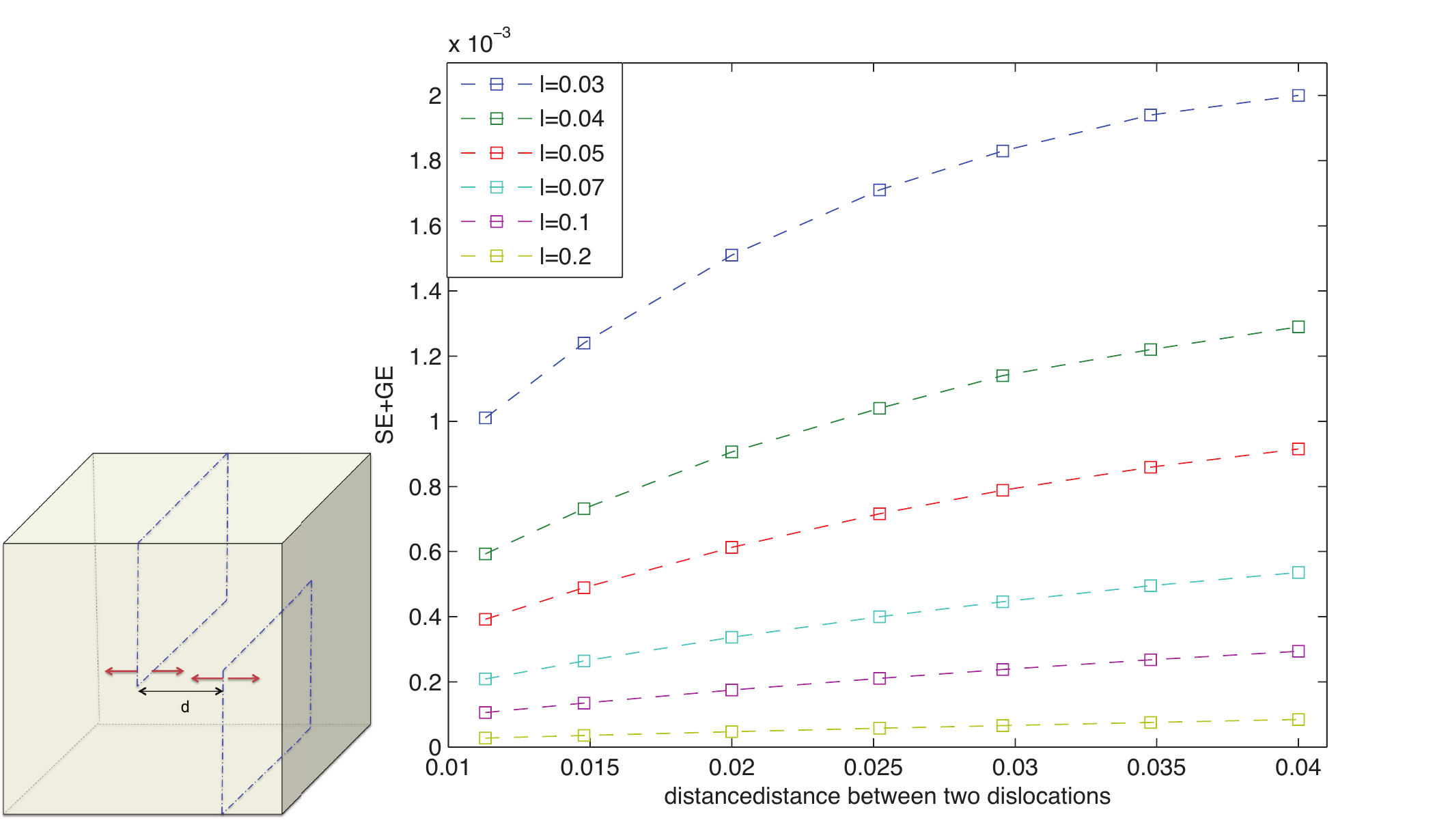}
\caption{Energy of a pair of interacting edge dislocations with opposite signs; gradient elasticity at finite strain.}
\label{fig:gradEnergy1}
\end{figure}

\begin{figure}[hbtp]
\centering
\includegraphics[scale=0.7]{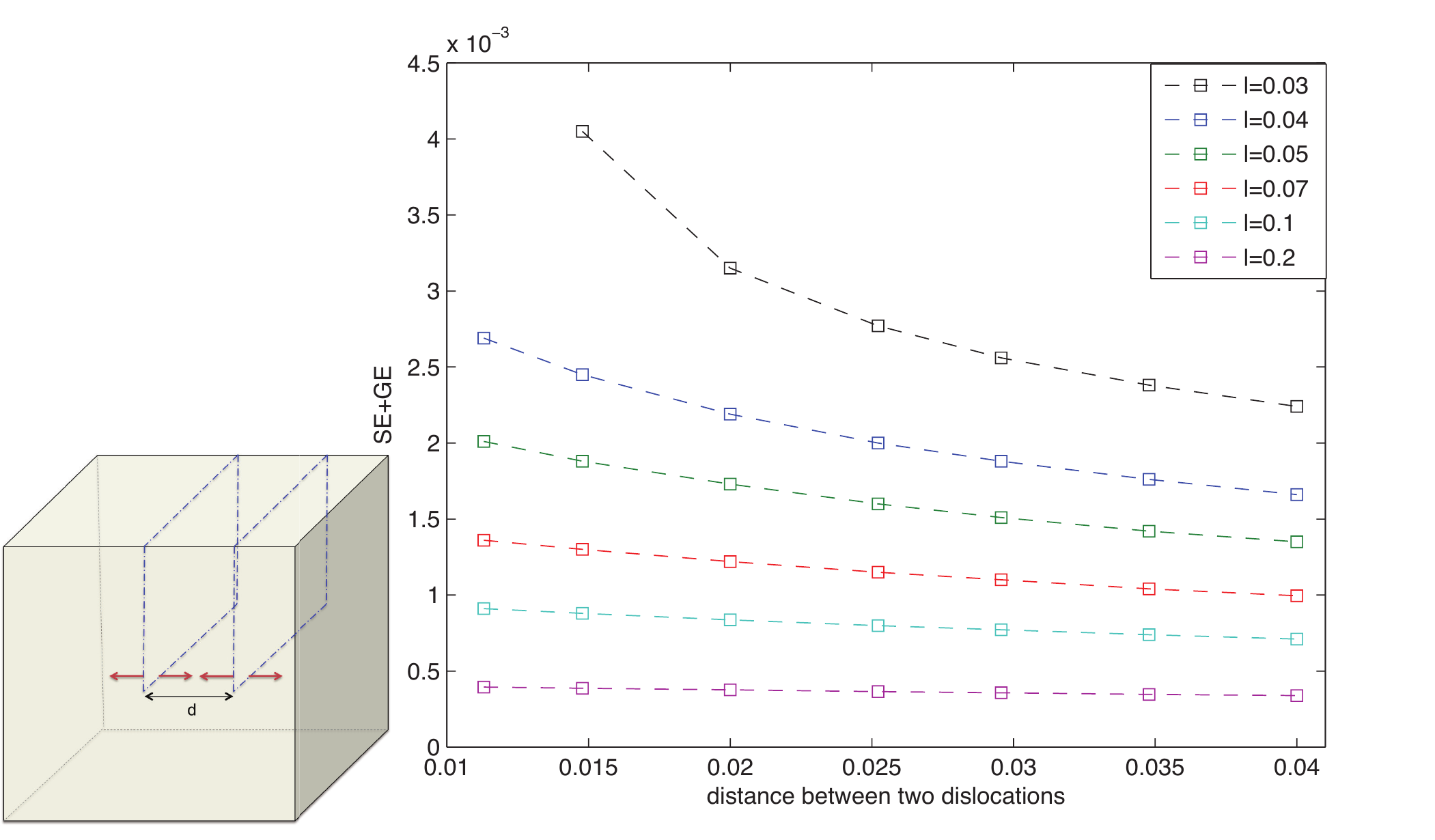}
\caption{Energy of a pair of interacting edge dislocations with like signs; gradient elasticity at finite strain.}
\label{fig:gradEnergy2}
\end{figure}

\begin{figure}[hbtp]
\centering
\includegraphics[scale=0.7]{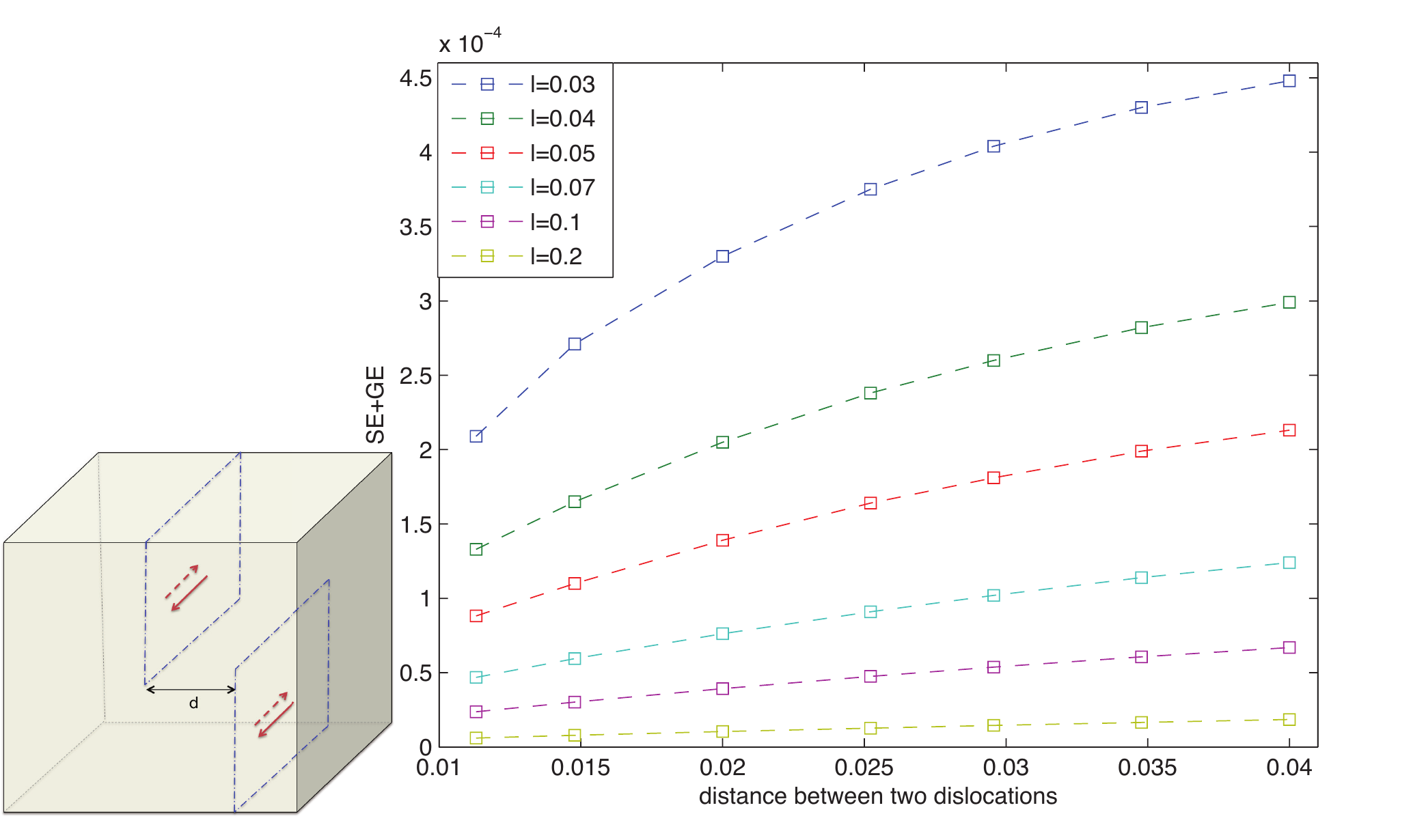}
\caption{Energy of a pair of interacting screw dislocations with opposite signs; gradient elasticity at finite strain.}
\label{fig:gradEnergy3}
\end{figure}

\begin{figure}[hbtp]
\centering
\includegraphics[scale=0.7]{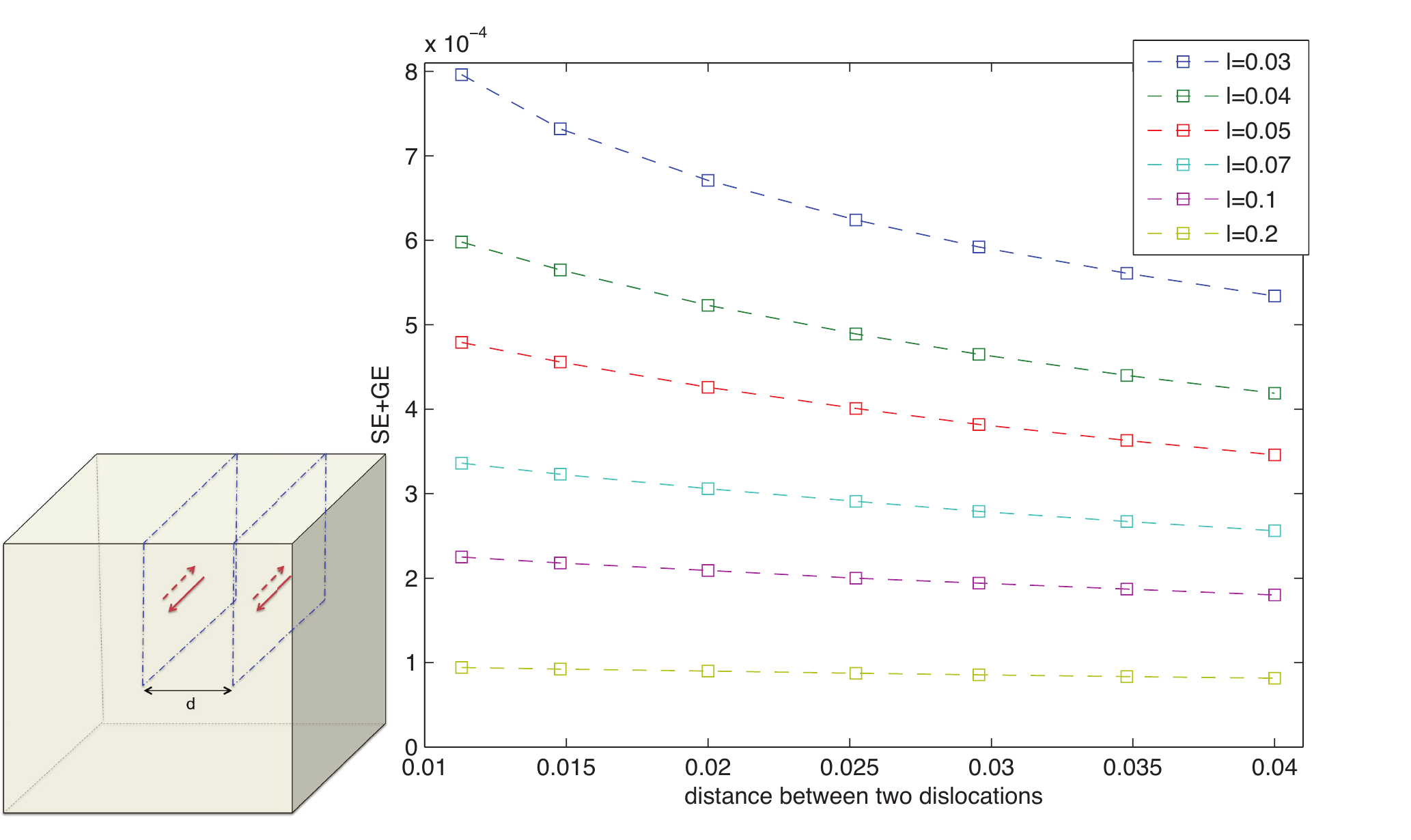}
\caption{Energy of a pair of interacting screw dislocations with like signs; gradient elasticity at finite strain.}
\label{fig:gradEnergy4}
\end{figure}

Finally, Figures \ref{fig:gradEnergy1}--\ref{fig:gradEnergy4} are equivalent computations for pairs of interacting, oppositely signed edge, like signed edge, oppositely signed screw and like signed screw dislocations, respectively, computed with gradient elasticity at finite strains, using $b = 0.01$ and $b/h = 11.5$. Here also, rather than attempt to define an interaction energy for nonlinear, finite strain elasticity, we simply present the total strain energy of the configuration versus the dislocation separation. No core cutoff is necessary in these computations because of the regularization of the singularity by gradient elasticity. Note the increased suppression of variation of the energy as $l$ increases, which is a result of the regularization.

\subsection{Edge dislocation near a free surface}
\begin{figure}[hbtp]
\centering
\includegraphics[scale=0.8]{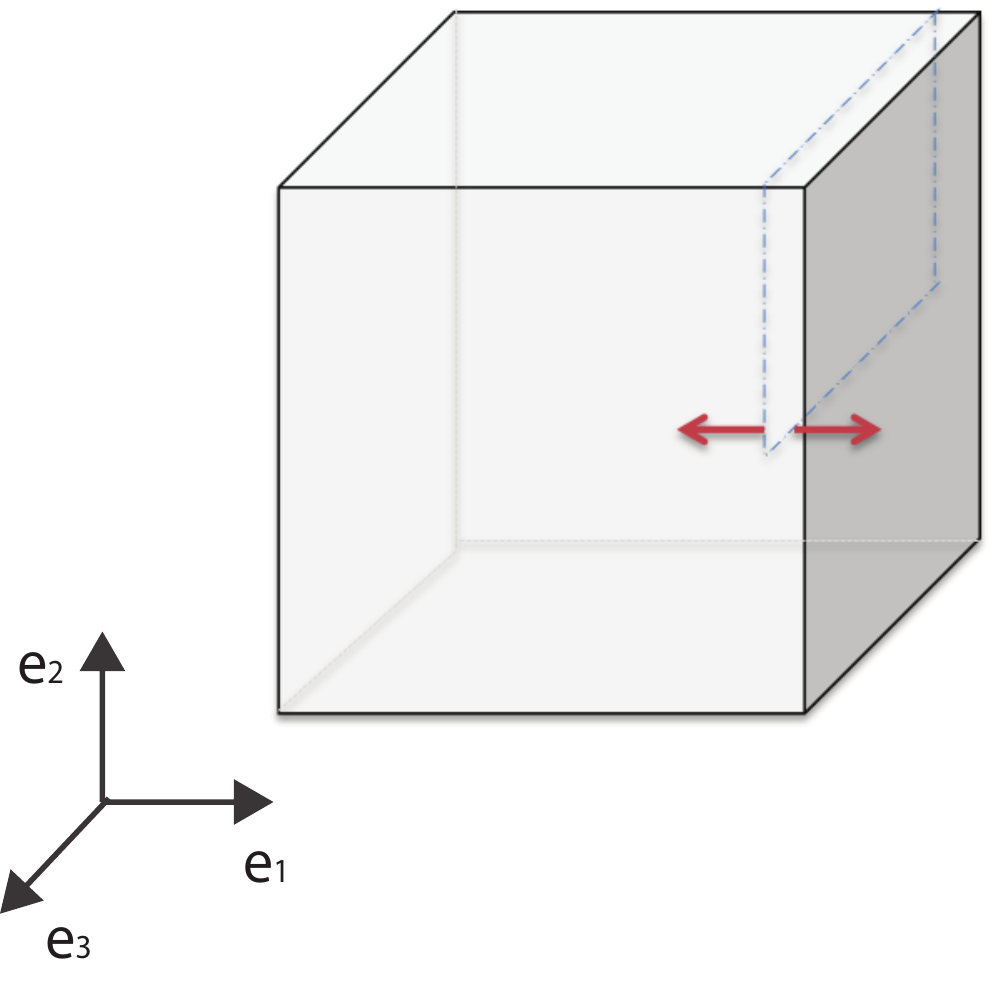}
\caption{Schematic of an edge dislocation near a free surface, showing the half plane and force dipole.}
\label{fig:schemefreesurf}
\end{figure}

\begin{figure}[hbtp]
\centering
\includegraphics[scale=0.35]{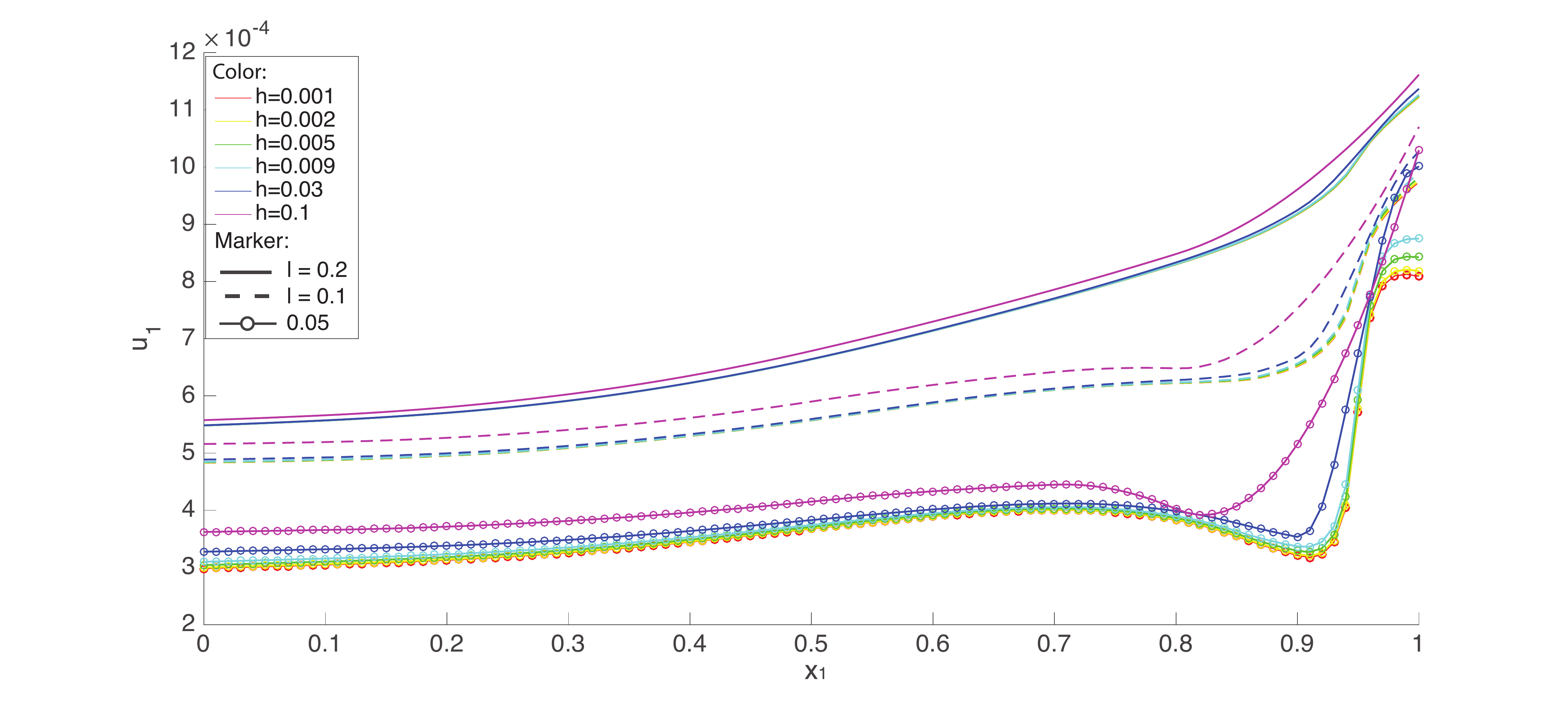}
\caption{Displacement component $u_1$ plotted along the line segment between $\bX = \{0,0.5,0.5 \}$ and $\bX = \{1,0.5,0.5 \}$. Note the convergence with $h$ and the increasing regularization with $l$.}
\label{fig:dispfreesurf}
\end{figure}

\begin{figure}[hbtp]
\centering
\includegraphics[scale=0.35]{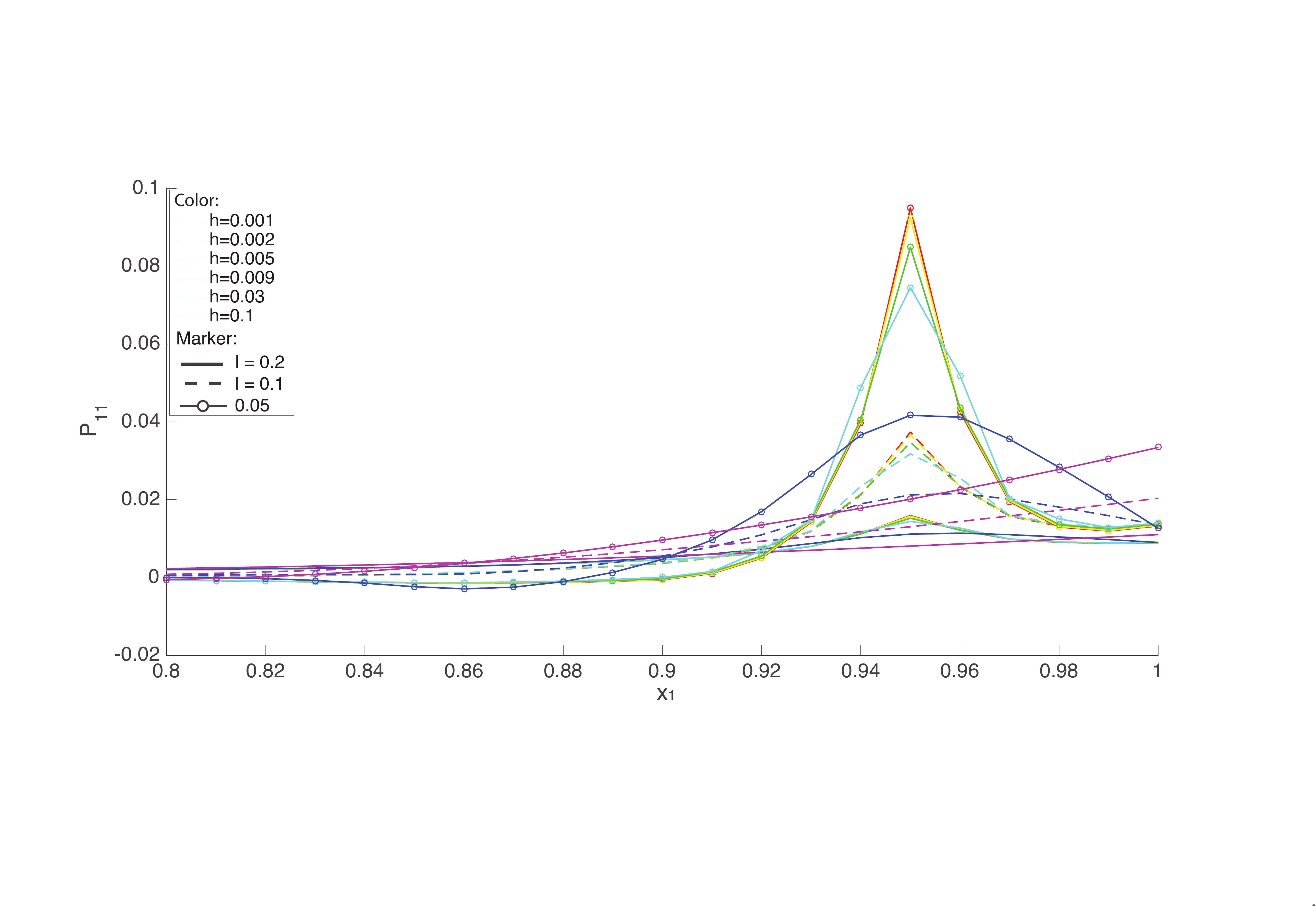}
\caption{Stress component $P_{11}$ plotted along the line segment between $\bX = \{0,0.5,0.5 \}$ and $\bX = \{1,0.5,0.5 \}$. Note the convergence with $h$ and the increasing regularization with $l$.}
\label{fig:stressfreesurf}
\end{figure}

Consider an edge dislocation in the unit cube $\Omega_0 = (0,1)^3$, with half plane $\Gamma^\prime$ a subset of the $X_{2}-X_{3}$ plane. The dislocation line and core are aligned with $\be_3$, lie at $X_1 = 0.95,\;  X_2 = 0.5$ and have Burgers vector $\bb = b\be_1$, where $b=0.01$. Dirichlet boundary conditions $\bu=0$ were applied at $X_2 = 0$, and the remaining surfaces are traction free. The schematic is shown in Figure \ref{fig:schemefreesurf}. Our studies show convergence of the $u_1$ displacement field with mesh refinement for different gradient length scales (Figure \ref{fig:dispfreesurf}), and of the $P_{11}$ stress component (Figure \ref{fig:stressfreesurf}) as well. For both fields, $u_1$ and $P_{11}$, an increased regularization is observed for larger $l$. For the Burgers vector used here, a knot span $h > 0.1$ fails to resolve the force dipole in the core. Note that $P_{11}$ does not constitute the full traction component along $\be_1$, and therefore does not vanish at the traction free edge. 


\subsection{The elastic field of a nonplanar dislocation loop}
\begin{figure}[hbtp]
\centering
\includegraphics[scale=0.35]{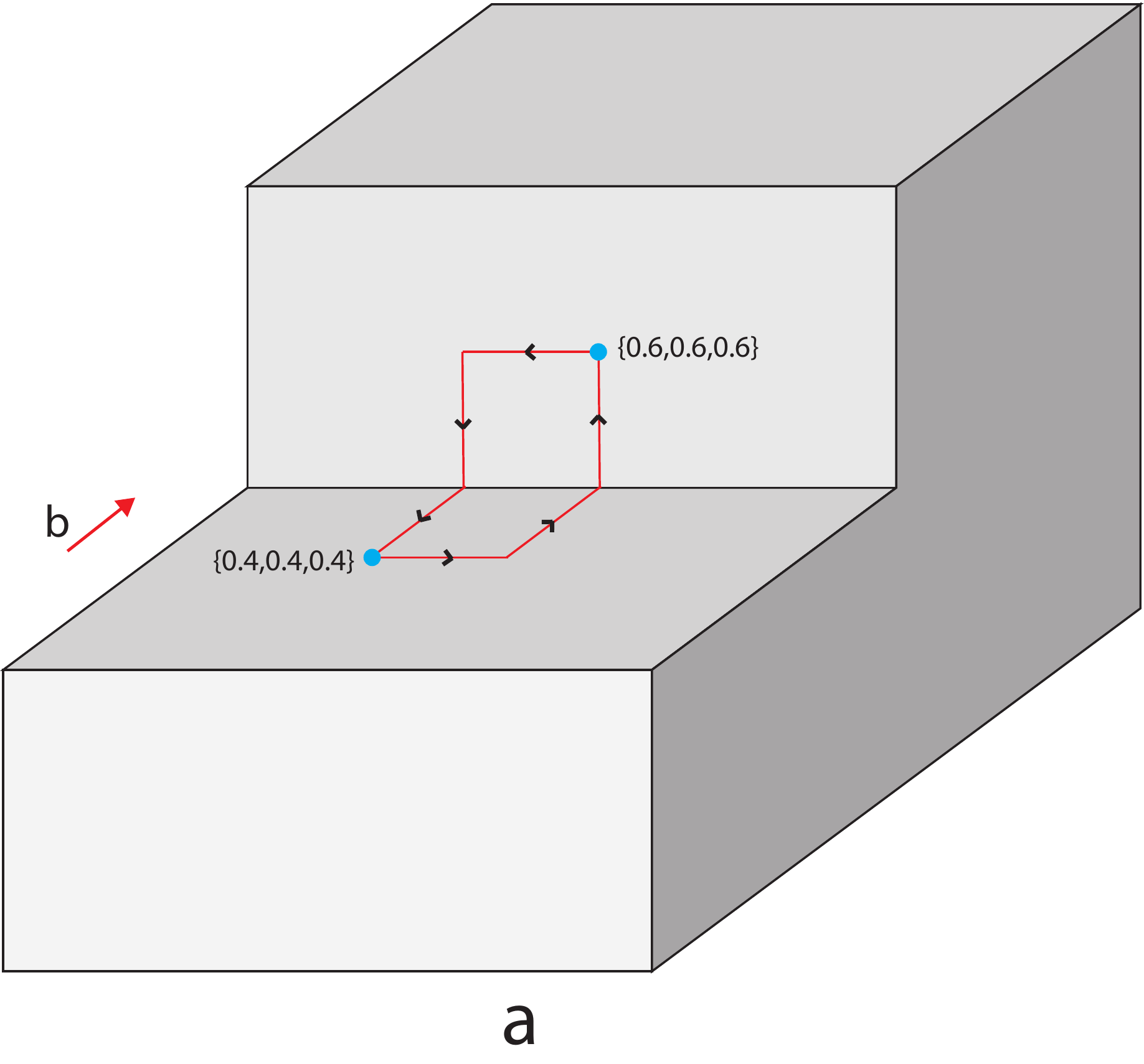}
\includegraphics[scale=0.3]{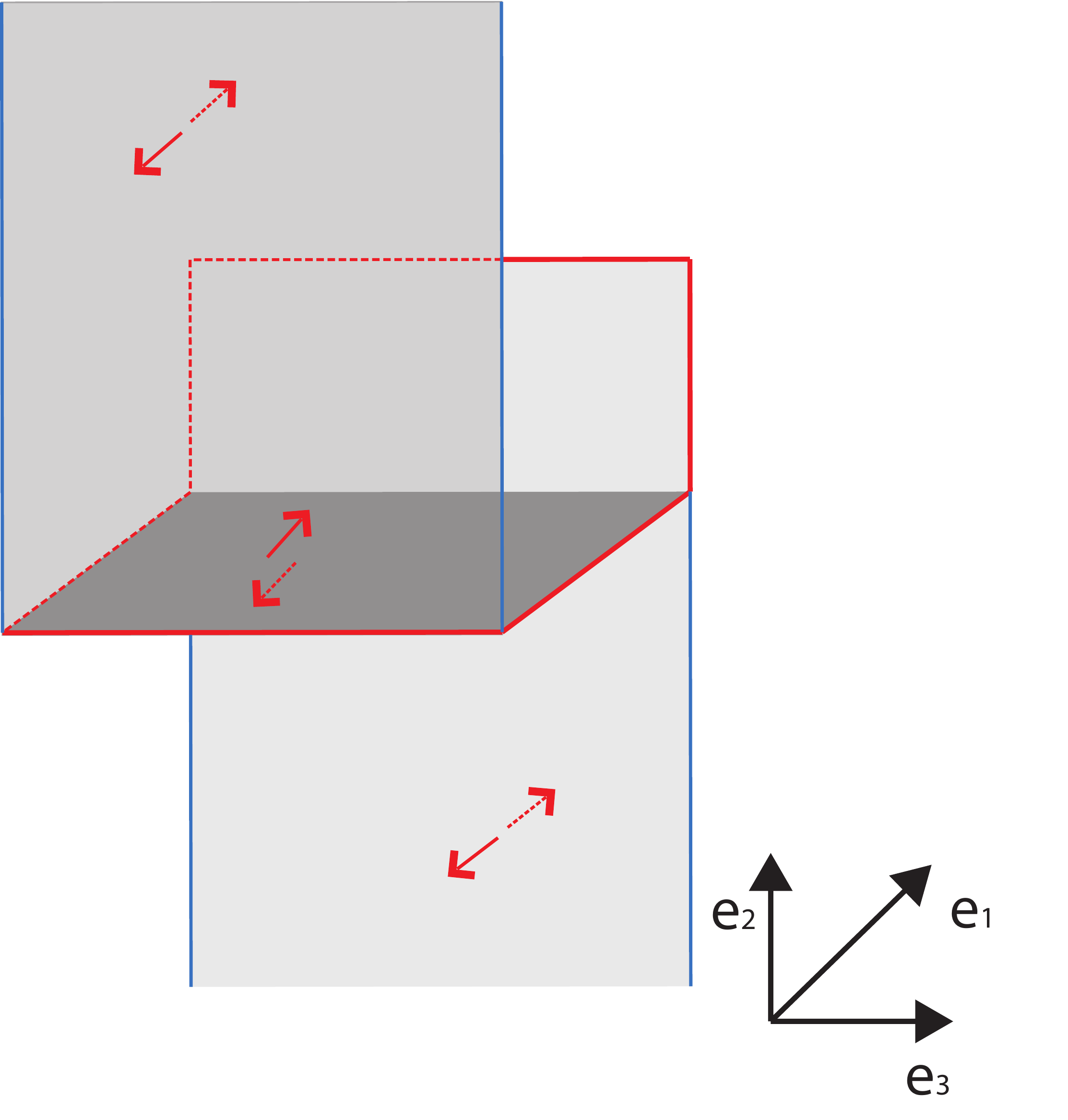}
\caption{Schematic of a non-planar dislocation loop with (a) the Burgers vector, edge and screw segments and (b) corresponding half planes with force dipoles. The two vertical half planes are semi-infinite.}
\label{fig:loop}
\end{figure}

The schematic of a non-planar dislocation loop in a finite domain is shown in Figure \ref{fig:loop} with the corresponding half-planes of force dipoles. The Burgers vector of the dislocation loop is $\bb = b\be_1$, where $b=0.001$. Each linear segment has length $0.2$, symmetrically located about the point $\bX = \{0.5,0.5,0.5\}$, with edge and screw segments discernible by the line direction relative to the Burgers vector. Dirichlet boundary conditions $\bu=0$ are applied at $X_2 = 0$, and the remaining surfaces are traction free. 

\begin{figure}[hbtp]
\centering
\includegraphics[scale=0.25]{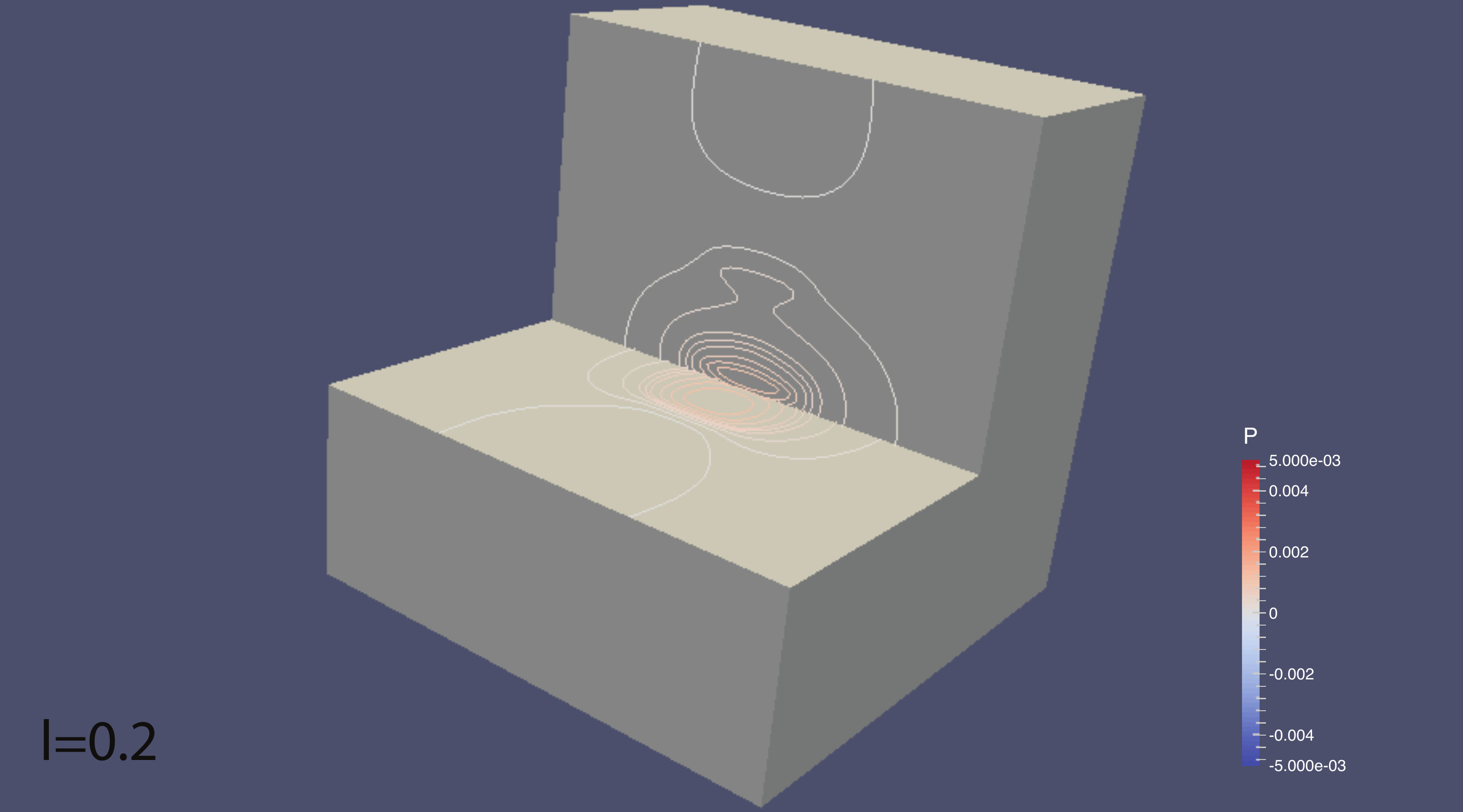}
\includegraphics[scale=0.25]{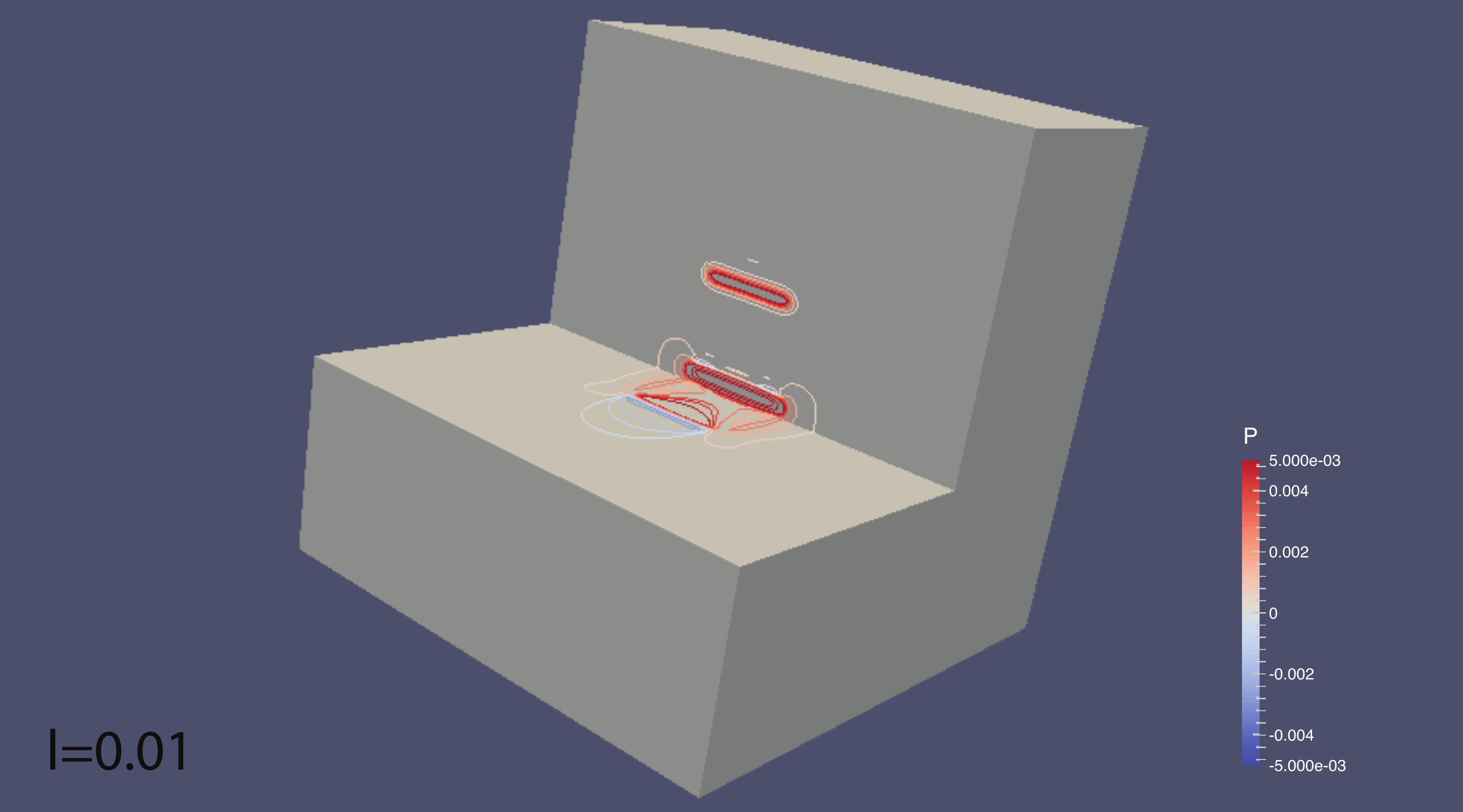}
\includegraphics[scale=0.25]{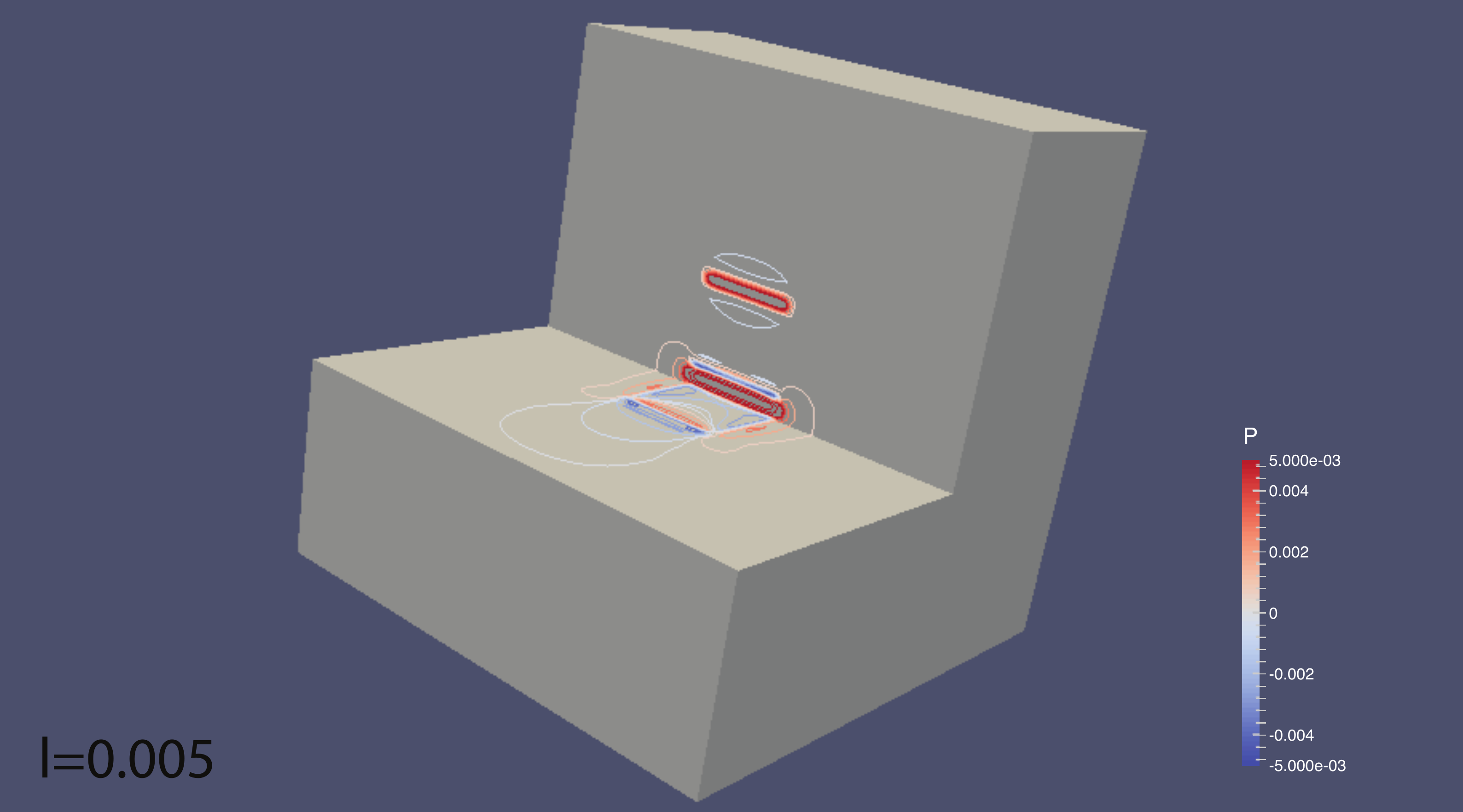}
\includegraphics[scale=0.25]{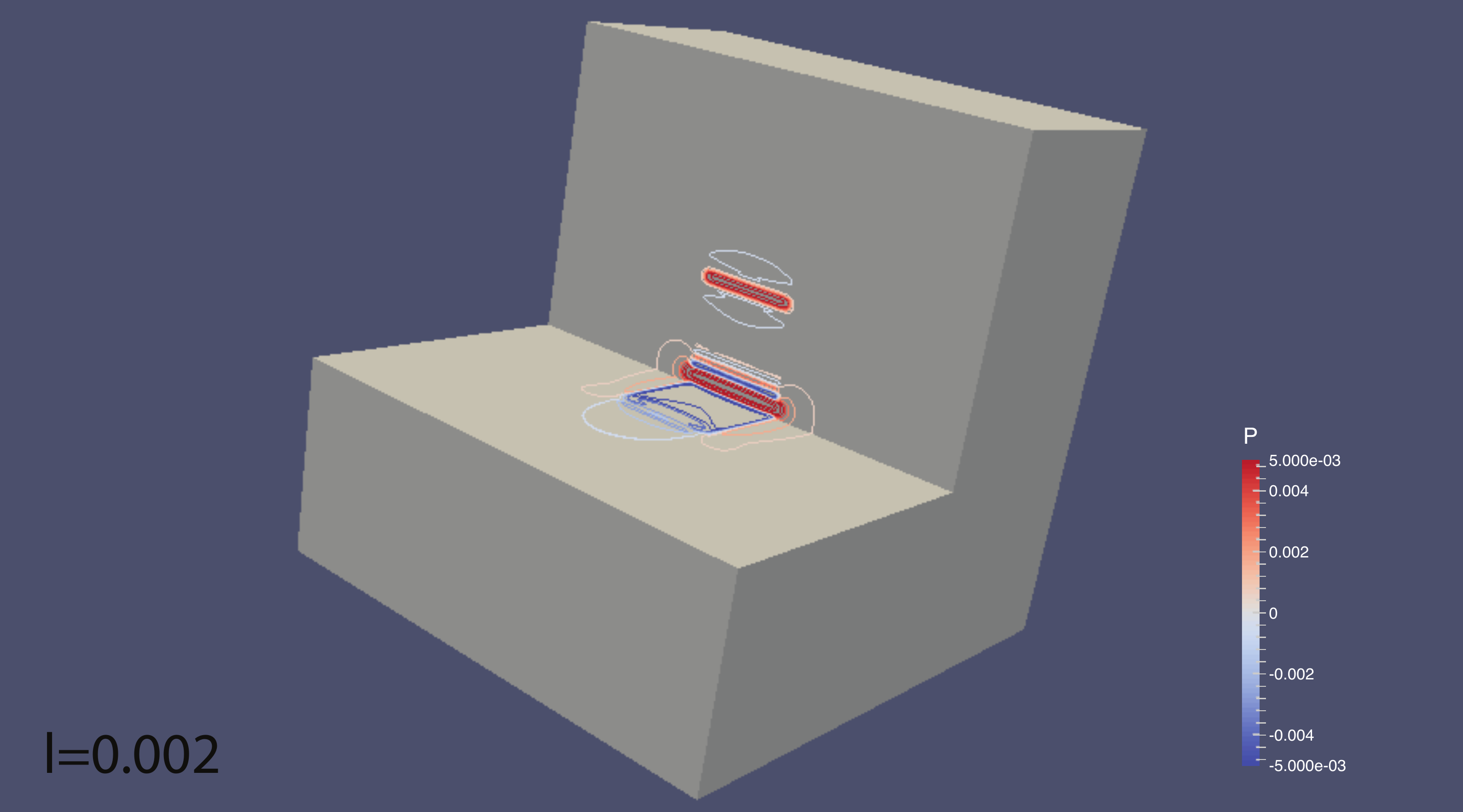}
\includegraphics[scale=0.25]{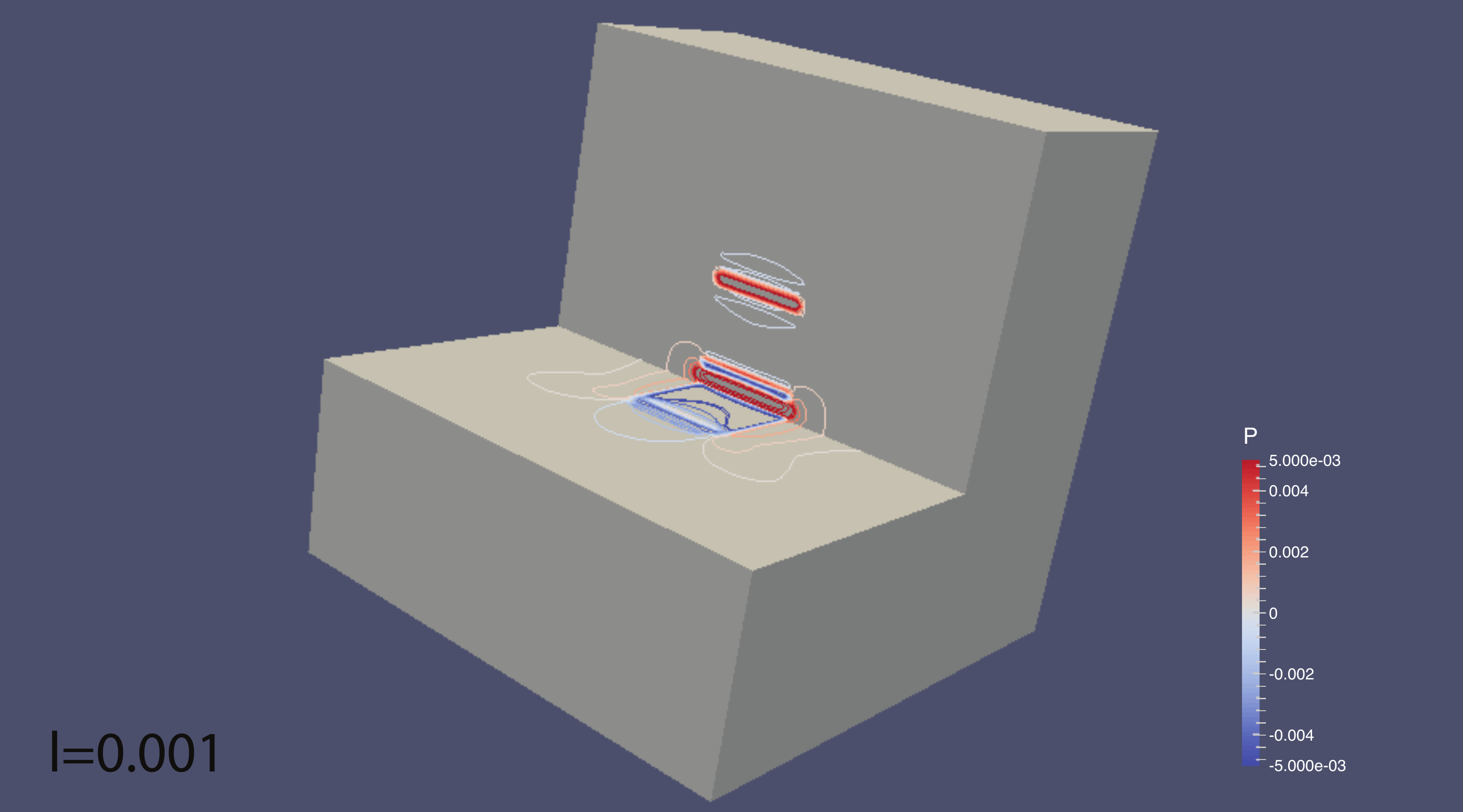}
\caption{Contours of $P_{12}$ induced by the non-planar dislocation loop.}
\label{fig:p12l2}
\end{figure}

Figure \ref{fig:p12l2} shows contours of the $P_{12}$ stress component created by the non-planar dislocation loop. Note the higher stresses induced for smaller values of gradient length scale $l$, due to the loss of regularization. This example is motivated by a similar computation by \citet{RoyAcharya2005} using a continuum theory based on the dislocation density tensor.

\subsection{Low angle grain boundary represented by a distribution of edge dislocations}
\label{sec:grainboundary}
\begin{figure}[hbtp]
\centering
\includegraphics[scale=0.5]{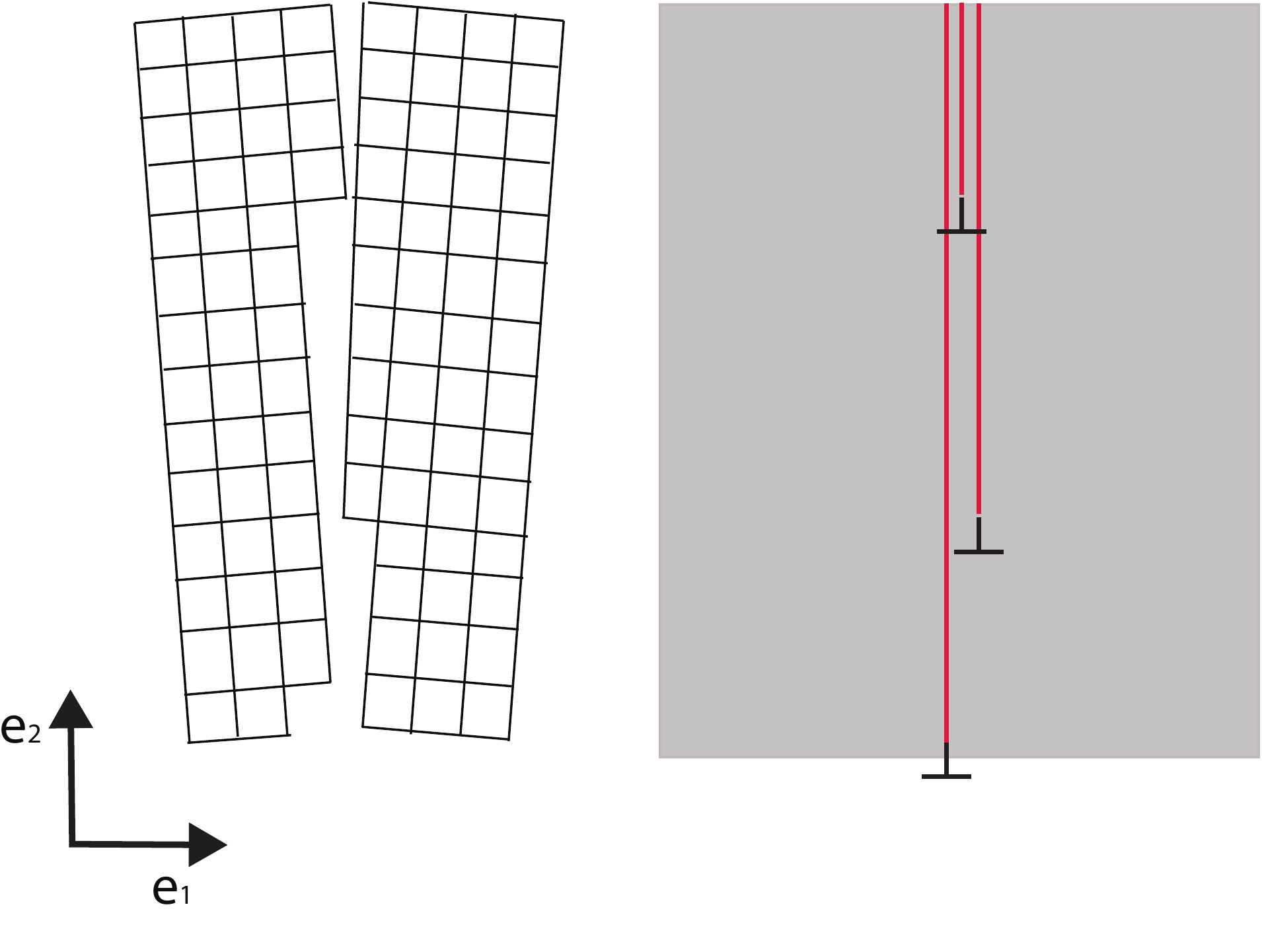}
\caption{(a) A distribution of dislocations modelling a low angle grain boundary. (b) Half planes (red) corresponding to the three edge dislocations.}
\label{fig:grain}
\end{figure}

\begin{figure}[hbtp]
\centering
\includegraphics[scale=0.5]{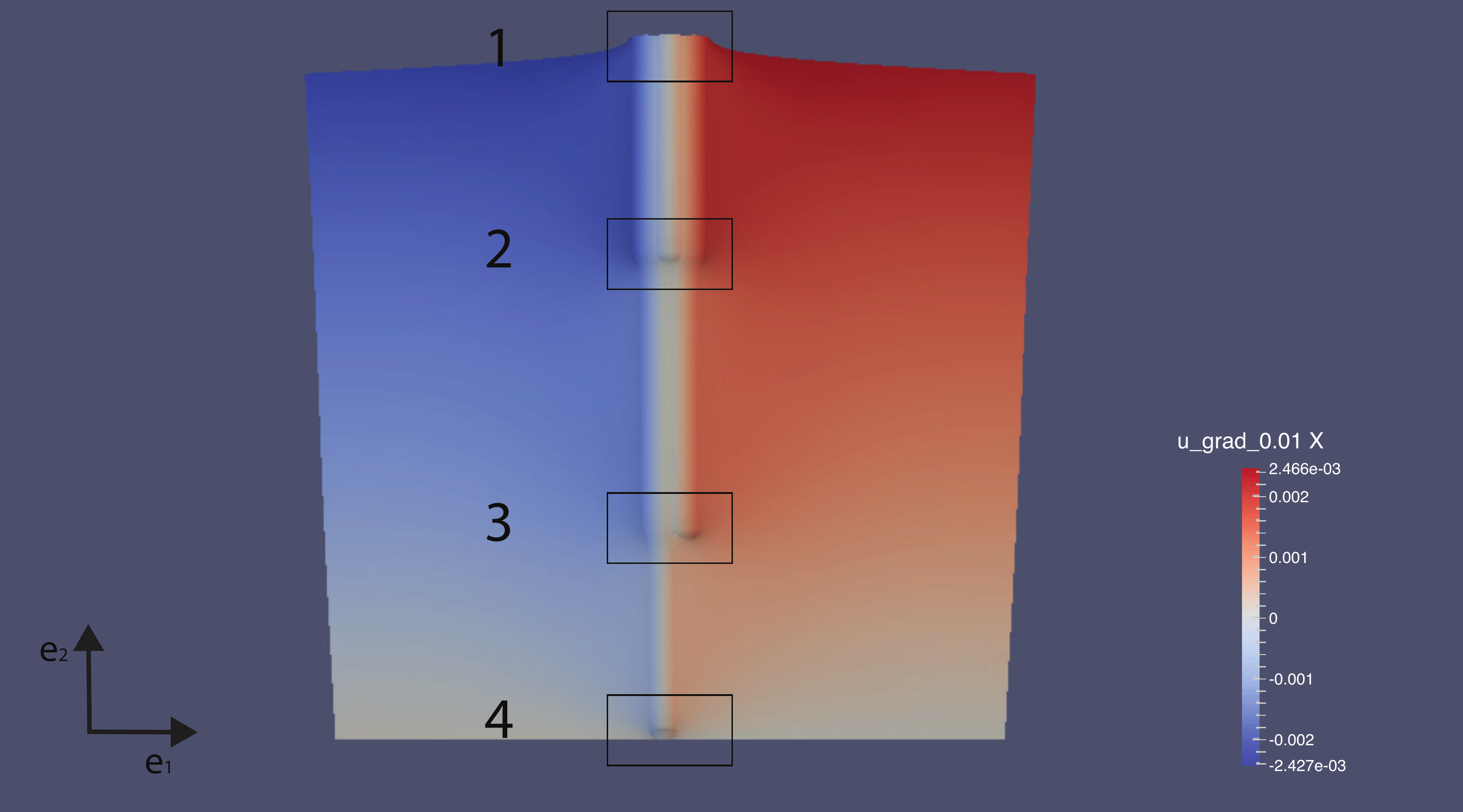}
\caption{The $u_{1}$ field in neighboring grains. The deformation has been scaled by a factor of 20 to make the grain misorientation apparent. Locations 1-4 are magnified in Figure \ref{fig:grainMesh}.}
\label{fig:grainFull}
\end{figure}

\begin{figure}[hbtp]
\centering
\includegraphics[scale=0.25]{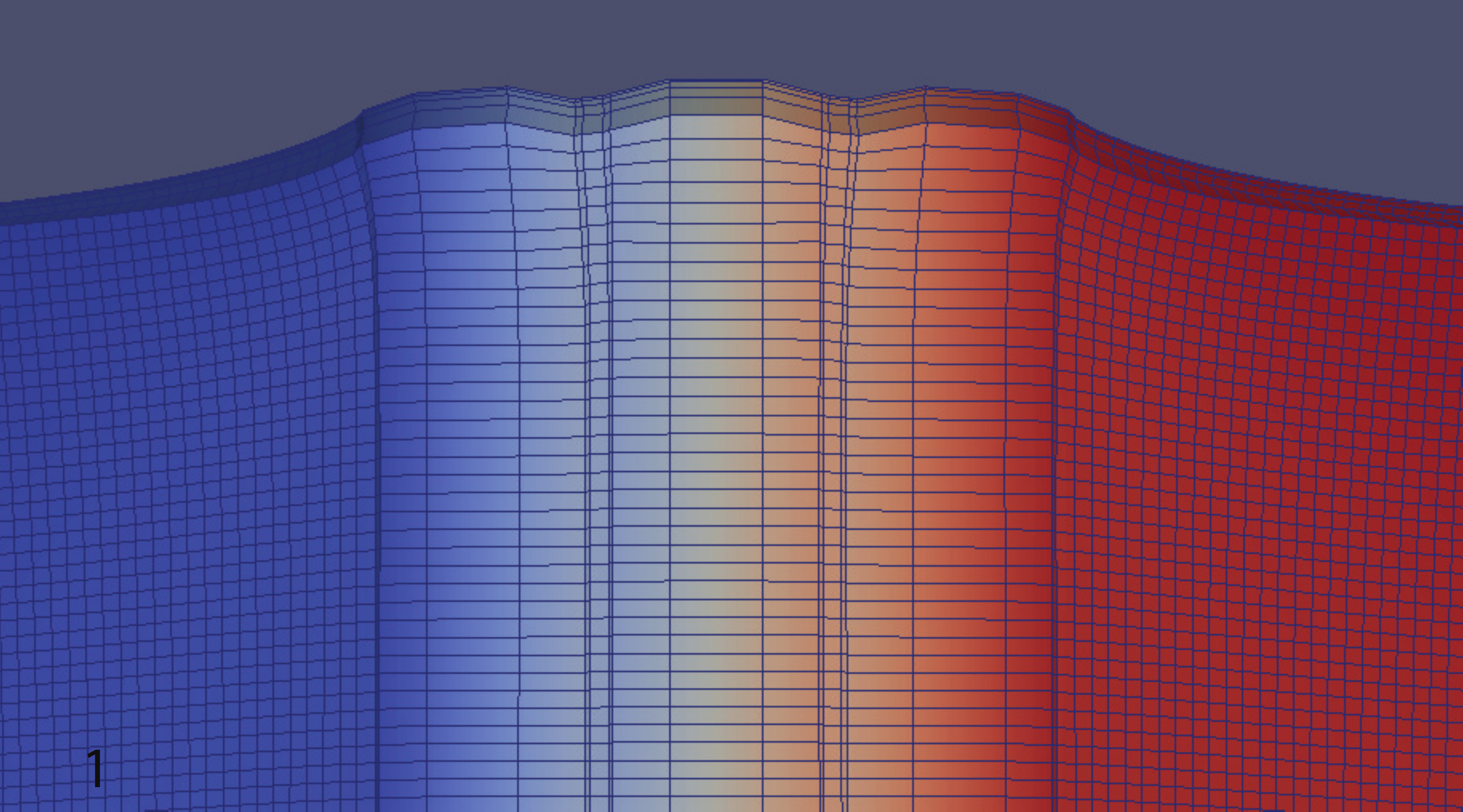}
\includegraphics[scale=0.25]{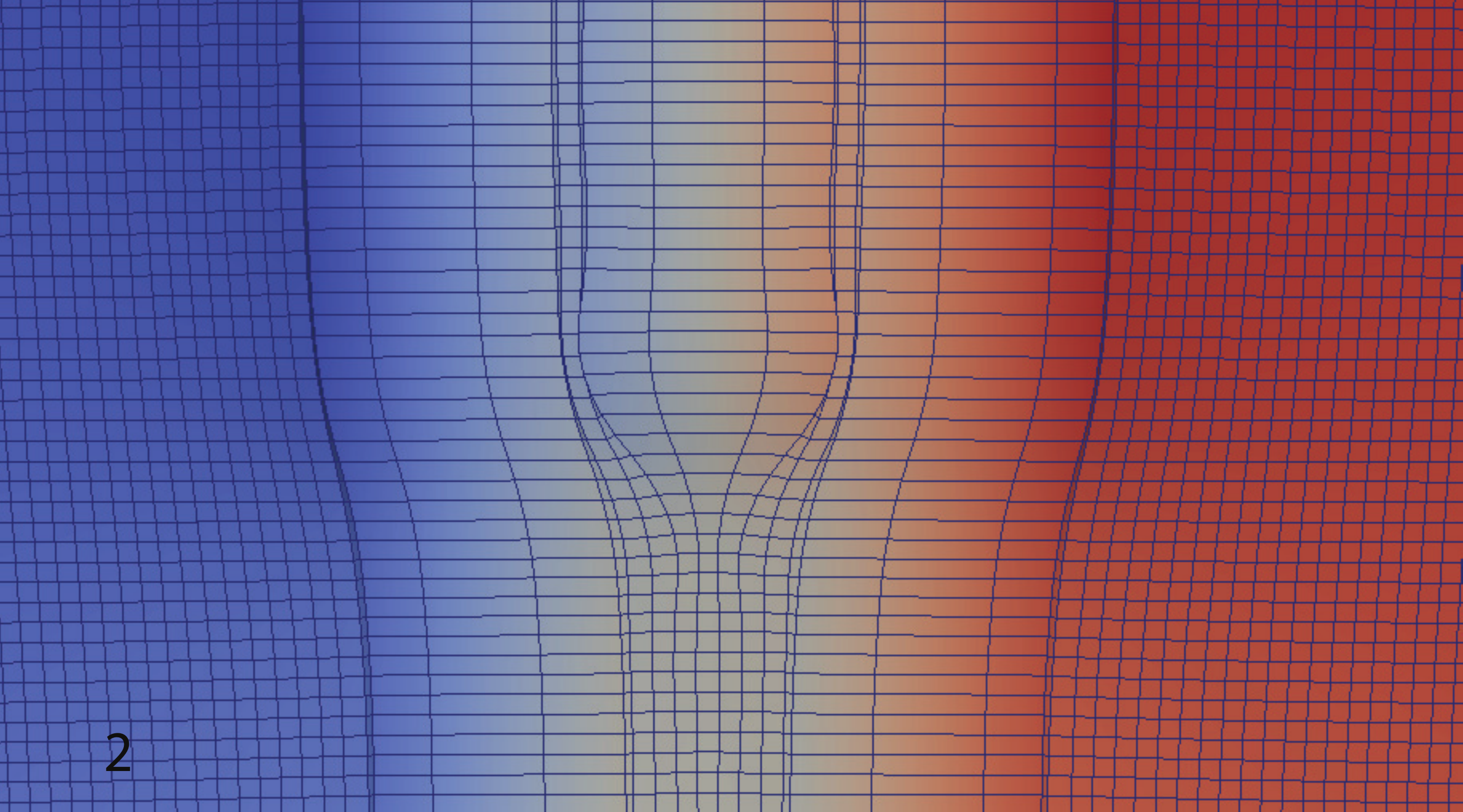}\\
\includegraphics[scale=0.25]{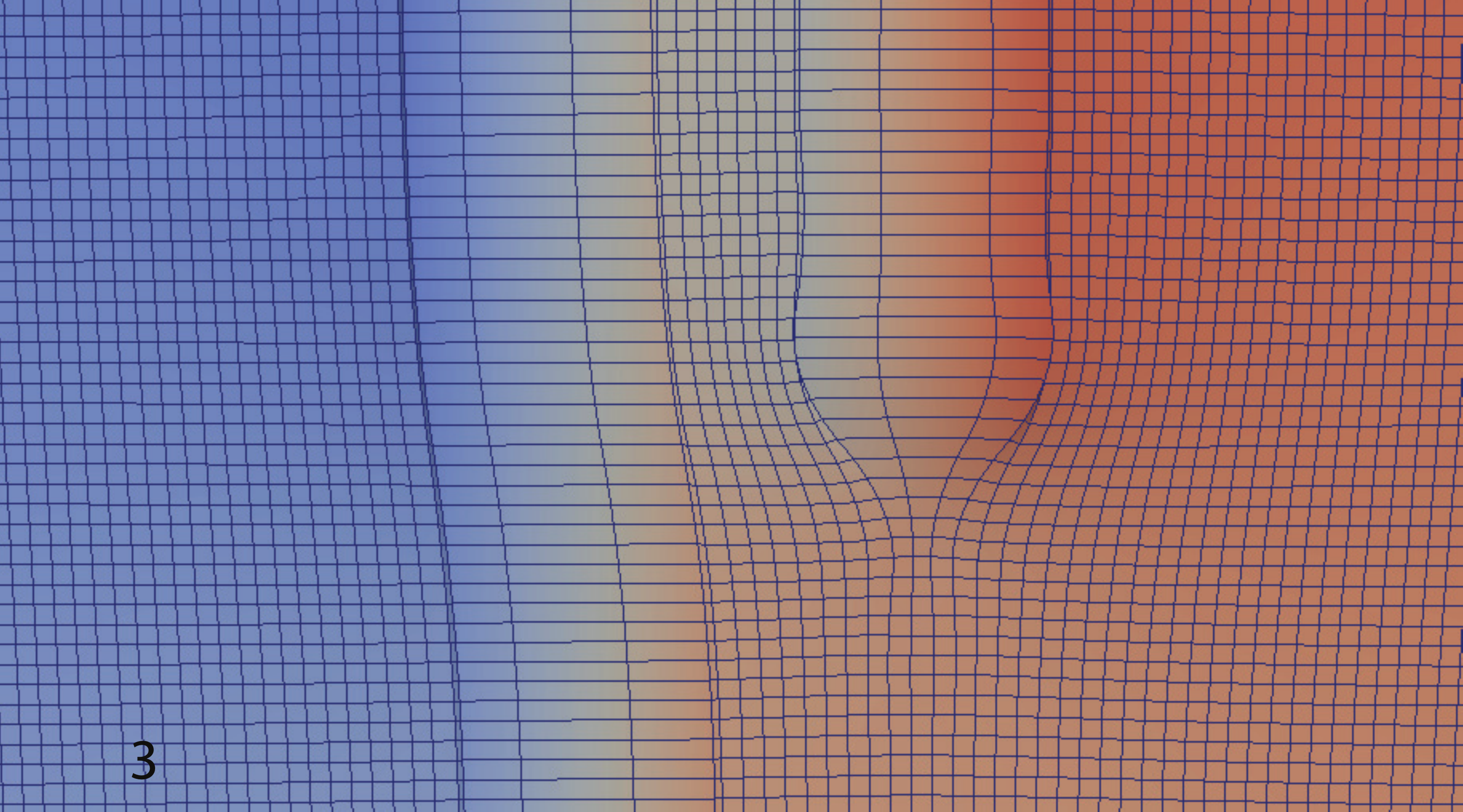}
\includegraphics[scale=0.25]{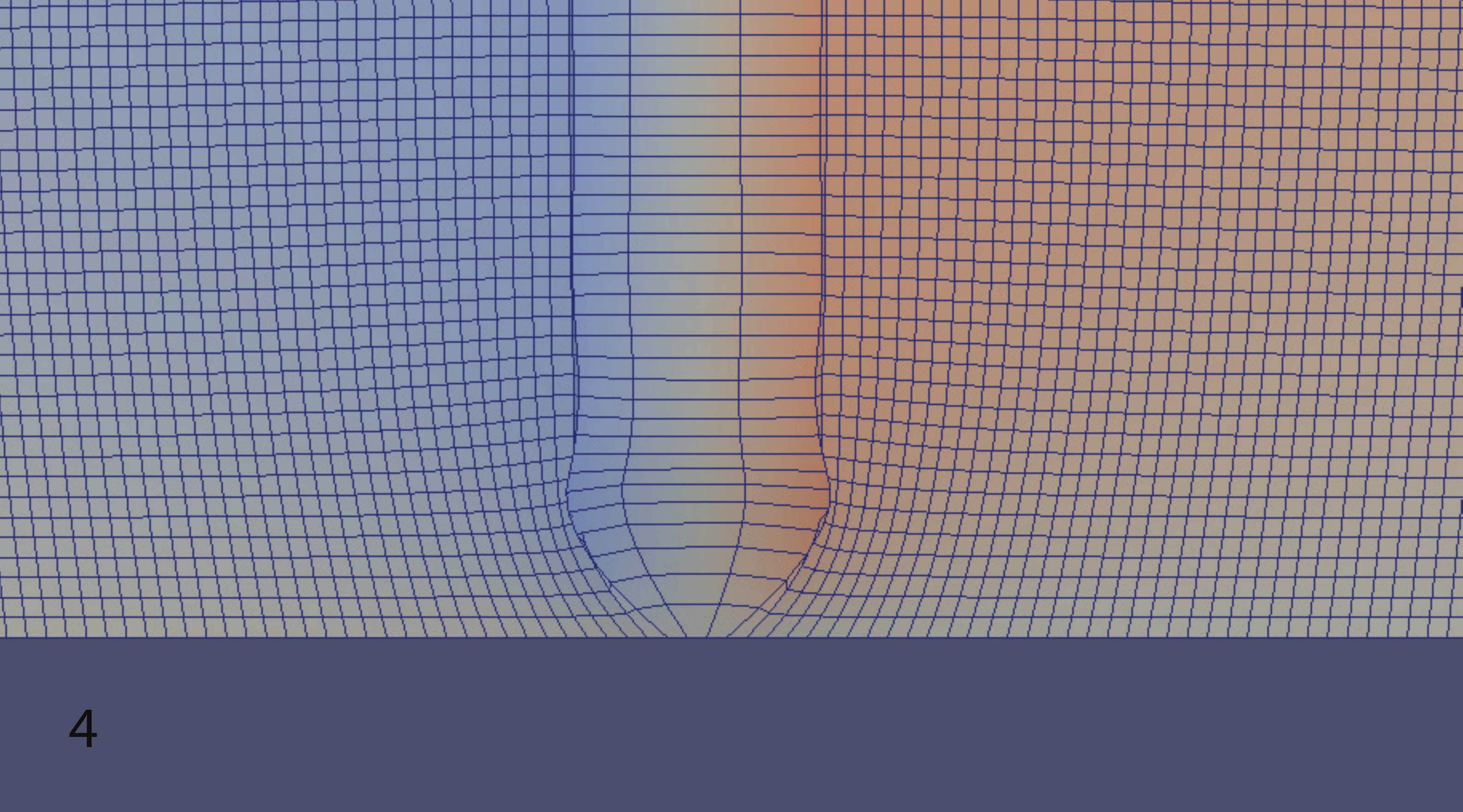}
\caption{The mesh lines demonstrate the distortion around the grain boundary. The deformation has been scaled by a factor of 10 for ease of visualization. The four sub-figures correspond to the locations in Figure \ref{fig:grainFull}.}
\label{fig:grainMesh}
\end{figure}

Figure \ref{fig:grain} is a schematic of a low angle grain boundary with a distribution of three edge dislocations to represent the misorientation. The dislocations are parallel, along $\be_3$, with cores located at $\{X_1,X_2\} = \{0.51,0.3\},\;\{0.5,0.7\}$ and $\{0.49,0\}$. The Burgers vector is $\bb = b\be_1$ , with $b = 0,0025$. The gradient length scale is $l = 0.0005$. Dirichlet boundary conditions $\bu=0$ are applied at $X_2 = 0$, and the remaining surfaces are traction free. The grain misorientation is shown in Figure \ref{fig:grainFull}, and the distortion around the grain boundary in Figure \ref{fig:grainMesh}.

\section{Discussion and conclusions}
\label{concl}
The crux of this work is that variationally based numerical methods are well-suited to representing the elastic fields of defects. The key, enabling insight is that the force dipoles used classically to model defect cores give rise to singular distributions, which can only be manipulated in the variational setting; i.e., in weak form. Our numerical results match well with analytic expressions for defect fields in the limit of classical (non-gradient) linearized elasticity, when the dipole tensor is suitably parameterized. The extension to nonlinear elasticity poses no special difficulty, and when gradient elasticity is included, the expected regularization of the elastic fields is obtained. This cluster of results establishes the viability of  representing defect fields in a fully numerical framework by modelling their cores as dipoles within distributional theory, and exploiting the variational framework. Beyond this, the interaction energy calculations indicate that the approach has potential for modelling defect interactions. Finally, the free surface, dislocation loop and grain boundary examples suggest that computations on more complex defect configurations will also prove successful.

It is important to note that by equating the core force dipole tensor with the effective force dipole tensor that emerges from linearized elasticity for point defects and dislocations, we have, in effect, matched those classical solutions. For the point defect, the comparison was made against the relaxation volume tensor, which is defined in the far-field. However, for dislocations, the dipole tensor was matched to the effective body force of Volterra's solution. Nevertheless, these are not the best fits, since the linearized elasticity solutions are poor representations in the core. In ongoing work we are exploring the extraction of the dipole tensor by comparison with density functional theory calculations of defect configurations. Another front on which we see potential for our approach is in computations involving large numbers of dislocations. The full elastic fields, including the influence of dislocation interactions will result from a single numerical solve without the need for multipole expansions. This could offset the computational expense of mesh resolution around defects that our method does require. Our method also works seamlessly to represent the interactions of point and line defects such as considered by \cite{YavariGoriely2014}. Even more attractively, these computations will be with gradient elasticity at finite strains. 

The evolution of large ensembles of dislocations is, of course, central to an understanding of mesoscopic material response. Our approach can be extended in this direction: The Peach-Koehler on a dislocation can be obtained via a variational calculation and emerges as a configurational force. The approach is similar to \cite{Steinmann2002}, with the additional feature of gradient elastic contributions. A formulation for dislocation dynamics would then be at hand with our formulation if the velocity of each dislocation were to be written via a first-order rate law driven by the Peach-Koehler force. The expense of multipole expansions could be avoided because a single numerical solve of the nonlinear elasticity problem would provide the driving forces on all dislocations, as already mentioned in the previous paragraph. 

There remain at least a few open questions in the context of our approach: (a) Can a variational calculation demonstrate a dipole tensor representing a single dislocation splitting into multiple dipoles, each representing partials? (b) The dipole tensor representation excludes any explicit representation of the separation of delta functions. For this reason, another open question is how we may model core spreading, such as in \cite{Zhang2015}. One possible approach would be to switch to an explicit representation of Dirac delta monopoles with a finite separation, and carry out a variational calculation seeking to minimize the free energy with respect to the separation as a configurational variable. (c) As suggested by a reviewer, there remains the problem of whether a solid passing from a stress-free state to a stressed state upon nucleation of a dislocation represents a global minimum of total free energy, or whether the non-convexity of realistic free energy functions implies that the dislocated solid is in a local, not global, minimum. In this regard, we have carried out calculations based on earlier work \citep{Rudrarajuetal2014}, which raise the possibility that non-convex free energy functions may allow such local, but not global minima of dislocated crystals. Another problem of interest is the interaction of defects with grain and phase boundaries. A continuum defect-based treatment was recently presented for this problem using dislocation and disclination densities \citep{Fressengeas2015}. A straightforward extension of our treatment would be possible in which the energy stored in grain boundary configurations such as those in Section \ref{sec:grainboundary} would interact with neighboring defects, thus repelling or attracting them. A similar treatment could also be established for diffuse interface models of phase boundaries based on the approach in \cite{Rudrarajuetal2015}, where strain-derived order parameters establish the change in crystal structure between phases. The elastic energy interactions between these interfaces and defects modelled as shown here could be a viable representation of coupling of defects with phase boundaries. The sharp interface problem would require different approaches to include interface energies.

We note that finite element computations of a full, field theory of the continuum dislocations have appeared in \cite{RoyAcharya2005}, including comparisons of stress fields and a number of dislocation configurations. Their formulation consists of the governing equilibrium equations, a divÐ-curl system for elastic incompatibility arising from the definition of the dislocation density tensor, and a first-order wave propagation equation for the evolution of the dislocation density. The incompatibility and wave propagation equations are solved with a least-squares finite element approach. Our formulation, based on the distributional dipole tensor and its rigorous variational treatment, may be considered somewhat simpler since a standard finite element method will do---at least for classical, non-gradient elasticity---with extra quadrature points. It is for the $C^1$ requirement of gradient elasticity that we invoke isogeometric analysis, although we have used it also for classical, non-gradient elasticity. 

\clearpage
\section*{Acknowledgements}
The mathematical formulation for this work was carried out under an NSF CDI Type I grant: CHE1027729 ``Meta-Codes for Computational Kinetics'', and an NSF DMREF grant: DMR1436154 ``DMREF: Integrated Computational Framework for Designing Dynamically Controlled Alloy-Oxide Heterostructures''. The numerical formulation and computations have been carried out as part of research supported by the U.S. Department of Energy, Office of Basic Energy Sciences, Division of Materials Sciences and Engineering under Award \#DE-SC0008637 that funds the PRedictive Integrated Structural Materials Science (PRISMS) Center at University of Michigan.

%
%
\bibliographystyle{elsart-harv}
\bibliography{references}

\begin{thebibliography}{39}
\expandafter\ifx\csname natexlab\endcsname\relax\def\natexlab#1{#1}\fi
\providecommand{\url}[1]{\texttt{#1}}
\providecommand{\href}[2]{#2}
\providecommand{\path}[1]{#1}
\providecommand{\DOIprefix}{doi:}
\providecommand{\ArXivprefix}{arXiv:}
\providecommand{\URLprefix}{URL: }
\providecommand{\Pubmedprefix}{pmid:}
\providecommand{\doi}[1]{\href{http://dx.doi.org/#1}{\path{#1}}}
\providecommand{\Pubmed}[1]{\href{pmid:#1}{\path{#1}}}
\providecommand{\bibinfo}[2]{#2}
\ifx\xfnm\relax \def\xfnm[#1]{\unskip,\space#1}\fi
\bibitem[{Acharya(2004)}]{Acharya2004}
\bibinfo{author}{Acharya, A.}, \bibinfo{year}{2004}.
\newblock \bibinfo{title}{Constitutive analysis of finite deformation field
  dislocation mechanics}.
\newblock \bibinfo{journal}{J. Mech. Phys. Sol.} \bibinfo{volume}{52},
  \bibinfo{pages}{301--316}.
\bibitem[{Acharya(2011)}]{Acharya2011}
\bibinfo{author}{Acharya, A.}, \bibinfo{year}{2011}.
\newblock \bibinfo{title}{Microcanonical entropy and mesoscale dislocation
  mechanics and plasticity}.
\newblock \bibinfo{journal}{J. Elast.} \bibinfo{volume}{104},
  \bibinfo{pages}{23--44}.
\bibitem[{Acharya and Fressegeas(2015)}]{Fressengeas2015}
\bibinfo{author}{Acharya, A.}, \bibinfo{author}{Fressegeas, C.},
  \bibinfo{year}{2015}.
\newblock \bibinfo{title}{Continuum mechanics of the interaction of phase
  boundaries and dislocations in solids}, in: \bibinfo{editor}{Chen, G.Q.G.},
  \bibinfo{editor}{Grinfield, M.}, \bibinfo{editor}{Knops, R.J.} (Eds.),
  \bibinfo{booktitle}{Differential Geometry and Continuum Mechanics}.
  \bibinfo{publisher}{Springer}. Springer Proceedings in Mathematics \&
  Statistics 137, pp. \bibinfo{pages}{123--1666}.
\bibitem[{Clouet(2011)}]{Clouet2011}
\bibinfo{author}{Clouet, E.}, \bibinfo{year}{2011}.
\newblock \bibinfo{title}{{Dislocation core field. I. Modeling in anisotropic
  linear elasticity theory}}.
\newblock \bibinfo{journal}{Phys. Reb. B} \bibinfo{volume}{84},
  \bibinfo{pages}{224111}.
\bibitem[{Clouet et~al.(2011)Clouet, Ventelon and Willaime}]{Clouetetal2011}
\bibinfo{author}{Clouet, E.}, \bibinfo{author}{Ventelon, L.},
  \bibinfo{author}{Willaime, F.}, \bibinfo{year}{2011}.
\newblock \bibinfo{title}{{Dislocation core field. II. Screw dislocation in
  iron}}.
\newblock \bibinfo{journal}{Phys. Rev. B} \bibinfo{volume}{84},
  \bibinfo{pages}{224107}.
\bibitem[{Cosserat and Cosserat(1909)}]{Cosserats1909}
\bibinfo{author}{Cosserat, E.}, \bibinfo{author}{Cosserat, F.},
  \bibinfo{year}{1909}.
\newblock \bibinfo{title}{Th\'{e}orie des corps deformables}.
\newblock \bibinfo{publisher}{Libraire Sci\'{e}ntifique A. Hermann et Fils,
  Paris}.
\bibitem[{Cottrell et~al.(2009)Cottrell, Hughes and
  Bazilevs}]{CottrellHughesBazilevs2009}
\bibinfo{author}{Cottrell, J.}, \bibinfo{author}{Hughes, T.},
  \bibinfo{author}{Bazilevs, Y.}, \bibinfo{year}{2009}.
\newblock \bibinfo{title}{Isogeometric Analysis: Toward Integration of CAD and
  FEA}.
\newblock \bibinfo{publisher}{Wiley, Chichester}.
\bibitem[{Engel et~al.(2002)Engel, Garikipati, Hughes, Larson, Mazzei and
  Taylor}]{Engeletal2002}
\bibinfo{author}{Engel, G.}, \bibinfo{author}{Garikipati, K.},
  \bibinfo{author}{Hughes, T.}, \bibinfo{author}{Larson, M.},
  \bibinfo{author}{Mazzei, L.}, \bibinfo{author}{Taylor, R.},
  \bibinfo{year}{2002}.
\newblock \bibinfo{title}{Continuous/discontinuous finite element
  approximations of fourth-order elliptic equations in structural and continuum
  mechanics with applications to thin beams and plates, and strain gradient
  elasticity}.
\newblock \bibinfo{journal}{Comp. Meth. App. Mech. Engrg.}
  \bibinfo{volume}{191}, \bibinfo{pages}{3669--3750}.
\bibitem[{Eringen and Edelen(1972)}]{EringenEdelen1972}
\bibinfo{author}{Eringen, A.}, \bibinfo{author}{Edelen, D.},
  \bibinfo{year}{1972}.
\newblock \bibinfo{title}{Nonlocal elasticity}.
\newblock \bibinfo{journal}{Int. J. Engr. Sci.} \bibinfo{volume}{10},
  \bibinfo{pages}{233}.
\bibitem[{Eshelby et~al.(1953)Eshelby, Read and Shockley}]{Eshelby1953}
\bibinfo{author}{Eshelby, J.}, \bibinfo{author}{Read, W.},
  \bibinfo{author}{Shockley, W.}, \bibinfo{year}{1953}.
\newblock \bibinfo{title}{Anisotropic elasticity with applications to
  dislocation theory}.
\newblock \bibinfo{journal}{Acta Met.} \bibinfo{volume}{1},
  \bibinfo{pages}{251}.
\bibitem[{Garikipati et~al.(2006)Garikipati, Falk, Bouville, Puchala and
  Narayanan}]{Garikipatietal2006}
\bibinfo{author}{Garikipati, K.}, \bibinfo{author}{Falk, M.},
  \bibinfo{author}{Bouville, M.}, \bibinfo{author}{Puchala, B.},
  \bibinfo{author}{Narayanan, H.}, \bibinfo{year}{2006}.
\newblock \bibinfo{title}{The continuum elastic and atomistic viewpoints on the
  formation volume and strain energy of a point defect}.
\newblock \bibinfo{journal}{J. Mech. Phys. Sol.} \bibinfo{volume}{54},
  \bibinfo{pages}{1929--1951}.
\bibitem[{Gehlen et~al.(1972)Gehlen, Hirth, Hoagland and
  Kannien}]{Gehlenetal1972}
\bibinfo{author}{Gehlen, P.}, \bibinfo{author}{Hirth, J.},
  \bibinfo{author}{Hoagland, R.}, \bibinfo{author}{Kannien, M.},
  \bibinfo{year}{1972}.
\newblock \bibinfo{title}{A new representation of the strain field associated
  with the cube-edge dislocation in a model of $\alpha$-iron}.
\newblock \bibinfo{journal}{J. App. Phys.} \bibinfo{volume}{43},
  \bibinfo{pages}{3921}.
\bibitem[{Gutkin and Aifantis(1999)}]{GutkinAifantis1999}
\bibinfo{author}{Gutkin, M.}, \bibinfo{author}{Aifantis, E.},
  \bibinfo{year}{1999}.
\newblock \bibinfo{title}{Dislocations and disclinations in gradient
  elasticity}.
\newblock \bibinfo{journal}{Phys. Stat. Sol.} \bibinfo{volume}{245},
  \bibinfo{pages}{214}.
\bibitem[{Hirth and Lothe(1973)}]{HirthLothe1973}
\bibinfo{author}{Hirth, J.}, \bibinfo{author}{Lothe, J.}, \bibinfo{year}{1973}.
\newblock \bibinfo{title}{Anisotropic elastic solutions for line defects in
  high-symmetry cases}.
\newblock \bibinfo{journal}{J. App. Phys.} \bibinfo{volume}{44},
  \bibinfo{pages}{1029--1032}.
\bibitem[{Hirth and Lothe(1982)}]{HirthLothe1982}
\bibinfo{author}{Hirth, J.}, \bibinfo{author}{Lothe, J.}, \bibinfo{year}{1982}.
\newblock \bibinfo{title}{Theory of dislocations}.
\newblock \bibinfo{publisher}{Krieger, Malabar, Florida}.
\bibitem[{Hughes et~al.(2005)Hughes, Cottrell and
  Bazilevs}]{HughesCottrellBazilevs2005}
\bibinfo{author}{Hughes, T.}, \bibinfo{author}{Cottrell, J.},
  \bibinfo{author}{Bazilevs, Y.}, \bibinfo{year}{2005}.
\newblock \bibinfo{title}{Isogeometricanalysis: Cad, finite elements, nurbs,
  exact geometry and mesh refinement}.
\newblock \bibinfo{journal}{Comp. Meth. App. Mech.} \bibinfo{volume}{194},
  \bibinfo{pages}{4135--4195}.
\bibitem[{Kessel(1970)}]{Kessel1970}
\bibinfo{author}{Kessel, S.}, \bibinfo{year}{1970}.
\newblock \bibinfo{title}{{Stress fields of a screw dislocation and and edge
  dislocation in Cosserat's continuum}}.
\newblock \bibinfo{journal}{Zeit. Ange. Math. Mech.} \bibinfo{volume}{50},
  \bibinfo{pages}{547}.
\bibitem[{Lazar and Maugin(2005)}]{LazarMaugin2005}
\bibinfo{author}{Lazar, M.}, \bibinfo{author}{Maugin, G.},
  \bibinfo{year}{2005}.
\newblock \bibinfo{title}{Nonsingular stress and strain fields of dislocations
  and disclinations in first strain gradient elasticity}.
\newblock \bibinfo{journal}{Int. J. Engr. Sci.} \bibinfo{volume}{43},
  \bibinfo{pages}{1157--1184}.
\bibitem[{Lazar et~al.(2005)Lazar, Maugin and Aifantis}]{Lazaretal2005}
\bibinfo{author}{Lazar, M.}, \bibinfo{author}{Maugin, G.},
  \bibinfo{author}{Aifantis, E.}, \bibinfo{year}{2005}.
\newblock \bibinfo{title}{On dislocations in a special class of generalized
  elasticity}.
\newblock \bibinfo{journal}{Phys. Stat. Sol.} \bibinfo{volume}{242},
  \bibinfo{pages}{2365--2390}.
\bibitem[{Lazar et~al.(2006)Lazar, Maugin and Aifantis}]{Lazaretal2006}
\bibinfo{author}{Lazar, M.}, \bibinfo{author}{Maugin, G.},
  \bibinfo{author}{Aifantis, E.}, \bibinfo{year}{2006}.
\newblock \bibinfo{title}{Dislocations in second gradient elasticity}.
\newblock \bibinfo{journal}{International Journal of Solids and Structures}
  \bibinfo{volume}{43}, \bibinfo{pages}{1787--1817}.
\bibitem[{Mindlin(1964)}]{Mindlin1964}
\bibinfo{author}{Mindlin, R.}, \bibinfo{year}{1964}.
\newblock \bibinfo{title}{Micro-structure in linear elasticity}.
\newblock \bibinfo{journal}{Archive for Rational Mechanics and Analysis}
  \bibinfo{volume}{16}, \bibinfo{pages}{51--78}.
\bibitem[{Molari et~al.(2006)Molari, Wells, Garikipati and
  Ubertini}]{Molarietal2006}
\bibinfo{author}{Molari, L.}, \bibinfo{author}{Wells, G.},
  \bibinfo{author}{Garikipati, K.}, \bibinfo{author}{Ubertini, F.},
  \bibinfo{year}{2006}.
\newblock \bibinfo{title}{A discontinuous galerkin method for strain
  gradient-dependent damage: Study of interpolations and convergence}.
\newblock \bibinfo{journal}{Comp. Meth. App. Mech. Engrg.}
  \bibinfo{volume}{195}, \bibinfo{pages}{1480--1498}.
\bibitem[{Papanicolopulos et~al.(2009)Papanicolopulos, Zervos and
  Vardoulakis}]{Papanicolopulos2009}
\bibinfo{author}{Papanicolopulos, S.}, \bibinfo{author}{Zervos, A.},
  \bibinfo{author}{Vardoulakis, I.}, \bibinfo{year}{2009}.
\newblock \bibinfo{title}{A three-dimensional c1 finite element for gradient
  elasticity}.
\newblock \bibinfo{journal}{Int. J. Numer. Meth. Engng.} \bibinfo{volume}{77},
  \bibinfo{pages}{1396?1415}.
\bibitem[{Piegl and Tiller(1997)}]{Piegl1997}
\bibinfo{author}{Piegl, L.}, \bibinfo{author}{Tiller, W.},
  \bibinfo{year}{1997}.
\newblock \bibinfo{title}{The NURBS book (2nd ed.)}.
\newblock \bibinfo{publisher}{Springer-Verlag New York, Inc.},
  \bibinfo{address}{New York, NY, USA}.
\bibitem[{Rosakis and Rosakis(1988)}]{Rosakis1988}
\bibinfo{author}{Rosakis, P.}, \bibinfo{author}{Rosakis, A.J.},
  \bibinfo{year}{1988}.
\newblock \bibinfo{title}{The screw dislocation problem in incompressible
  finite elastostatics: a discussion of nonlinear effects}.
\newblock \bibinfo{journal}{J. Elast.} \bibinfo{volume}{20},
  \bibinfo{pages}{3--40}.
\bibitem[{Roy and Acharya(2005)}]{RoyAcharya2005}
\bibinfo{author}{Roy, A.}, \bibinfo{author}{Acharya, A.}, \bibinfo{year}{2005}.
\newblock \bibinfo{title}{Finite element approximation of field dislocation
  mechanics}.
\newblock \bibinfo{journal}{J. Mech. Phys. Sol.} \bibinfo{volume}{53},
  \bibinfo{pages}{143--170}.
\bibitem[{Rudraraju et~al.(2015)Rudraraju, Van~der Ven and
  Garikipati}]{Rudrarajuetal2015}
\bibinfo{author}{Rudraraju, S.}, \bibinfo{author}{Van~der Ven, A.},
  \bibinfo{author}{Garikipati, K.}, \bibinfo{year}{2015}.
\newblock \bibinfo{title}{Mechano-chemical spinodal decomposition: {A}
  phenomenological theory of phase transformations in multi-component
  crystalline solids}.
\newblock \bibinfo{journal}{arXiv:1508.05930}
  \href{http://arxiv.org/abs/1508.05930}{{\tt arXiv:1508.05930}}.
  \bibinfo{note}{in review}.
\bibitem[{Rudraraju et~al.(2014)Rudraraju, der Ven and
  Garikipati}]{Rudrarajuetal2014}
\bibinfo{author}{Rudraraju, S.}, \bibinfo{author}{der Ven, A.V.},
  \bibinfo{author}{Garikipati, K.}, \bibinfo{year}{2014}.
\newblock \bibinfo{title}{Three-dimensional isogeometric solutions to general
  boundary value problems of toupinÕs gradient elasticity theory at finite
  strains}.
\newblock \bibinfo{journal}{Comp. Meth. App Mech. Engr.} \bibinfo{volume}{278},
  \bibinfo{pages}{705--728}.
\bibitem[{Sinclair et~al.(1978)Sinclair, Gehlen, Hoagland and
  Hirth}]{SinclairHirth1978}
\bibinfo{author}{Sinclair, J.}, \bibinfo{author}{Gehlen, P.},
  \bibinfo{author}{Hoagland, R.}, \bibinfo{author}{Hirth, J.P.},
  \bibinfo{year}{1978}.
\newblock \bibinfo{title}{Flexible boundary conditions and nonlinear geometric
  effects in atomic dislocation modeling}.
\newblock \bibinfo{journal}{J. App. Phys.} \bibinfo{volume}{49},
  \bibinfo{pages}{3890--3897}.
\bibitem[{Stakgold(1979)}]{Stakgold1979}
\bibinfo{author}{Stakgold, I.}, \bibinfo{year}{1979}.
\newblock \bibinfo{title}{Green's functions and boundary value problems}.
\newblock \bibinfo{publisher}{Wiley, North America}.
\bibitem[{Steinmann(2002)}]{Steinmann2002}
\bibinfo{author}{Steinmann, P.}, \bibinfo{year}{2002}.
\newblock \bibinfo{title}{On spatial and material settings of hyperelastostatic
  crystal defects}.
\newblock \bibinfo{journal}{J. Mech. Phys. Sol.} \bibinfo{volume}{50},
  \bibinfo{pages}{1743--1766}.
\bibitem[{Toupin(1962)}]{Toupin1962}
\bibinfo{author}{Toupin, R.}, \bibinfo{year}{1962}.
\newblock \bibinfo{title}{Elastic materials with couple-stresses}.
\newblock \bibinfo{journal}{Archive for Rational Mechanics and Analysis}
  \bibinfo{volume}{11}, \bibinfo{pages}{385--414}.
\bibitem[{Toupin(1964)}]{Toupin1964}
\bibinfo{author}{Toupin, R.A.}, \bibinfo{year}{1964}.
\newblock \bibinfo{title}{Theories of elasticity with couple stresses}.
\newblock \bibinfo{journal}{Arch. Rational Mech. Anal.} \bibinfo{volume}{17},
  \bibinfo{pages}{85--112}.
\bibitem[{Volterra(1907)}]{Volterra1907}
\bibinfo{author}{Volterra, V.}, \bibinfo{year}{1907}.
\newblock \bibinfo{title}{Sur l'\'{e}quilibre des corps \'{e}lastiques
  multiplement connex\'{e}s}.
\newblock \bibinfo{journal}{Ann. Sci\'{e}tif. \'{E}cole Norm. Sup.}
  \bibinfo{volume}{24}, \bibinfo{pages}{401--517}.
\bibitem[{Wells et~al.(2004)Wells, Garikipati and Molari}]{Wellsetal2004}
\bibinfo{author}{Wells, G.}, \bibinfo{author}{Garikipati, K.},
  \bibinfo{author}{Molari, L.}, \bibinfo{year}{2004}.
\newblock \bibinfo{title}{A discontinuous galerkin formulation for a strain
  gradient-dependent damage model}.
\newblock \bibinfo{journal}{Comp. Meth. App. Mech. Engrg.}
  \bibinfo{volume}{193}, \bibinfo{pages}{3633--3645}.
\bibitem[{Wells et~al.(2006)Wells, Kuhl and
  Garikipati}]{WellsKuhlGarikipati2006}
\bibinfo{author}{Wells, G.}, \bibinfo{author}{Kuhl, E.},
  \bibinfo{author}{Garikipati, K.}, \bibinfo{year}{2006}.
\newblock \bibinfo{title}{A discontinuous galerkin method for the cahn hilliard
  equation}.
\newblock \bibinfo{journal}{J. Comp. Phys.} \bibinfo{volume}{218},
  \bibinfo{pages}{860--877}.
\bibitem[{Yavari and Goriely(2012)}]{YavariGoriely2012}
\bibinfo{author}{Yavari, A.}, \bibinfo{author}{Goriely, A.},
  \bibinfo{year}{2012}.
\newblock \bibinfo{title}{RiemannÐcartan geometry of nonlinear dislocation
  mechanics}.
\newblock \bibinfo{journal}{Arch. Rational Mech. Anal.} \bibinfo{volume}{205},
  \bibinfo{pages}{59--118}.
\bibitem[{Yavari and Goriely(2013)}]{YavariGoriely2013}
\bibinfo{author}{Yavari, A.}, \bibinfo{author}{Goriely, A.},
  \bibinfo{year}{2013}.
\newblock \bibinfo{title}{Nonlinear elastic inclusions in isotropic solids}.
\newblock \bibinfo{journal}{Proc. Roy. Soc. A} \bibinfo{volume}{469},
  \bibinfo{pages}{20130415}.
\bibitem[{Zhang et~al.(2015)Zhang, Acharya, Wilkington and Bielak}]{Zhang2015}
\bibinfo{author}{Zhang, X.}, \bibinfo{author}{Acharya, A.},
  \bibinfo{author}{Wilkington, N.J.}, \bibinfo{author}{Bielak, J.},
  \bibinfo{year}{2015}.
\newblock \bibinfo{title}{A single theory for some quasi-static, supersonic,
  atomic, and tectonic scale applications of dislocations}.
\newblock \bibinfo{journal}{J. Mech. Phys. Sol.} \bibinfo{volume}{84},
  \bibinfo{pages}{145--195}.

\end{thebibliography}
\end{document}